\documentclass[superscriptaddress,letter]{revtex4}%
\usepackage{amssymb}
\usepackage{color}
\usepackage{graphicx}
\usepackage{dcolumn}
\usepackage{bm}
\usepackage[header,title,page,titletoc]{appendix}
\usepackage{amsmath}
\usepackage{amsfonts}%
\providecommand{\U}[1]{\protect \rule{.1in}{.1in}}
\begin{document}
\title{Kelvin Wave and Knot Dynamics on Entangled Vortices}
\author{Su-Peng Kou}
\thanks{Corresponding author}
\email{spkou@bnu.edu.cn}
\affiliation{Center for Advanced Quantum Studies, Department of Physics, Beijing Normal
University, Beijing, 100875, China}

\begin{abstract}
In this paper, starting from Biot-Savart mechanics for entangled
vortex-membranes, a new theory -- knot physics is developed to explore the
underlying physics of quantum mechanics. Owning to the conservation conditions
of the volume of knots on vortices in incompressible fluid, the shape of knots
will never be changed and the corresponding Kelvin waves cannot evolve
smoothly. Instead, the knot can only split. The knot-pieces evolves following
the equation of motion of Biot-Savart equation that becomes Schr\"{o}dinger
equation for probability waves of knots. The classical functions for Kelvin
waves become wave-functions for knots. The effective theory of perturbative
entangled vortex-membranes becomes a traditional model of relativistic quantum
field theory -- a massive Dirac model. As a result, this work would help
researchers to understand the mystery in quantum mechanics.

\end{abstract}
\maketitle

\section{Introduction}

One hundred years ago, Kelvin (Sir W. Thomson) studied the physical properties
of vortex-lines that consist of the rotating motion of fluid around a common
centerline. In an incompressible fluid, the vorticity of vortex-lines
manifests itself in the circulation $%
%TCIMACRO{\doint }%
%BeginExpansion
{\displaystyle \oint}
%EndExpansion
\mathbf{v}\cdot d\mathbf{l}=\kappa$ where $\kappa$ is a constant. For
classical hydrodynamic vortex-lines, Kelvin found a transverse and circularly
polarized wave\cite{Thomson1880a} (called Kelvin wave), in which the
vortex-lines twist around their equilibrium position forming a helical
structure\cite{Donnelly1991a}. For two vortex-lines, owing to the nonlocal
interaction, the leapfrogging motion has been predicted in classical fluids
from the works of Helmholtz and Kelvin\cite{dys93, hic22, bor13,wac14, cap14}.
Owing to Kolmogorov-like turbulence\cite{S95,V00}, a variety of approaches
have been used to study this phenomenon. In experiments, Kelvin waves has been
observed in uniformly rotating $^{4}$He superfluid (SF) \cite{hall1} and
Bose-Einstein condensates \cite{Bretin2003a}.

In addition, Kelvin and Toit tried to develop an early atomic theory that is
linked to the existence and dynamics of knotted vortex-rings in ether.
However, they failed -- the fundamental structure of atoms is irrelevant to
knots. Today, the failure reason is clear. The elementary particles in our
universe are not classical knots but quantum objects obeying quantum mechanics
and Einstein's relativity. Quantum mechanics (also known as quantum physics or
quantum theory) is a fundamental branch of modern physics that was proposed
from Planck's solution in 1900 to the black-body radiation problem and
Einstein's 1905 paper which offered a quantum-based theory to explain the
photoelectric effect. After several decades, quantum mechanics is established
and becomes a successful theory that agrees very well with experiments and
provides an accurate description of the dynamic behavior of microcosmic
objects. In quantum mechanics, the energy is quantized and can only change by
discrete amounts, i.e. $E=\hslash \omega$ where $\hslash$ is Planck constant.
There are several fundamental principles in quantum mechanics: wave-particle
duality, uncertainty principle, and superposition principle. The
Schr\"{o}dinger equation describes how wave-functions evolve, playing a role
similar to Newton's second law in classical mechanics. In quantum mechanics,
when considering special relativity, the Schr\"{o}dinger equation is replaced
by the Dirac equation.

However, quantum mechanics is far from being well understood. There are a lot
of unsolved mysteries in quantum mechanics including quantum entanglement
problem and quantum measurement problem. Einstein said, "\emph{Quantum
mechanics is certainly imposing. But an inner voice tells me that it is not
yet the real thing. The theory says a lot, but does not really bring us any
closer to the secret of the 'old one'. I, at any rate, am convinced that He
does not throw dice.}" The exploration of the underlying physics of quantum
mechanics is going on since its establishment\cite{jammer}. There are a lot of
attempts, such as De Broglie's pivot-wave theory\cite{de brogile}, the Bohmian
mechanics\cite{Bohm1}, the many-world theory\cite{many}, the Nelsonian
Mechanics\cite{nelson} and the idea of primary state diffusion\cite{pri}.
However, for all these interpretations of quantum mechanics, people focus on
the issue of quantum motion but miss another important issue, \emph{what is
matter (or reality)}?

In this paper, we develop a new theory towards understanding quantum mechanics
based on three-dimensional (3D) leapfrogging vortex-membranes in
five-dimensional (5D) incompressible fluid. We call the new theory --
\emph{knot physics}. According to knot physics, it is the 3D quantum Dirac
model that characterizes the knot dynamics of leapfrogging vortex-membranes.
The knot physics gives a complete interpretation on quantum mechanics. Fig.1
is an illustration of the framework structure of knot physics.

\begin{figure}[ptb]
\includegraphics[clip,width=0.75\textwidth]{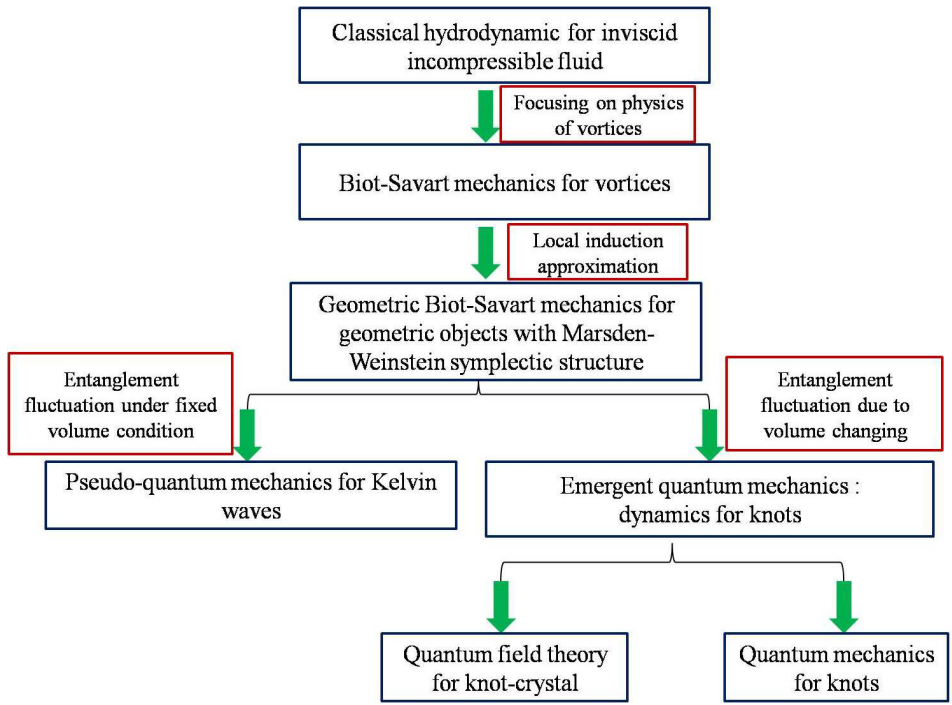}\caption{The framework
structure of knot physics}%
\end{figure}

The paper is organized as below. In Sec. II, we review Biot-Savart mechanics
for 3D vortex-membrane in a 5D incompressible fluid. In Sec. III, the Kelvin
waves of 3D vortex-membrane are studied. In this section, the Biot-Savart
mechanics for helical vortex-membranes is mapped onto "pseudo-quantum
mechanics" for a particle, by which we apply to the calculation of the
evolution of a deformed helical vortex-membrane. In Sec. IV, we develop the
Biot-Savart mechanics for entangled vortex-membranes. In addition, we
introduce tensor representation to describe the Kelvin waves. In Sec. V, we
develop an emergent quantum mechanics to describe the knot dynamics on
knot-crystal. In particular, a 3D massive Dirac model is obtained to
characterize the knot dynamics on the entangled vortex-membranes. We also
address the measurement theory and quantum entanglement of emergent quantum
mechanics. The emergent quantum mechanics help us understand the mysteries of
quantum mechanics. Finally, the conclusions are drawn in Sec. VI.

\section{Biot-Savart mechanics for vortices}

\subsection{The Euler equation on vorticity}

In this paper, we would study the vortex in an inviscid incompressible fluid.
In condensed matter physics, we may consider superfluid (SF) as an inviscid
incompressible fluid. For a divergence-free inviscid incompressible fluid
($\nabla \cdot \mathbf{v}=0$), the fluid motion is described by the classical
Euler equation:
\begin{equation}
\partial_{t}\mathbf{v}+\mathbf{\nabla}_{v}\mathbf{v}=-\frac{1}{\rho
}\mathbf{\nabla}p
\end{equation}
where $\mathbf{v}$ is the fluid velocity field, $\rho$ is the uniform density
and $p$ is the pressure function. $\mathbf{\nabla}_{v}\mathbf{v}$ is
Riemannian covariant derivative of the field $\mathbf{v}$ in the direction of
itself. On the other hand, to characterize vorticity, the Helmholtz form of
the Euler equation is written as
\begin{equation}
\partial_{t}\mathbf{\Omega}+L_{v}\mathbf{\Omega}=0\,,
\end{equation}
where $L_{v}$ is the Lie derivative along the velocity field $\mathbf{v}$ and
$\mathbf{\Omega}=\mathbf{\nabla}\times \mathbf{v}$ is the vorticity field.

Vortices are extended objects with singular vorticity in an inviscid
incompressible fluid that can be regarded as a closed oriented embedded
subvortex-membrane with Marsden-Weinstein (MW) symplectic structure. For
example, for two dimensional (2D) inviscid incompressible fluid, we have
0-dimensional point-vortices; For 3D case, we have one dimensional (1D)
vortex-lines; For 4D case, we have 2D vortex-surfaces; For 5D case, we have 3D
vortex-membranes. See the illustration of vortex-line in 3D space in Fig.2. In
this paper, we focus on 3D vortex-membranes in the 5D inviscid incompressible fluid.

\subsection{Vortices in inviscid incompressible fluid}

\subsubsection{Point-vortices in 2D inviscid incompressible fluid}

Firstly, we review the dynamics of point-vortices within the framework of
Marsden-Weinstein symplectic structure.

For $N$ point-vortices in a 2D inviscid incompressible fluid, we define scalar
vorticity field
\begin{equation}
\Omega=\sum_{j=1}^{N}\kappa_{j}\, \delta(\mathrm{z}-\mathrm{z}_{j})(1\leq
j\leq N),
\end{equation}
where the complex value $\mathrm{z}_{j}=\xi_{j}+i\eta_{j}$ denotes the
position of the $j$-th point-vortex and $\kappa_{j}$ denotes the constant
vorticity of $j$-th point-vortex. So we have a topological condition
\begin{equation}%
%TCIMACRO{\doint \limits_{C}}%
%BeginExpansion
{\displaystyle \oint \limits_{C}}
%EndExpansion
\Omega dl=\sum_{j=1}^{N}\kappa_{j}%
\end{equation}
with loop integral along a closed path $C$.

According to Kirchhoff's theorem, the Euler equation for point-vortices is
\begin{equation}
\kappa_{j}\dot{\xi}_{j}=\frac{\partial \mathrm{H}_{\mathrm{Kirchhoff}}%
}{\partial \eta_{j}},\qquad \kappa_{j}\dot{\eta}_{j}=-\frac{\partial
\mathrm{H}_{\mathrm{Kirchhoff}}}{\partial \xi_{j}}%
\end{equation}
or
\begin{equation}
i\kappa_{i}\frac{d\mathrm{z}_{i}}{dt}=\frac{\delta \mathrm{H}%
_{\mathrm{Kirchhoff}}}{\delta \mathrm{z}_{i}^{\ast}}%
\end{equation}
where $\mathrm{H}_{\mathrm{Kirchhoff}}$ is the Kirchhoff Hamiltonian given by
\begin{equation}
\mathrm{H}_{\mathrm{Kirchhoff}}=-\frac{1}{2\pi}\sum_{j<k}^{N}\kappa_{j}%
\kappa_{k}\, \ln|\mathrm{z}_{j}-\mathrm{z}_{k}|.
\end{equation}
The Kirchhoff Hamiltonian is really potential energy that characterizes the
long range interaction between two point-vortices: for two vortices with
vorticities of same signs, we have repulsive interaction; For two vortices
with vorticities of opposite signs, we have attractive interaction. In
addition, above Euler equation can be changed in term of Poisson bracket
$\{,\}=\sum_{j=1}^{N}\frac{1}{\kappa_{j}}(\frac{\partial}{\partial \xi_{j}%
}\frac{\partial}{\partial \eta_{j}}-\frac{\partial}{\partial \eta_{j}}%
\frac{\partial}{\partial \xi_{j}})$ to be\cite{point,leap}
\begin{equation}
\partial_{t}\Omega=\{ \mathrm{H}_{\mathrm{Kirchhoff}},\Omega \}.
\end{equation}

\subsubsection{Vortex-lines in 3D inviscid incompressible fluid}

Secondly, we review the dynamics of vortex-lines in 3D inviscid incompressible fluid.

\paragraph{Biot-Savart equation}

vortex-lines are 1D topological objects in 3D inviscid incompressible fluid
with $\nabla \cdot \mathbf{v}\equiv0$. For a fluid with a vortex-line, a
rotationless superfluid component flow $\mathbf{\nabla \times v}=0$ is violated
on 1D singularities $\mathbf{s}(\zeta,t)$, which depends on the variables --
arc length $\zeta$ and the time $t$. Away from the singularities, the velocity
increases to infinity so that the circulation $¦Ê $ of the fluid velocity
remains constant,
\begin{equation}%
%TCIMACRO{\doint }%
%BeginExpansion
{\displaystyle \oint}
%EndExpansion
\mathbf{v}\cdot d\mathbf{l}=\kappa
\end{equation}
where $\mathbf{v=\dot{r}}$. As a result, the vortex filament could be
described by the Biot-Savart equation
\begin{equation}
\dot{\mathbf{r}}={\frac{\kappa}{4\pi}}\int{\frac{d\mathbf{s}\times
(\mathbf{r}-\mathbf{s})}{|\mathbf{r}-\mathbf{s}|^{3}}},
\end{equation}
where $\kappa$ is the circulation and $\mathbf{s}$ is the vector that denotes
the position of vortex filament. For a 3D superfluid, $\kappa=h/m$ is the
discreteness of the circulation owing to its quantum nature. $h$ is Planck
constant and $m$ is particle mass of SF.

\paragraph{Impulse and angular impulse}

In fluid with a vortex-line, the conventional momentum of the fluid motion
cannot be well defined. Instead, the hydrodynamic impulse (the Lamb impulse)
plays the role of the effective momentum that denotes the total mechanical
impulse of the non-conservative body force applied to a limited fluid volume
to generate instantaneously from rest the given motion of the whole of the
fluid at time $t$. In general, the (effective) momentum for a vortex-line from
Lamb impulse density is defined by
\begin{align}
\mathbf{P}_{\mathrm{Lamb}}  &  =\rho_{0}\int \mathbf{v}dV_{3D}\\
&  \Longrightarrow \frac{1}{2}\rho_{0}\int(\mathbf{s}\times \mathbf{\Omega
)}dV_{3D}\nonumber
\end{align}
where $dV_{3D}$ is an infinitesimal volume in 3D fluid and $\rho_{0}$ is the
superfluid mass density. For a vortex-line with a global velocity
$\mathbf{U},$ we have
\begin{equation}
\delta H-\mathbf{U}\cdot \delta \mathbf{P}_{\mathrm{Lamb}}=0.
\end{equation}
\ As a result, the Lamb impulse becomes the right physical quantity describing
the momentum of the vortex filament.

By using the definition of a vortex-line\cite{la1,la2,la3,la4,la5}
\begin{equation}
\mathbf{\Omega}=\kappa \delta_{\mathbf{s}},
\end{equation}
we have the (Lamb impulse) momentum
\begin{equation}
\mathbf{P}_{\mathrm{Lamb}}=\frac{\rho_{0}\kappa}{2}\int \mathbf{s\times
}d\mathbf{s}%
\end{equation}
and the (Lamb impulse) angular momentum for a vortex-line
\begin{align}
\mathbf{J}_{\mathrm{Lamb}}  &  =\rho_{0}\int(\mathbf{s\times v)}%
dV_{3D}\nonumber \\
&  =\frac{1}{2}\rho_{0}\int \mathbf{s\times(s\times \Omega)}dV_{3D}.
\end{align}
It is obvious that $\mathbf{P}_{\mathrm{Lamb}}$ and $\mathbf{J}_{\mathrm{Lamb}%
}$ are conserved quantities, i.e.,
\begin{equation}
\frac{d\mathbf{P}_{\mathrm{Lamb}}}{dt}\equiv0
\end{equation}
and
\begin{equation}
\frac{d\mathbf{J}_{\mathrm{Lamb}}}{dt}\equiv0.
\end{equation}

\paragraph{Geometric Biot-Savart mechanics for vortex-lines with 1D
Marsden-Weinstein symplectic structure under localized induction
approximation}

Localized induction approximation (LIA) of the vorticity motion is to keep the
local terms in the vorticity Euler equation. The Biot-Savart mechanics under
localized induction approximation for 1D vortex-lines is reduced to a special
classical mechanics -- geometric Biot-Savart mechanics for geometric objects
with 1D Marsden-Weinstein symplectic structure. The corresponding evolution
becomes the vortex filament equation
\begin{equation}
\partial_{t}\mathbf{s}=({\frac{\kappa}{4\pi}}\ln \epsilon){(}\mathbf{s}%
^{\prime}\times \mathbf{s}^{\prime \prime}), \label{binormal}%
\end{equation}
where $\epsilon={\frac{\ell}{a_{0}},}$ $\ell$ is the length of the order of
the curvature radius (or inter-vortex distance when the considered vortex
filament is a part of a vortex tangle) and $a_{0}$ denotes the vortex filament
radius which is much smaller than any other characteristic size in the system.

For an arc-length parametrization, the tangent vectors $\mathbf{t}%
=\partial \mathbf{s}/\partial \zeta=\mathbf{s}^{\prime}$ have unit length and
the acceleration vectors are $\mathbf{s}^{\prime \prime}=\partial
\mathbf{t}/\partial \zeta=\varkappa \cdot \mathbf{n}$, i.e., $\partial
_{t}\mathbf{s}=({\frac{\kappa}{4\pi}}\ln \epsilon)(\mathbf{s}^{\prime}%
\times \mathbf{s}^{\prime \prime})$ becomes
\begin{equation}
\partial_{t}\mathbf{s}=({\frac{\kappa}{4\pi}}\ln \epsilon){(}\varkappa
\cdot \mathbf{b)}%
\end{equation}
where $\varkappa$ and $\mathbf{b=t\times n}$ denote the curvature value and
bi-normal unit vector of the curve $\mathbf{s}$ at the corresponding point,
respectively. See the illustration in Fig.2.

\begin{figure}[ptb]
\includegraphics[clip,width=0.63\textwidth]{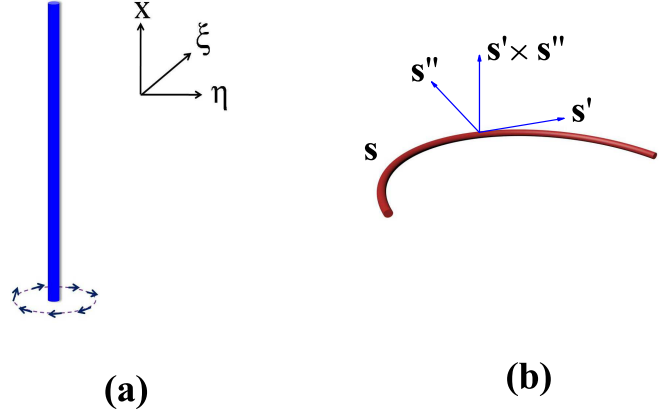}\caption{(a) An
illustration of a flat 1D vortex-line in 3D space; (b) An illustration of a
curved 1D vortex-line in 3D space. The local orthogonal coordinate system on
vortex-line described by the local vectors $\mathbf{s}^{\prime}$,
$\mathbf{s}^{\prime \prime}$, $\mathbf{s}^{\prime}\times \mathbf{s}%
^{\prime \prime}$ is shown.}%
\end{figure}

In addition, in term of the Hamiltonian $\mathrm{H}_{\text{\textrm{length}}%
}(\mathbf{s})$, the Biot-Savart equation\ can be written into
\begin{equation}
\mathbf{\dot{s}}=\hat{J}[\frac{\delta \mathrm{H}_{\text{\textrm{length}}%
}(\mathbf{s})}{\delta \mathbf{s}}]
\end{equation}
or%
\begin{equation}
\dot{\xi}=\frac{\partial \mathrm{H}_{\text{\textrm{length}}}(\xi,\eta
)}{\partial \eta},\qquad \dot{\eta}=-\frac{\partial \mathrm{H}%
_{\text{\textrm{length}}}(\xi,\eta)}{\partial \xi}.
\end{equation}
where $\hat{J}$ is an operator rotating the plane by $\pi/2$. The Hamiltonian
$\mathrm{H}_{\text{\textrm{length}}}(\mathbf{s})$ is defined by
\begin{equation}
\mathrm{H}_{\text{\textrm{length}}}(\mathbf{s})={\frac{\kappa \ln \epsilon}%
{4\pi}}\cdot \mathrm{length}(\mathbf{s}),
\end{equation}
where $\mathrm{length}(\mathbf{s})=\int dx\sqrt{1+\left \vert \frac{dz}%
{dx}\right \vert ^{2}}$ is the length of the vortex-line.

We introduce a complex description on the vortex-line, $\mathrm{z}=\xi
+i\eta=\left \vert \mathrm{z}\right \vert e^{i\phi}$ where $\left \vert
\mathrm{z}\right \vert $ is the amplitude and $\phi$ is the angle in the
complex plane. Under LIA\cite{AH65} and a simple geometrical constraint
$z^{\prime}\ll1$, in terms of the complex canonical coordinate $\mathrm{z}%
(x)$, the Biot-Savart equation becomes \cite{S95}
\begin{equation}
i\frac{d\mathrm{z}}{dt}=\frac{\delta \mathrm{H}_{\text{\textrm{length}}%
}(\mathrm{z})}{\delta \mathrm{z}^{\ast}}.
\end{equation}
When the vortex-line is rotating $\frac{dz}{dt}\neq0$, it becomes longer,
$\Delta \mathrm{H}_{\text{\textrm{length}}}(z)\neq0$.

\subsubsection{3D vortex-membranes in 5D inviscid incompressible fluid}

Thirdly, we develop the physics of 3D vortex-membranes in 5D inviscid
incompressible fluid. The Biot-Savart mechanics under localized induction
approximation for 3D vortex-membranes is reduced to 3D geometric Biot-Savart
mechanics for geometric objects with 3D Marsden-Weinstein symplectic structure.

For 5D case, we have 3D vortex-membranes with MW symplectic
structure\cite{leap}. The 3D vortex-membrane is defined by a given singular
vorticity
\begin{equation}
\mathbf{\Omega}=\kappa \delta_{P}%
\end{equation}
where the singular $\delta$-type vorticity denotes the sub-manifold $P$ in 5D
space, and $\kappa$ is the constant circulation strength.

\paragraph{Generalized Biot-Savart equation in 5D fluid}

A generalized Biot-Savart equation in 5D fluid is given by\cite{leap}%
\begin{equation}
\mathbf{v}(q)=\kappa \int \limits_{p\in P}\hat{J}\left(  \mathrm{Proj}%
_{N}\mathbf{\nabla}_{p}G(q,p)\right)  \,dV_{5D},
\end{equation}
where $\mathrm{Proj}_{N}\mathbf{A}$ is the orthogonal projection of
$\mathbf{A}$ to the fiber $NpP$ of the normal bundle to $P$ at $p\in P$, and
the operator $\hat{J}$ is the positive rotation around $p$ by $\pi/2$ in extra
dimensional space $NpP$. $G(q,p)=\frac{\alpha}{\left \vert q-p\right \vert
^{-3}}$ is the Green's function for the Laplace operator, i.e.
\begin{equation}
\Delta_{p}G(q,p)=\delta_{q}(p),
\end{equation}
the $\delta$-function supported at $q$. $dV_{5D}$ is an infinitesimal volume
in 5D fluid and $\alpha=\frac{\Gamma(\frac{5}{2})}{6\pi^{\frac{5}{2}}}$ where
$\Gamma(z)$ is the Gamma function.

\paragraph{Impulse and angular impulse}

We then consider the (effective) momentum for a vortex-membrane from Lamb
impulse density
\begin{align}
\mathbf{P}_{\mathrm{Lamb}} &  =\rho_{0}\int \mathbf{v}dV_{5D}\\
&  =\frac{1}{2}\rho_{0}\int(\mathbf{s}\left[  \times \right]  _{5D}%
\mathbf{\Omega)}dV_{5D}.\nonumber
\end{align}
Here, "$\left[  \times \right]  _{5D}$" denotes positive rotation around $p$
after vector product in 5D space. On the other hand, the (Lamb impulse)
angular momentum for a vortex-membrane is defined by
\begin{align}
\mathbf{J}_{\mathrm{Lamb}} &  =\int(\mathbf{s}\left[  \times \right]
_{5D}\mathbf{v)}dV_{P}\nonumber \\
&  =\frac{1}{2}\rho_{0}\int \mathbf{s}\left[  \times \right]  _{5D}%
\mathbf{(s}\left[  \times \right]  _{5D}\mathbf{\Omega)}dV_{5D}.
\end{align}
$\mathbf{P}_{\mathrm{Lamb}}$ and $\mathbf{J}_{\mathrm{Lamb}}$ are also
conserved quantities, i.e., $\frac{d\mathbf{P}_{\mathrm{Lamb}}}{dt}\equiv0$
and $\frac{d\mathbf{J}_{\mathrm{Lamb}}}{dt}\equiv0$.

\paragraph{Geometric Biot-Savart mechanics for vortex-membranes with 3D
Marsden-Weinstein symplectic structure under localized induction
approximation}

Under LIA, the generalized Biot-Savart equation for a 3D vortex-piece is
reduced into\cite{leap}
\begin{equation}
\mathbf{v}(q)\simeq(\kappa \alpha \ln \epsilon)\cdot J(\mathbf{M}_{\mathbf{C}%
}(p))
\end{equation}
where $\mathbf{M}_{\mathbf{C}}(p)$ is the mean curvature vector to $P$ at the
point $p$. The mean curvature vector $\mathbf{M}_{\mathbf{C}}(p)\in NpP$ is
the mean value of the curvature vectors of geodesics in $P$ passing through
the point $p$ when we average over the sphere $S^{4}$ of all possible unit
tangent vectors in $TpP$ for these geodesics. Thus, the generalized
Biot-Savart equation under LIA is given by the skew-mean-curvature flow
\begin{equation}
\frac{\partial P(p)}{\partial t}=-(\kappa \alpha \ln \epsilon)J(\mathbf{M}%
_{\mathbf{C}}(p)).
\end{equation}

\begin{figure}[ptb]
\includegraphics[clip,width=0.75\textwidth]{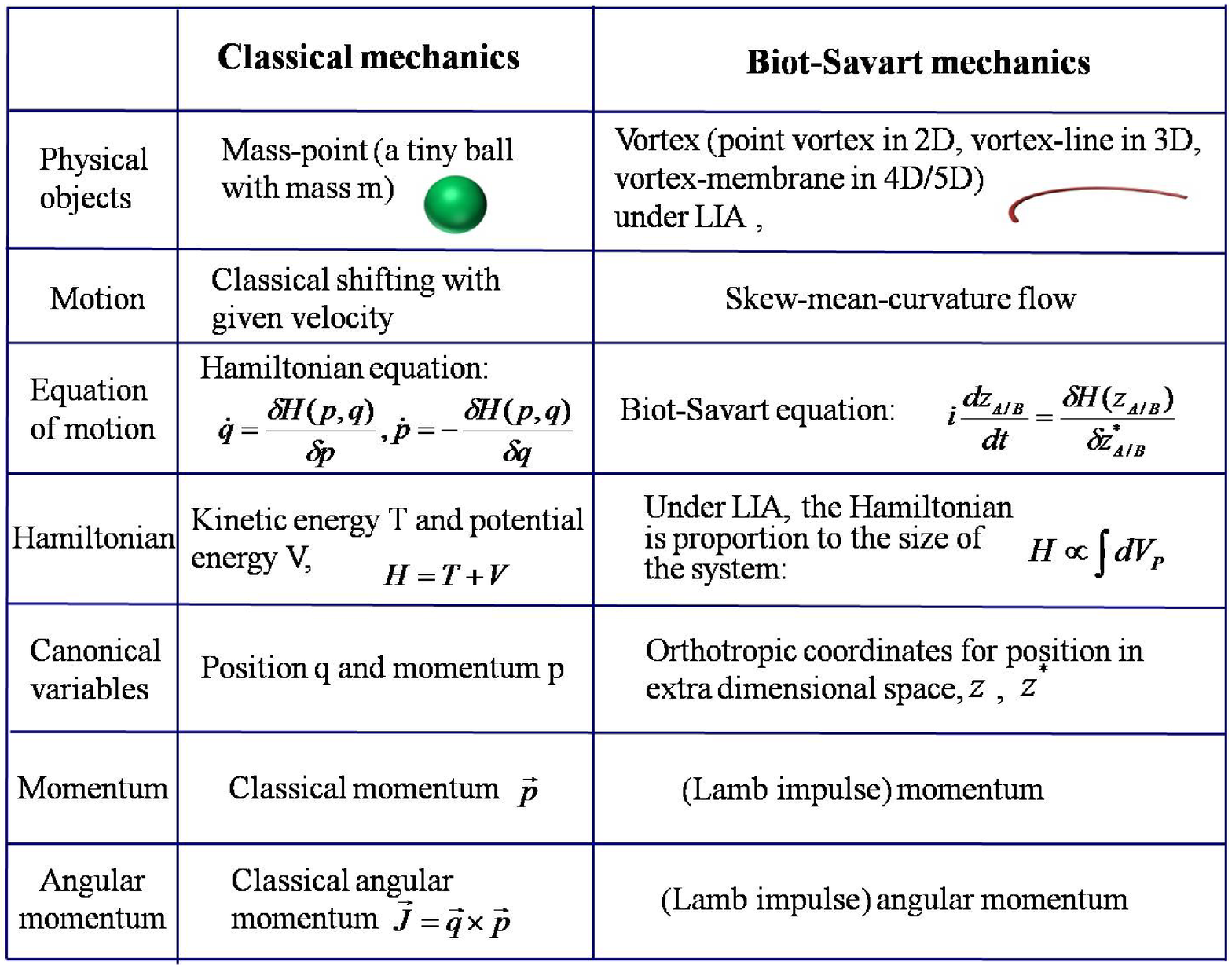}\caption{Comparision
between classical mechanics for mass-point and (geometric) Biot-Savart
mechanics for vortices}%
\end{figure}

According to the fact that the mean curvature vector field is the gradient to
the volume functional, the generalized Biot-Savart equation for a 3D
vortex-piece under LIA can be described by Hamiltonian formula and the
Hamiltonian on the vortex-membranes is just $3$-volume
\begin{equation}
\mathrm{H}_{\text{\textrm{volume}}}(P)=(\kappa \alpha \ln \epsilon)\cdot
\mathrm{volume}(P)
\end{equation}
with $\mathrm{volume}(P)=\int_{P}dV_{P}$. In term of the Hamiltonian
$\mathrm{H}(P)$ for 3D vortex-membrane, the generalized Biot-Savart
equation\ becomes
\begin{equation}
\frac{\partial P(p)}{\partial t}=(\kappa \alpha \ln \epsilon)\hat{J}[\frac
{\delta \mathrm{H}(P)}{\delta P}].
\end{equation}
The volume of the subvortex-membrane $P$ becomes a conserved quantity that
plays the role of Hamiltonian function of the corresponding dynamics. In
addition, the generalized Biot-Savart equation\ becomes
\begin{equation}
\dot{\xi}=\frac{\partial \mathrm{H}_{\text{\textrm{volume}}}(\xi,\eta
)}{\partial \eta},\qquad \dot{\eta}=-\frac{\partial \mathrm{H}%
_{\text{\textrm{volume}}}(\xi,\eta)}{\partial \xi}. \label{34}%
\end{equation}
In complex representation, $\mathrm{z}=\xi+i\eta,$ Eq.(\ref{34}) can also be
written into\cite{leap}
\begin{equation}
i\frac{d\mathrm{z}}{dt}=\frac{\delta \mathrm{H}_{\text{\textrm{volume}}}%
(P)}{\delta \mathrm{z}^{\ast}}.
\end{equation}

In particular, we emphasize that the evolution equation of the
subvortex-membrane $P$ is really mechanics of geometric objects -- the
Hamiltonian $\mathrm{H}_{\text{\textrm{volume}}}(P)$ is $3$-volume
\begin{equation}
\mathrm{volume}(P)=\int_{P}dV_{P}%
\end{equation}
that is a geometric quantity and the dynamic variables $(\xi(t),\eta(t))$ or
$\mathrm{z}(t)$ denote the position in extra dimensions that are also
geometric quantities. The generalized Biot-Savart equation is to minimize the
$3$-volume $\mathrm{volume}(P)=\int_{P}dV_{P}$. As a result, the
vortex-membranes always do the skew-mean-curvature flow that differs by the
$\pi/2$-rotation from the mean-curvature vector. The skew-mean-curvature flow
does not stretch the subvortex-membrane while moving its points orthogonally
to the mean curvatures. In Fig.3, we compare Newton mechanics for mass-point
with (geometric) Biot-Savart mechanics for vortices.

\section{Biot-Savart mechanics for a vortex-membrane: pseudo-quantum mechanics
for single particle}

The issue of Biot-Savart mechanics for a vortex-membrane is about the dynamic
evolution of windings for a vortex-membrane. We may ask a question --
\emph{how to characterize the evolution of a non-uniform helical
vortex-membrane?} For example, a non-uniform helical vortex-line is shown in
Fig.4.(b). We point out that the corresponding Biot-Savart mechanics is
reduced to pseudo-quantum mechanics for single particle. Here, "pseudo"
indicates that this mechanics just has similar structure to the traditional
quantum mechanics but is different.

\begin{figure}[ptb]
\includegraphics[clip,width=0.65\textwidth]{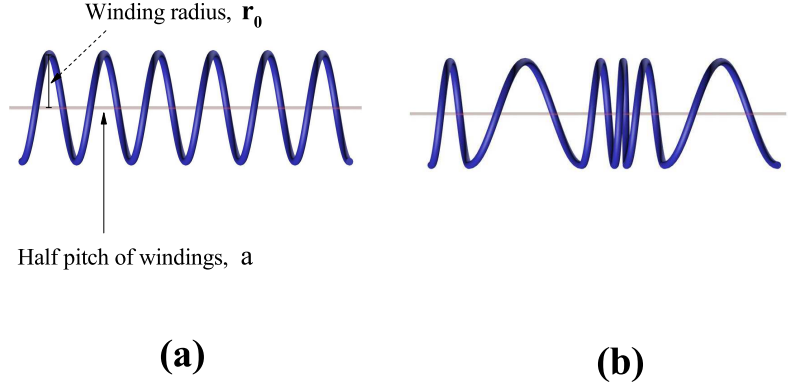}\caption{(a) A uniform
helical vortex-line; (b) A non-uniform helical vortex-line}%
\end{figure}

\subsection{Kelvin wave and its dispersion}

For 1D vortex-lines, there exists Kelvin wave, a transverse and circularly
polarized wave, in which the vortex-line twists around its equilibrium
position forming a helical structure. The plane-wave ansatz of Kelvin waves is
described by a complex field\cite{AH65,epl},
\begin{equation}
\mathrm{z}(x,t)=r_{0}e^{\pm ik\cdot x-i\omega t+i\phi_{0}}%
\end{equation}
where $r_{0}$ is the winding radius of vortex-line, $k=\frac{\pi}{a}>0$ and
$a$ is a fixed length that denotes the half pitch of the windings. $\phi_{0}$
is a constant angle. $\pm$ denotes two possible winding directions: left-hand
with clockwise winding, or right-hand with counterclockwise winding. For
Kelvin waves on vortex-lines, under LIA, the dispersion is obtained as
\begin{equation}
\omega \simeq(\frac{\kappa}{4\pi}\ln \epsilon)k^{2}%
\end{equation}
where $\epsilon \sim \frac{1}{ka_{0}}.$ In mathematics, we can generate a Kelvin
wave by an operator $\mathrm{\hat{U}}(x,t)$ on a flat vortex-line
\begin{equation}
\mathrm{z}(x,t)=\mathrm{\hat{U}}(x,t)\mathrm{z}_{0},
\end{equation}
where%
\begin{equation}
\mathrm{\hat{U}}(x,t)=e^{i\int[\phi(x,t)\cdot \hat{K}]dx}\cdot \mathrm{\hat
{\digamma}}(r_{0}).
\end{equation}
Here, $\mathrm{z}_{0}=0$ denotes a vortex-membrane with $r_{0}=0.$
$\mathrm{\hat{U}}(x,t)=e^{i\int[\phi(x,t)\cdot \hat{K}]dx}\cdot \mathrm{\hat
{\digamma}}(r_{0})$ is winding operator with $\hat{K}=-i\frac{d}{d\phi}$,
$\phi(x,t)=\pm kx-\omega t$. $\mathrm{\hat{\digamma}}(r_{0})$ is an expanding
operator by shifting radius from $0$ to $r_{0}$ on the membrane, i.e.,
$\mathrm{\hat{\digamma}}(r_{0})\cdot0=r_{0}$. So $\left(  \mathrm{\hat
{\digamma}}(r_{0})\right)  ^{-1}$ is a shrinking operator by shifting radius
from $r_{0}$ to $0$ on the membrane, i.e., $\left(  \mathrm{\hat{\digamma}%
}(r_{0})\right)  ^{-1}\cdot r_{0}=0$.

We then calculate the physical quantities of the Kelvin waves along winding
direction (x-direction), i.e., the projected angular momentum $J_{x}$ and the
projected momentum $p_{x}$. The projected (Lamb impulse) momentum along the
x-direction $p_{x}$ of a vortex-line with a plane Kelvin wave is obtained
as\cite{epl}%
\begin{equation}
p_{x}=\mathbf{P}_{\mathrm{Lamb}}\cdot \mathbf{e}_{x}=\pm \frac{1}{2}\rho
_{0}\kappa lr_{0}^{2}k
\end{equation}
that leads to the projected (Lamb impulse) momentum density $\rho_{p_{x}%
}=\frac{1}{2}\rho_{0}\kappa r_{0}^{2}\cdot k$. $l$ is the length of the system
in x-direction. The projected (Lamb impulse) angular momentum along
x-direction of a vortex-line with a plane Kelvin wave is given by\cite{epl}
\begin{equation}
J_{x}=\mathbf{J}_{\mathrm{Lamb}}\cdot \mathbf{e}_{x}=\mp \frac{1}{2}\rho
_{0}\kappa lr_{0}^{2}%
\end{equation}
that leads to the projected (Lamb impulse) angular momentum density along
x-direction $\rho_{J_{x}}=\left \vert \frac{J_{x}}{l}\right \vert =\rho
_{0}\kappa \frac{r_{0}^{2}}{2}$. Here, we use a mathematic result,
\begin{equation}
\int \mathbf{(s\times}d\mathbf{\mathbf{s})\times s}\cdot \mathbf{e}_{x}\equiv
V_{\mathrm{cylinder}}%
\end{equation}
with a (uniform or non-uniform) helical curve on a cylinder with volume
$V_{\mathrm{cylinder}}=r_{0}^{2}\cdot l$ (the symmetric axis is along
x-direction, the radius of cross-section is $r$, the length along x-direction
is $l$).

For Kelvin waves on a 3D helical vortex-membrane, the plane-wave ansatz is
described by a complex field,
\begin{equation}
\mathrm{z}(\vec{x},t)=r_{0}e^{\pm i\vec{k}\cdot \vec{x}-i\omega t+i\phi_{0}}%
\end{equation}
where $\vec{k}$ is the winding wave vector on 3D vortex-membrane with
$\left \vert \vec{k}\right \vert =\frac{\pi}{a}$ and $a$ is a fixed length that
denotes the half pitch of the windings.

For the plane Kelvin waves on a 3D helical vortex-membrane, we have the
Hamiltonian
\begin{align}
\mathrm{H}_{\text{\textrm{volume}}}(P)  &  =(\kappa \alpha \ln \epsilon
)\mathrm{volume}(P)=(\kappa \alpha \ln \epsilon)\int_{P}dV_{P}\nonumber \\
&  =(\kappa \alpha \ln \epsilon)%
%TCIMACRO{\diiint }%
%BeginExpansion
{\displaystyle \iiint}
%EndExpansion
dxdydz\nonumber \\
&  \cdot \sqrt{1+\left \vert \frac{d\mathrm{z}}{dx}\right \vert ^{2}+\left \vert
\frac{d\mathrm{z}}{dy}\right \vert ^{2}+\left \vert \frac{d\mathrm{z}}%
{dz}\right \vert ^{2}}.
\end{align}
Under local induction approximation and a simple geometrical constraint
$\left \vert \frac{d\mathrm{z}}{dx}\right \vert ^{2}+\left \vert \frac
{d\mathrm{z}}{dy}\right \vert ^{2}+\left \vert \frac{d\mathrm{z}}{dz}\right \vert
^{2}\ll1$, the Hamiltonian is reduced into
\begin{align}
\mathrm{H}_{\text{\textrm{volume}}}(P)  &  \simeq \text{\textrm{constant}%
}+\frac{(\kappa \alpha \ln \epsilon)}{2}%
%TCIMACRO{\diiint }%
%BeginExpansion
{\displaystyle \iiint}
%EndExpansion
dxdydz\nonumber \\
&  \cdot(\left \vert \frac{d\mathrm{z}}{dx}\right \vert ^{2}+\left \vert
\frac{d\mathrm{z}}{dy}\right \vert ^{2}+\left \vert \frac{d\mathrm{z}}%
{dz}\right \vert ^{2}).
\end{align}
Thus, we derive the dispersion of Kelvin waves as
\begin{equation}
\omega \simeq \frac{(\alpha \kappa \ln \epsilon)}{2}\vec{k}^{2}%
\end{equation}
where $\vec{k}^{2}=k_{x}^{2}+k_{y}^{2}+k_{z}^{2}$.

The (Lamb impulse) momentum along $\mathbf{\vec{e}}$-direction on
vortex-membrane with a plane Kelvin wave $\mathrm{z}(x,t)=r_{0}\exp
(-i\omega \cdot t+i\vec{k}\cdot \vec{x})$ is obtained as
\begin{equation}
\vec{p}_{\mathrm{Lamb}}=\mathbf{P}_{\mathrm{Lamb}}\cdot \mathbf{\vec{e}=}%
\pm \frac{1}{2}\rho_{0}\kappa V_{P}r_{0}^{2}k
\end{equation}
where $V_{p}$ is the total volume of the vortex-membrane. The (Lamb impulse)
angular momentum of a plane Kelvin wave along $\mathbf{\vec{e}}$-direction is
given by
\begin{equation}
\left \vert J_{\mathrm{Lamb}}\right \vert =\left \vert \mathbf{J}_{\mathrm{Lamb}%
}\cdot \mathbf{\vec{e}}\right \vert =\frac{1}{2}\rho_{0}\kappa V_{P}r_{0}^{2}.
\end{equation}
The projected (Lamb impulse) angular momentum is a constant on the vortex-membrane.

\subsection{Mapping to pseudo-quantum mechanics}

We then map the Biot-Savart mechanics for a (constraint) helical
vortex-membrane to a pseudo-quantum mechanics.

Firstly, we define the state-vector $\mathrm{z}_{\vec{k}}(\vec{x},t)$ to
denote the helical vortex-membrane as%
\begin{equation}
\mathrm{z}_{\vec{k}}(\vec{x},t)=r_{0}\exp(-i\omega \cdot t+i\vec{k}\cdot \vec
{x}).
\end{equation}
With the help of the operator $\mathrm{\hat{U}}(\vec{x},t)=e^{i\int[\phi
(\vec{x},t)\cdot \hat{K}]dx}\cdot \mathrm{\hat{\digamma}}(r_{0})$, the Kelvin
waves can also be generated from a straight one,%
\begin{equation}
\mathrm{z}_{\vec{k}}(\vec{x},t)=\mathrm{\hat{U}}(\vec{x},t)\cdot \mathrm{z}%
_{0}.
\end{equation}
The Hilbert space of pseudo-quantum mechanics for a helical vortex-membrane
consists of different states $\mathrm{z}_{\vec{k}}(\vec{x},t).$ We use the
state vector $\left[  \vec{k}\right \rangle $ to denote the state of Kelvin
waves of vortex-membranes and the state vector $\left \vert \vec{k}%
\right \rangle $ to denote the state of knots in the following parts.

In general, under LIA, according to superposition principle, we construct the
Kelvin wave as%
\begin{equation}
\mathrm{z}(\vec{x},t)=%
%TCIMACRO{\dsum \nolimits_{\vec{k}}}%
%BeginExpansion
{\displaystyle \sum \nolimits_{\vec{k}}}
%EndExpansion
r_{\vec{k}}\exp(-i\omega \cdot t+i\vec{k}\cdot \vec{x})
\end{equation}
where $r_{\vec{k}}$ is the radius of a partial wave $\exp(-i\omega \cdot
t+i\vec{k}\cdot \vec{x}).$ There exists a normalized volume condition,
\begin{align}%
%TCIMACRO{\dint }%
%BeginExpansion
{\displaystyle \int}
%EndExpansion
\mathrm{z}^{\ast}(\vec{x},t)\mathrm{z}(\vec{x},t)dV_{P}  &  =%
%TCIMACRO{\dint }%
%BeginExpansion
{\displaystyle \int}
%EndExpansion
(%
%TCIMACRO{\dsum \nolimits_{\vec{k}}}%
%BeginExpansion
{\displaystyle \sum \nolimits_{\vec{k}}}
%EndExpansion
\left \vert r_{\vec{k}}\right \vert ^{2})dV_{P}\nonumber \\
&  =V_{P}\cdot r_{0}^{2}.
\end{align}
We may consider the value $V_{P}\cdot r_{0}^{2}$ to be the fixed volume of the
system. For example, a 1D standing Kelvin wave denoted by $\frac{1}{\sqrt{2}%
}(\left[  k\right \rangle +\left[  -k\right \rangle )$ is a superposition Kelvin
wave of two plane Kelvin waves with opposite wave vectors
\begin{align}
\mathrm{z}(x,t)  &  =\frac{1}{\sqrt{2}}(r_{0}e^{ik\cdot x-i\omega t+i\phi_{0}%
}+r_{0}e^{-ik\cdot x-i\omega t+i\phi_{0}})\\
&  =\sqrt{2}r_{0}\cos(kx)e^{-i\omega t+i\phi_{0}}.\nonumber
\end{align}

For a plane wave, $\mathrm{z}_{\vec{k}}(\vec{x},t)=r_{0}\exp(-i\omega \cdot
t+i\vec{k}\cdot \vec{x})$, the (Lamb impulse) momentum is%
\begin{equation}
\vec{p}=\hbar_{\mathrm{eff}}\vec{k}%
\end{equation}
where the (Lamb impulse) angular momentum $J_{\mathrm{Lamb}}$ is obtained as
the effective Planck constant $\hbar_{\mathrm{eff}}$,
\[
\hbar_{\mathrm{eff}}=J_{\mathrm{Lamb}}=\frac{1}{2}\rho_{0}\kappa V_{P}%
r_{0}^{2}.
\]
The effective Planck constant is proportional to the total volume of the
system. The effective energy of a Kelvin wave is
\begin{equation}
E_{\mathrm{Lamb}}=\hbar_{\mathrm{eff}}\omega.
\end{equation}
As a result, we define the "wave-function" of a plane Kelvin wave as%
\begin{align}
\psi(\vec{x},t)  &  =\sqrt{\frac{1}{V_{P}r_{0}^{2}}}\mathrm{z}(\vec
{x},t)=\frac{1}{\sqrt{V_{P}}}\exp(-i\omega \cdot t+i\vec{k}\cdot \vec
{x})\nonumber \\
&  =\frac{1}{\sqrt{V_{P}}}\exp(\frac{-iE_{\mathrm{Lamb}}t+i\vec{p}%
_{\mathrm{Lamb}}\cdot \vec{x}}{\hbar_{\mathrm{eff}}})
\end{align}
and a generalized constraint Kelvin waves as
\begin{align}
\psi(\vec{x},t)  &  =%
%TCIMACRO{\dsum \nolimits_{\vec{k}}}%
%BeginExpansion
{\displaystyle \sum \nolimits_{\vec{k}}}
%EndExpansion
\sqrt{\frac{1}{V_{P}r_{0}^{2}}}r_{\vec{k}}\exp(-i\omega \cdot t+i\vec{k}%
\cdot \vec{x})\nonumber \\
&  =%
%TCIMACRO{\dsum \nolimits_{\vec{k}}}%
%BeginExpansion
{\displaystyle \sum \nolimits_{\vec{k}}}
%EndExpansion
a_{\vec{k}}\exp(\frac{-iE_{\mathrm{Lamb}}t+i\vec{p}_{\mathrm{Lamb}}\cdot
\vec{x}}{\hbar_{\mathrm{eff}}})
\end{align}
where $a_{\vec{k}}=\sqrt{\frac{1}{V_{P}r_{0}^{2}}}r_{\vec{k}}$.

For the average value of the effective energy $\left \langle E_{\mathrm{Lamb}%
}\right \rangle $, we have
\begin{align}
\left \langle E_{\mathrm{Lamb}}\right \rangle  &  =%
%TCIMACRO{\dint }%
%BeginExpansion
{\displaystyle \int}
%EndExpansion
\psi^{\ast}(\vec{x},t)E_{\mathrm{Lamb}}\psi(\vec{x},t)dV_{p}\\
&  =\int[%
%TCIMACRO{\dsum \nolimits_{\vec{p}}}%
%BeginExpansion
{\displaystyle \sum \nolimits_{\vec{p}}}
%EndExpansion
c_{\vec{p}}^{\ast}\exp(\frac{i(E_{\mathrm{Lamb}}t-\vec{p}_{\mathrm{Lamb}}%
\cdot \vec{x})}{\hbar_{\mathrm{eff}}})]\nonumber \\
&  E_{\mathrm{Lamb}}[%
%TCIMACRO{\dsum \nolimits_{\vec{p}^{\prime}}}%
%BeginExpansion
{\displaystyle \sum \nolimits_{\vec{p}^{\prime}}}
%EndExpansion
c_{\vec{p}^{\prime}}\exp(\frac{-i(E_{\mathrm{Lamb}}t-\vec{p}_{\mathrm{lamb}%
}^{\prime}\cdot \vec{x})}{\hbar_{\mathrm{eff}}})]dV_{p}\nonumber \\
&  =\int[%
%TCIMACRO{\dsum \nolimits_{\vec{p}}}%
%BeginExpansion
{\displaystyle \sum \nolimits_{\vec{p}}}
%EndExpansion
c_{\vec{p}}^{\ast}\exp(\frac{i(E_{\mathrm{Lamb}}t-\vec{p}_{\mathrm{Lamb}}%
\cdot \vec{x})}{\hbar_{\mathrm{eff}}})]\nonumber \\
&  (i\hbar_{\mathrm{eff}}\frac{d}{dt})[%
%TCIMACRO{\dsum \nolimits_{\vec{p}^{\prime}}}%
%BeginExpansion
{\displaystyle \sum \nolimits_{\vec{p}^{\prime}}}
%EndExpansion
c_{\vec{p}^{\prime}}\exp(\frac{-i(E_{\mathrm{Lamb}}t-\vec{p}_{\mathrm{Lamb}%
}^{\prime}\cdot \vec{x})}{\hbar_{\mathrm{eff}}})]dV_{p}\nonumber \\
&  =\int \psi^{\ast}(\vec{x},t)(i\hbar_{\mathrm{eff}}\frac{d}{dt})\psi(\vec
{x},t)dV_{p}.\nonumber
\end{align}
This result indicates that the effective energy becomes an operator%
\begin{equation}
E_{\mathrm{Lamb}}\rightarrow \hat{H}_{\mathrm{Lamb}}=i\hbar_{\mathrm{eff}}%
\frac{d}{dt}.
\end{equation}
Using the similar approach, we derive
\begin{align}
\left \langle \vec{p}_{\mathrm{Lamb}}\right \rangle  &  =\int \psi^{\ast
}(x,t)\vec{p}_{\mathrm{Lamb}}\psi(x,t)dx\\
&  =\int \psi^{\ast}(x,t)(-i\hbar_{\mathrm{eff}}\frac{d}{d\vec{x}}%
)\psi(x,t)dx.\nonumber
\end{align}
As a result, the (Lamb impulse) momentum also becomes an operator%
\begin{equation}
\vec{p}_{\mathrm{Lamb}}\rightarrow \hat{p}_{\mathrm{Lamb}}=-i\hbar
_{\mathrm{eff}}\frac{d}{d\vec{x}}.
\end{equation}

From the dispersion of Kelvin waves, we have%
\[
\omega \simeq \frac{(\alpha \kappa \ln \epsilon)}{2}\vec{k}^{2}.
\]
The energy-momentum relationship $E_{\mathrm{Lamb}}=H(\vec{p}_{\mathrm{Lamb}%
})=\frac{(\alpha \kappa \ln \epsilon)}{2\hbar_{\mathrm{eff}}}\vec{p}%
_{\mathrm{Lamb}}^{2}$ determines the equation of motion for vortex-membranes
that becomes the "Schr\"{o}dinger equation",
\begin{equation}
\hat{H}_{\mathrm{Lamb}}\psi(\vec{x},t)=\frac{\hat{p}_{\mathrm{Lamb}}^{2}%
}{2m_{\mathrm{pseudo}}}\psi(\vec{x},t)
\end{equation}
or
\begin{equation}
i\hbar_{\mathrm{eff}}\frac{d\psi(\vec{x},t)}{dt}=\frac{\hat{p}_{\mathrm{lamb}%
}^{2}}{2m_{\mathrm{pseudo}}}\psi(\vec{x},t)
\end{equation}
with an effective mass
\begin{equation}
m_{\mathrm{pseudo}}=\frac{\hbar_{\mathrm{eff}}}{\alpha \kappa \ln \epsilon}.
\end{equation}

\begin{figure}[ptb]
\includegraphics[clip,width=0.75\textwidth]{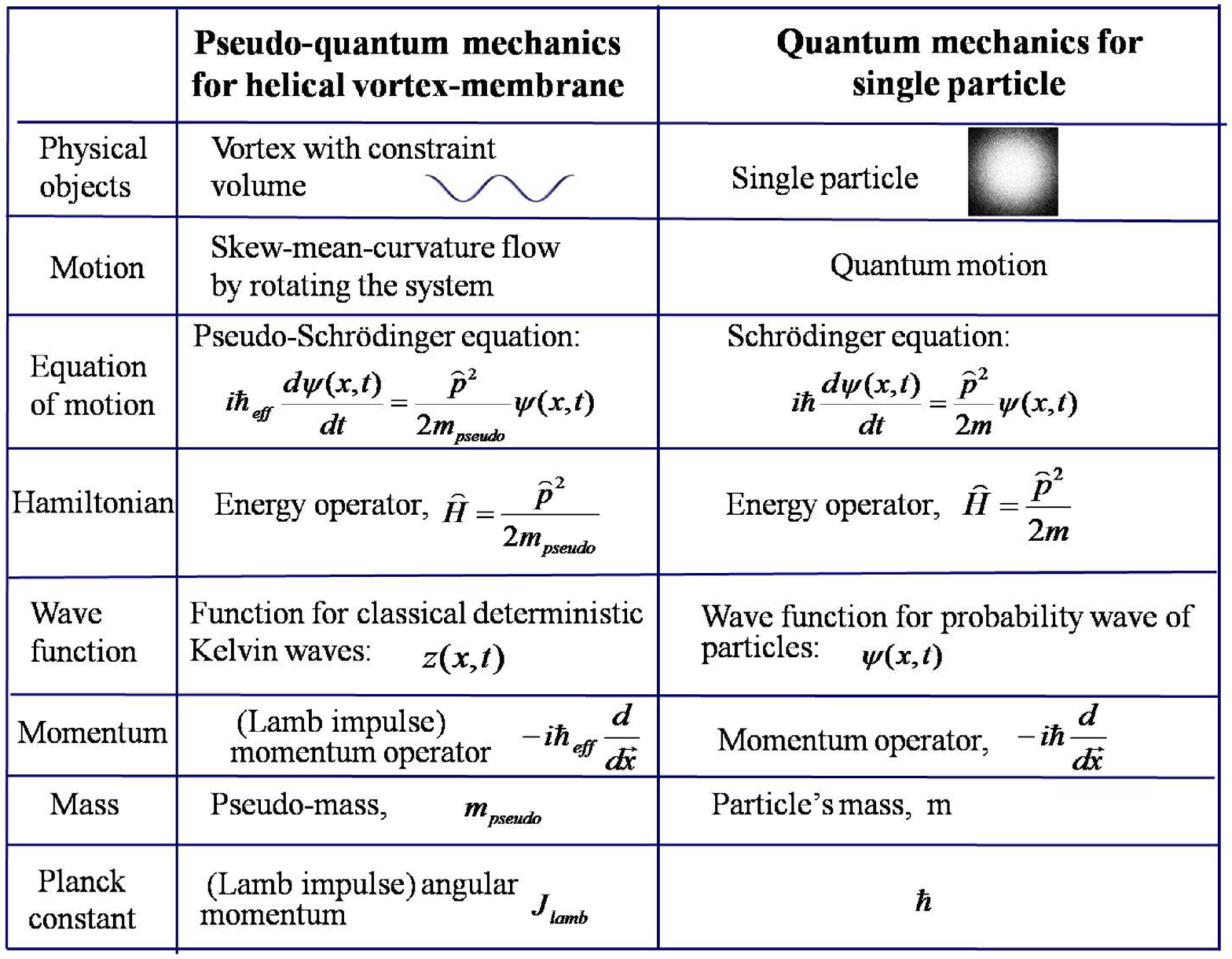}\caption{Comparision
between pseudo-quantum mechanics for helical vortex-membrane and quantum
mechanics for single quantum particle}%
\end{figure}

For the eigenstate with eigenvalue $E_{\mathrm{Lamb}},$ the wave-function is
given by
\begin{equation}
\psi(\vec{x},t)=f(\vec{x})\exp(\frac{iE_{\mathrm{Lamb}}t}{\hbar_{\mathrm{eff}%
}})
\end{equation}
where $f(\vec{x})$ is spatial function. This corresponds to a twisting motion
with fixed twisting angular velocity, $\omega=\frac{E_{\mathrm{Lamb}}}%
{\hbar_{\mathrm{eff}}}$. That means the excitations of the quantum state must
have the quantized projected (Lamb impulse) energy,
\begin{equation}
\Delta E_{\mathrm{lamb}}=nE_{\mathrm{lamb}}=n\hbar_{\mathrm{eff}}\omega
\end{equation}
where $n$ is a positive integer number.

In pseudo-quantum mechanics, there are three conserved physical quantities for
helical vortex-membranes: the energy $H(\vec{p}_{\mathrm{Lamb}})$ that is
proportional to the volume of the vortex-membrane $V_{P}=\mathrm{volume}(P)$;
the momentum $\vec{p}_{\mathrm{Lamb}}$ that is proportional to the winding
number along given direction $\vec{w}_{1D}$ (see below discussion); and the
(Lamb impulse) angular momentum (the effective Planck constant $\hbar
_{\mathrm{eff}}$) that is proportional to the volume of the vortex-membrane in
the 5D fluid $V_{P}\cdot r_{0}^{2}$.

As a result, the pseudo-quantum mechanics describes the dynamics of a
vortex-membrane with fixed volume in the 5D fluid. Under the constraint, the
total degree of freedom of a vortex-membrane is reduced to $1$. In Fig.5, we
show the comparison between pseudo-quantum mechanics for helical
vortex-membrane and quantum mechanics for single quantum particle. The
quantization of (constraint) Kelvin waves is similar to quantization of a
matter wave in quantum mechanics: a constant $J_{\mathrm{Lamb}}$ plays the
role of the Planck constant $\hbar_{\mathrm{eff}}$, $m_{\mathrm{pseudo}}$
plays the role of mass in Schr\"{o}dinger equation, and so on. However, they
are different. The Kelvin waves are classical waves that obey deterministic
classical mechanics, not probabilistic quantum mechanics. This is why we call
it pseudo-quantum mechanics. In the following parts, we will show that the
knots rather than vortex-membranes obey true quantum mechanics.

\subsection{Winding number and winding-number density}

In above part, we derive the equation of motion of Kelvin waves. In principle,
the evolution of the system can be solved. We then introduce winding number
and winding-number density to describe the deformed helical vortex-membrane.

The winding number $w_{1D}$ of two 1D vortex-lines is defined as
\begin{equation}
w_{1D}=\frac{-i}{2\pi r_{0}^{2}}\oint_{C}\mathrm{z}^{\ast}(x,t)d\mathrm{z}%
(x,t).
\end{equation}
To locally characterize the winding behavior, we define the winding-number
density $\rho_{\mathrm{wind}},$ $\rho_{\mathrm{wind}}=\frac{\Delta w_{1D}%
}{\Delta x}.$ For example, for a vortex-line described by a plane Kelvin wave
$\mathrm{z}(x,t)=r_{0}e^{-i\omega \cdot t+ik\cdot x}$, the winding number
$w_{1D}=\frac{k}{2\pi}l$ is proportional to the length of the system $l$ along
the Kelvin wave and the density of winding number $\rho_{\mathrm{wind}}$ is
uniform, i.e., $\rho_{\mathrm{wind}}=\frac{w_{1D}}{l}=\frac{k}{2\pi}.$ From
this equation, the momentum $p_{\mathrm{Lamb}}$ is proportional of the winding
number $w_{1D}.$

According to pseudo-quantum mechanics, the local winding can be described by
the density operator of winding number,
\begin{align}
\rho_{\mathrm{wind}}  &  \rightarrow \hat{\rho}_{\mathrm{wind}}=\frac{\hat{k}%
}{2\pi}\\
&  =\frac{1}{2\pi \hbar_{\mathrm{eff}}}\hat{p}=-i\frac{1}{2\pi}\frac{d}%
{dx}.\nonumber
\end{align}
For a vortex-line described by an arbitrary plane Kelvin wave $\mathrm{z}%
(x,t)=%
%TCIMACRO{\dsum \nolimits_{k}}%
%BeginExpansion
{\displaystyle \sum \nolimits_{k}}
%EndExpansion
c_{k}e^{-i\omega \cdot t+ik\cdot x}$, the total 1D winding number is obtained
by the following equation,
\begin{align}
w_{1D}  &  =\left \langle x,t\right]  \frac{l\cdot \hat{k}}{2\pi}\left[
x,t\right \rangle \\
&  =\int \psi^{\ast}(x,t)\frac{l\cdot \hat{k}}{2\pi}\psi(x,t)dx\nonumber \\
&  =\frac{1}{r_{0}^{2}}\int \mathrm{z}^{\ast}(x,t)\frac{\hat{k}}{2\pi
}\mathrm{z}(x,t)dx\nonumber \\
&  =\frac{1}{r_{0}^{2}}\int \mathrm{z}^{\ast}(x,t)(-i\frac{1}{2\pi}\frac{d}%
{dx})\mathrm{z}(x,t)dx.\nonumber
\end{align}
The winding-number density becomes
\begin{equation}
\rho_{\mathrm{wind}}(x,t)=\mathrm{z}^{\ast}(x,t)(-i\frac{1}{2\pi}\frac
{1}{r_{0}^{2}}\frac{d}{dx})\mathrm{z}(x,t).
\end{equation}
For example, a local winding at $x=x_{0}$ is defined by $\rho_{\mathrm{wind}%
}(x,t)=\delta(x-x_{0})$ that corresponds to a sharp changing of vortex-line;
for the case of $\mathrm{z}(x,t)=$ \textrm{constant}, there is no winding at
all $\rho_{\mathrm{wind}}(x,t)=0$.

In 5D space, we also define the winding number and the winding-number density
for a helical vortex-membrane along different directions.

Along the given direction $\vec{e}^{I},$ winding number of the vortex-membrane
($r_{0}$ is a constant) described by plane Kelvin wave $\mathrm{z}%
(x^{I},t)=r_{0}e^{-i\omega \cdot t+ik^{I}\cdot x^{I}}$ is well defined. Because
an arbitrary Kelvin wave can be described by $\mathrm{z}(\vec{x},t)=%
%TCIMACRO{\dsum \limits_{k}}%
%BeginExpansion
{\displaystyle \sum \limits_{k}}
%EndExpansion
c_{k}e^{-i\omega \cdot t+i\vec{k}\cdot \vec{x}},$ we can use a vector of
winding-number density to describe local winding of a vortex-membrane
$\mathrm{z}(\vec{x},t)$. So In 3D space, there are three winding-numbers
$w_{1D}^{I}$ ($I=X,Y,Z$) along different directions as
\begin{equation}
w_{1D}^{I}=\frac{-i}{2\pi r_{0}^{2}}\oint_{C^{I}}\mathrm{z}^{\ast}(\vec
{x},t)d\mathrm{z}(\vec{x},t).
\end{equation}
where $C^{I}$ denotes the closed path around a flat membrane along different
directions. For a vortex-membrane described by a plane Kelvin wave
$\mathrm{z}(\vec{x},t)=r_{0}e^{-i\omega \cdot t+i\vec{k}\cdot \vec{x}}$, the
winding number along direction $\vec{e}^{I}$ is given by
\begin{equation}
w_{1D}^{I}=\frac{k^{I}}{2\pi}L^{I}%
\end{equation}
that is proportional to the length of the system $L^{I}$ along $\vec{e}^{I}%
$-direction and the winding-number density is $\rho_{\mathrm{wind}}^{I}%
=\frac{k^{I}}{2\pi}.$

Thus, the local winding along direction $\vec{e}^{I}$ can be described by the
operator of winding-number density,
\begin{equation}
\mathbf{\hat{\rho}}_{\mathrm{wind}}=\frac{\hat{k}^{I}}{2\pi}=-i\frac{1}{2\pi
}\frac{d}{dx^{I}}%
\end{equation}
where $\mathbf{\hat{\rho}}_{\mathrm{wind}}=\hat{\rho}_{\mathrm{wind}}\cdot
\vec{e}^{I}$. With the help of $\hat{\rho}_{\mathrm{wind}}$, the winding
number of a vortex-membrane is obtained as
\begin{align}
\vec{w}_{1D}  &  =\left \langle x,t\right]  l\cdot \mathbf{\hat{\rho}%
}_{\mathrm{wind}}\left[  x,t\right \rangle \\
&  =\int \mathrm{z}^{\ast}(\vec{x},t)(-i\frac{1}{2\pi r_{0}^{2}}\frac{d}%
{d\vec{x}})\mathrm{z}(\vec{x},t)dx.\nonumber
\end{align}
The vector of winding-number density becomes
\begin{equation}
\vec{\rho}_{\mathrm{wind}}=\mathrm{z}^{\ast}(\vec{x},t)(-i\frac{1}{2\pi
r_{0}^{2}}\frac{d}{d\vec{x}})\mathrm{z}(\vec{x},t).
\end{equation}

Therefore, we answer the question -- \emph{how to characterize the evolution
of a non-uniform helical vortex-membrane?} We consider a deformed
vortex-membrane $\mathrm{z}(\vec{x})$ as the initial condition. We expand
$\mathrm{z}(\vec{x})$ by the eigenstates of Kelvin-waves as $\mathrm{z}%
(\vec{x})=%
%TCIMACRO{\dsum \nolimits_{k}}%
%BeginExpansion
{\displaystyle \sum \nolimits_{k}}
%EndExpansion
c_{k}e^{i\vec{k}\cdot \vec{x}}$. Under time evolution, the function of
vortex-membranes becomes $\mathrm{z}(\vec{x},t)=%
%TCIMACRO{\dsum \nolimits_{k}}%
%BeginExpansion
{\displaystyle \sum \nolimits_{k}}
%EndExpansion
c_{k}e^{i\vec{k}\cdot \vec{x}}e^{-i\omega \cdot t}.$ Finally, the vector of
winding-number density for the deformed vortex-membrane is obtained as
\begin{align}
\vec{\rho}_{\mathrm{wind}}(\vec{x},t)  &  =\mathrm{z}^{\ast}(\vec
{x},t)(-i\frac{1}{2\pi r_{0}^{2}}\frac{d}{d\vec{x}})\mathrm{z}(\vec{x},t)\\
&  =\frac{1}{2\pi r_{0}^{2}}%
%TCIMACRO{\dsum \nolimits_{k,k^{\prime}}}%
%BeginExpansion
{\displaystyle \sum \nolimits_{k,k^{\prime}}}
%EndExpansion
\vec{k}\cdot c_{k^{\prime}}^{\ast}c_{k}\cdot e^{-i(\vec{k}^{\prime}-\vec
{k})\cdot \vec{x}}\cdot e^{i(\omega^{\prime}-\omega)\cdot t}.\nonumber
\end{align}
We emphasize that under time evolution, the energy $H(\vec{p}_{\mathrm{Lamb}%
})$, the momentum $\vec{p}_{\mathrm{Lamb}}$ (the total winding number along
given direction $w_{1D}$) and the (Lamb impulse) angular momentum (the
effective Planck constant $\hbar_{\mathrm{eff}}$) are all conserved.

In particular, we have the topological representation on action $S,$ i.e.,
\begin{align}
S  &  =%
%TCIMACRO{\dint }%
%BeginExpansion
{\displaystyle \int}
%EndExpansion
p_{\mathrm{Lamb}}\cdot dx=\hbar_{\mathrm{eff}}%
%TCIMACRO{\dint }%
%BeginExpansion
{\displaystyle \int}
%EndExpansion
k\cdot dx\\
&  =\hbar_{\mathrm{eff}}%
%TCIMACRO{\dint }%
%BeginExpansion
{\displaystyle \int}
%EndExpansion
2\pi \rho_{\mathrm{wind}}\cdot dx=2\pi \hbar_{\mathrm{eff}}\cdot w_{1D}%
\nonumber \\
&  =h_{\mathrm{eff}}\cdot w_{1D}.\nonumber
\end{align}
From the point view of topology, we explain the Sommerfeld quantization
condition
\begin{equation}%
%TCIMACRO{\doint }%
%BeginExpansion
{\displaystyle \oint}
%EndExpansion
p_{\mathrm{Lamb}}\cdot dx=nh_{\mathrm{eff}}%
\end{equation}
where $h_{\mathrm{eff}}=2\pi \hbar_{\mathrm{eff}}$ and $n$ is an integer
number. The Sommerfeld quantization condition is just the topological
condition of the winding number, i.e.,%
\[%
%TCIMACRO{\doint }%
%BeginExpansion
{\displaystyle \oint}
%EndExpansion
p_{\mathrm{Lamb}}\cdot dx=h_{\mathrm{eff}}\cdot w_{1D}%
\]
where $w_{1D}$ is the winding number of the helical vortex-membrane.

\subsection{Path integral formulation for winding evolution}

In 1948, Feynman derived path integral formulation of quantum mechanics, based
on the fact that the propagator can be written as a sum over all possible
paths between the initial point and the final point. With the help of the path
integral, people can calculate the probability amplitude $K(\vec{x}^{\prime
},t_{f};\vec{x},t_{i})$ from an initial position $\vec{x}$ at time $t=t_{i}$
(that is described by a state $\left \vert t_{i},\vec{x}\right \rangle $) to
position $\vec{x}^{\prime}$ at a later time $t=t_{f}$ ($\left \vert t_{f}%
,\vec{x}^{\prime}\right \rangle $).

Using similar approach, we derive the path integral formulation for
pseudo-quantum mechanics. A state $\left[  t,\vec{x}\right \rangle $ in
pseudo-quantum mechanics denotes a local winding at point $\vec{x}$ and time
$t$. Letting the state evolve with time and project on the state $\left[
\vec{x}^{\prime}\right \rangle $, we have the transition amplitude of the
process as
\begin{align}
K(\vec{x}^{\prime},t_{f};\vec{x},t_{i})  &  =\left \langle t_{f},\vec
{x}^{\prime}\right]  \left[  t_{i},\vec{x}\right \rangle =%
%TCIMACRO{\dsum \limits_{n}}%
%BeginExpansion
{\displaystyle \sum \limits_{n}}
%EndExpansion
e^{iS_{n}/\hbar_{\mathrm{eff}}}\nonumber \\
&  =\int \mathcal{D}\vec{p}_{\mathrm{Lamb}}(t)\mathcal{D}\vec{x}(t)e^{iS/\hbar
_{\mathrm{eff}}}%
\end{align}
where $S=%
%TCIMACRO{\dint }%
%BeginExpansion
{\displaystyle \int}
%EndExpansion
[\vec{p}_{\mathrm{Lamb}}\cdot \frac{d\vec{x}}{dt}-\hat{H}_{\mathrm{Lamb}}%
(\vec{p}_{\mathrm{Lamb}},\vec{x})]dt$ and $\hat{E}_{\mathrm{Lamb}}(\vec
{p}_{\mathrm{Lamb}},\vec{x})=\frac{\vec{p}_{\mathrm{Lamb}}^{2}}%
{2m_{\mathrm{pseudo}}}$ Each path contributes $e^{iS_{n}/\hbar_{\mathrm{eff}}%
}$ where $S_{n}$ is the $n$-th classical action i.e., $%
%TCIMACRO{\dsum \limits_{n}}%
%BeginExpansion
{\displaystyle \sum \limits_{n}}
%EndExpansion
e^{iS_{n}/\hbar_{\mathrm{eff}}}$.

Let us discuss the implication of path-integral formulation in pseudo-quantum mechanics.

We firstly consider a state with a local winding at the position $(\vec{x}%
_{i},\phi_{0})$ and time $t=t_{i}$ where $\vec{x}_{i}$ denotes the original
position and $\phi_{0}$ the phase angle. To calculate the transition amplitude
$K(\vec{x}^{\prime},t_{f};\vec{x},t_{i})=\left \langle t_{f},\vec{x}^{\prime
}\right]  \left[  t_{i},\vec{x}\right \rangle $, the winding splits into
pieces. For a winding-piece, the action is denoted by $S_{n}$ for an arbitrary
possible classical path from $(\vec{x}_{i},\phi_{0})$ to $(\vec{x}_{f}%
,\phi_{n})$ within time $T=t_{f}-t_{i}$. In particular, the phase angle $\phi$
may be different for a winding-piece moving along different classical pathes.
The phase-changing of a possible classical path is given by
\begin{equation}
\Delta \phi=\phi_{n}-\phi_{0}=\frac{S_{n}}{\hbar_{\mathrm{eff}}}.
\end{equation}
Because the path may be not closed, $\frac{S_{n}}{\hbar_{\mathrm{eff}}}$ may
be not an integer number. As a result, we have the transition amplitude as
$e^{iS_{n}/\hbar_{\mathrm{eff}}}$. We summarize the contribution from all
windings and get the final transition amplitude as
\begin{equation}%
%TCIMACRO{\dsum \limits_{n}}%
%BeginExpansion
{\displaystyle \sum \limits_{n}}
%EndExpansion
e^{i\Delta \phi_{n}}=%
%TCIMACRO{\dsum \limits_{n}}%
%BeginExpansion
{\displaystyle \sum \limits_{n}}
%EndExpansion
e^{iS_{n}/\hbar_{\mathrm{eff}}}%
\end{equation}
that is just the Feynman's path integral formulation.

In summary, pseudo-quantum mechanics becomes a toy mechanics to learn the
mechanism of true quantum mechanics.

\section{Biot-Savart mechanics for two entangled vortex-membranes:
pseudo-quantum mechanics for a two-component particle}

In this section, we discuss the Biot-Savart mechanics for two entangled
vortex-membranes. The issue of Biot-Savart mechanics for two entangled
vortex-membranes is about the dynamic evolution of entanglement between them.
Here, we ask a question -- \emph{how to characterize the evolution of
non-uniform entangled vortex-membranes?} For example, a non-uniform entangled
vortex-line is shown in Fig.6.(b). We point out that to characterize the two
entangled vortex-membranes, the corresponding Biot-Savart mechanics is mapped
to pseudo-quantum mechanics for a two-component particle.

\begin{figure}[ptb]
\includegraphics[clip,width=0.63\textwidth]{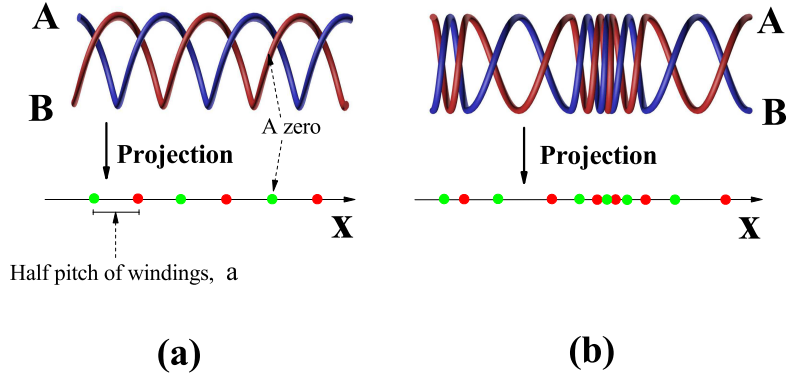}\caption{(a) Uniform
entangled vortex-lines; (b) non-uniform entangled vortex-lines. The index
$\mathrm{A/B}$ denotes vortex-line-A or vortex-line-B. The red/green spots
denote the zeros between two projected vortex-lines.}%
\end{figure}

\subsection{Biot-Savart mechanics for two entangled vortex-membranes with
leapfrogging motion}

In this part, we discuss the Biot-Savart mechanics for two entangled vortex-membranes.

In five dimensional space $\{ \vec{x},\xi,\eta \}$ ($\vec{x}=(x,y,z)$), we
introduce a complex description on $\xi$-$\eta$ complex plane, $\mathrm{z}%
_{\mathrm{A/B}}(\vec{x},t)=\xi_{\mathrm{A/B}}(\vec{x},t)+i\eta_{\mathrm{A/B}%
}(\vec{x},t)=r_{\mathrm{A/B}}(\vec{x},t)e^{i\phi_{\mathrm{A/B}}(\vec{x},t)}$
where $r_{\mathrm{A/B}}(\vec{x},t)$ is amplitude and $\phi_{\mathrm{A/B}}%
(\vec{x},t)$ is angle in $\xi$-$\eta$ complex plane. The index $\mathrm{A/B}$
denotes vortex-line-\textrm{A} or vortex-line-\textrm{B}. Under a simple
geometrical constraint, $\mathrm{z}_{\mathrm{A/B}}^{\prime}(\vec{x})\sim
\frac{|\mathrm{z}_{\mathrm{A/B}}(\vec{x}_{1})-\mathrm{z}_{\mathrm{A/B}}%
(\vec{x}_{2})|}{|\vec{x}_{1}-\vec{x}_{2}|}\ll1,$ the generalized Biot-Savart
equation can be written into
\begin{equation}
i\frac{d\mathrm{z}_{\mathrm{A/B}}}{dt}=\frac{\delta \mathrm{H}(z_{\mathrm{A/B}%
})}{\delta \mathrm{z}_{\mathrm{A/B}}^{\ast}}%
\end{equation}
where the Hamiltonian is given by\cite{1}
\begin{align}
\mathrm{H}  &  =\mathrm{Kinetic}\text{ }\mathrm{term}+\mathrm{interaction}%
\text{ }\mathrm{term}\nonumber \\
&  =\mathrm{H}_{\text{\textrm{volume}}}+\mathrm{H}_{\mathrm{Kirchhoff}%
}\nonumber \\
&  \simeq \int[\frac{(\alpha \kappa \ln \epsilon)}{2}(\left \vert \mathrm{z}%
_{\mathrm{A}}^{\prime}\right \vert ^{2}+\left \vert \mathrm{z}_{\mathrm{B}%
}^{\prime}\right \vert ^{2})\nonumber \\
&  -(\alpha \kappa \ln \epsilon)\ln{|\mathrm{z}_{\mathrm{A}}(\vec{x}%
)-\mathrm{z}_{\mathrm{B}}(\vec{x})|]dV}_{P}{+}\text{ \textrm{constant}}{.}%
\end{align}
Here $r_{0}$ is the inter-vortex distance and $a_{0}$ denotes the vortex
filament radius, i.e., $r_{0}\gg a_{0}$.

Because the vortex-membranes are almost straight, the nonlocal interactions,
varying on vortex-membranes, are approximated to be similar to those of two
straight vortex-membranes that is described by Kirchhoff Hamiltonian,
$\mathrm{H}_{\mathrm{Kirchhoff}}$.

Now, the Biot-Savart equation of motion for each vortex-membrane is given
by\cite{1}
\begin{align}
i\frac{\partial \mathrm{z}_{\mathrm{A}}}{\partial t}  &  =\frac{\delta
\mathrm{H}[\mathrm{z}_{\mathrm{A}},\mathrm{z}_{\mathrm{B}}]}{\delta
\mathrm{z}_{\mathrm{A}}^{\ast}},\nonumber \\
i\frac{\partial \mathrm{z}_{\mathrm{B}}}{\partial t}  &  =\frac{\delta
\mathrm{H}[\mathrm{z}_{\mathrm{A}},\mathrm{z}_{\mathrm{B}}]}{\delta
\mathrm{z}_{\mathrm{B}}^{\ast}},
\end{align}
where
\begin{align}
\frac{\partial \mathrm{z}_{\mathrm{A}}}{\partial t}  &  =i\frac{(\alpha
\kappa \ln \epsilon)}{2}\frac{\partial^{2}\mathrm{z}_{\mathrm{A}}}{\partial
\vec{x}^{2}}+i(\alpha \kappa \ln \epsilon)\frac{\mathrm{z}_{\mathrm{A}%
}-\mathrm{z}_{\mathrm{B}}}{|\mathrm{z}_{\mathrm{A}}-\mathrm{z}_{\mathrm{B}%
}|^{2}},\nonumber \\
\frac{\partial z_{\mathrm{B}}}{\partial t}  &  =i\frac{(\alpha \kappa
\ln \epsilon)}{2}\frac{\partial^{2}\mathrm{z}_{\mathrm{B}}}{\partial \vec{x}%
^{2}}-i(\alpha \kappa \ln \epsilon)\frac{\mathrm{z}_{\mathrm{A}}-\mathrm{z}%
_{\mathrm{B}}}{|\mathrm{z}_{\mathrm{A}}-\mathrm{z}_{\mathrm{B}}|^{2}}.
\end{align}
By introducing $\mathrm{z}_{\mathrm{A}}=\frac{\mathrm{z}_{u}+\mathrm{z}_{v}%
}{2}$ and $\mathrm{z}_{\mathrm{B}}=\frac{\mathrm{z}_{v}-\mathrm{z}_{u}}{2},$
above equations are transform into\cite{1}
\begin{align}
\frac{\partial \mathrm{z}_{u}}{\partial t}  &  =i\frac{(\alpha \kappa \ln
\epsilon)}{2}\frac{\partial^{2}\mathrm{z}_{u}}{\partial \vec{x}^{2}}%
+i(\alpha \kappa \ln \epsilon)\frac{2\mathrm{z}_{u}}{|\mathrm{z}_{u}|^{2}%
},\nonumber \\
\frac{\partial \mathrm{z}_{v}}{\partial t}  &  =i\frac{(\alpha \kappa \ln
\epsilon)}{2}\frac{\partial^{2}\mathrm{z}_{v}}{\partial \vec{x}^{2}}\text{.}%
\end{align}
The plane Kelvin wave solutions are obtained as
\begin{equation}
\mathrm{z}_{u}=r_{0}e^{i(\vec{k}\cdot \vec{x}-\omega_{u}t+\phi_{u})}%
,\quad \mathrm{z}_{v}=r_{0}e^{i(\vec{k}\cdot \vec{x}-\omega_{v}t+\phi_{v})},
\end{equation}
where $\phi_{u}$ and $\phi_{v}$ are constant phase angles, and the angular
frequencies are
\begin{equation}
\omega_{u}=\frac{(\alpha \kappa \ln \epsilon)}{2}\vec{k}^{2}-(\alpha \kappa
\ln \epsilon)\frac{2}{r_{0}^{2}}%
\end{equation}
$\quad$and$\quad$%
\begin{equation}
\omega_{v}=\frac{(\alpha \kappa \ln \epsilon)}{2}\vec{k}^{2}.
\end{equation}
In this paper, we set $\phi_{u}=0$ and $\phi_{v}=0$.

For two entangled vortex-membranes, the nonlocal interaction leads to
leapfrogging motion. Above solutions of the Biot-Savart equation can be
written into
\begin{align}
\left(
\begin{array}
[c]{c}%
\mathrm{z}_{\mathrm{A}}(\vec{x},t)\\
\mathrm{z}_{\mathrm{B}}(\vec{x},t)
\end{array}
\right)   &  =\frac{r_{0}}{2}\left(
\begin{array}
[c]{c}%
1+e^{i\omega^{\ast}t}\\
1-e^{i\omega^{\ast}t}%
\end{array}
\right)  e^{i\vec{k}\cdot \vec{x}-i\omega_{0}t}\nonumber \\
&  =\left(
\begin{array}
[c]{c}%
r_{\mathrm{A}}\\
r_{\mathrm{B}}%
\end{array}
\right)  e^{i\vec{k}\cdot \vec{x}-i\omega_{0}t+i\omega^{\ast}t/2}%
\end{align}
where the winding radii of two vortex-membranes are $r_{\mathrm{A}}=r_{0}%
\cos(\frac{\omega^{\ast}t}{2})$ and $r_{\mathrm{B}}=-r_{0}i\sin(\frac
{\omega^{\ast}t}{2})$, respectively. $\omega_{0}=\omega_{v}=\frac
{(\alpha \kappa \ln \epsilon)}{2}\vec{k}^{2}$ and$\quad \omega^{\ast}=\omega
_{v}-\omega_{u}=(\alpha \kappa \ln \epsilon)\frac{2}{r_{0}^{2}}$ denote (global)
rotating velocity frequency and (internal) leapfrogging angular frequency,
respectively. From above solutions, we have a constraint condition of total
volume, i.e.,
\begin{equation}
\left \vert r_{\mathrm{A}}\right \vert ^{2}+\left \vert r_{\mathrm{B}}\right \vert
^{2}\equiv r_{0}^{2}.
\end{equation}

For global rotating motion with finite $\omega_{0}$, the entangled
vortex-membranes are described by plane waves $e^{i\vec{k}\cdot \vec{x}%
-i\omega_{0}t}.$ For leapfrogging motion with finite $\omega^{\ast}$, the
entangled vortex-membranes exchange energy in a periodic fashion. The winding
radii of two vortex-membranes oscillate with a period $T=\frac{2\pi}%
{\omega^{\ast}}$: At $t=0,$ helical vortex-membrane-A winds around straight
vortex-membrane-B clockwise (the state $\left[  \mathrm{A}\right \rangle $); At
$t=\frac{T}{4},$ the system becomes a symmetric double helix vortex-membranes
(the state $((i+1)\left[  \mathrm{A}\right \rangle -(i-1)\left[  \mathrm{B}%
\right \rangle )/2$); At $t=\frac{T}{2},$ helical vortex-membrane-B winds
around straight vortex-membrane-A (the state $\left[  \mathrm{B}\right \rangle
$), ...

During leapfrogging process, the Lamb impulse (momentum) and the angular Lamb
impulse (angular momentum) are conserved. The effective Planck constant is
obtained as projected (Lamb impulse) angular momentum as $\hbar_{\mathrm{eff}%
}=J_{\mathrm{Lamb}}=\frac{1}{2}\rho_{0}\kappa r_{0}^{2}\cdot V_{P}$ where
$V_{P}$ is the total volume of the system.

We then map the Biot-Savart mechanics to the pseudo-quantum mechanics. The
system with two entangled vortex-membranes is mapped to a two-component
particle (not a two-particle case), i.e.,
\begin{align}
\left(
\begin{array}
[c]{c}%
i\hbar_{\mathrm{eff}}\frac{d\mathrm{z}_{\mathrm{A}}(\vec{x},t)}{dt}\\
i\hbar_{\mathrm{eff}}\frac{d\mathrm{z}_{\mathrm{B}}(\vec{x},t)}{dt}%
\end{array}
\right)   &  =\hat{H}_{\mathrm{Lamb}}\left(
\begin{array}
[c]{c}%
\mathrm{z}_{\mathrm{A}}(\vec{x},t)\\
\mathrm{z}_{\mathrm{B}}(\vec{x},t)
\end{array}
\right) \nonumber \\
&  =[\frac{\hat{p}_{\mathrm{Lamb}}^{2}}{2m_{\mathrm{pseudo}}}+\frac
{\hbar_{\mathrm{eff}}\omega^{\ast}}{2}\cdot(\tau_{x}-\vec{1})]\left(
\begin{array}
[c]{c}%
\mathrm{z}_{\mathrm{A}}(\vec{x},t)\\
\mathrm{z}_{\mathrm{B}}(\vec{x},t)
\end{array}
\right)
\end{align}
with a constraint
\begin{equation}
\left \vert \mathrm{z}_{\mathrm{A}}(\vec{x},t)\right \vert ^{2}+\left \vert
\mathrm{z}_{\mathrm{B}}(\vec{x},t)\right \vert ^{2}=1.
\end{equation}
Here, $\tau_{x}=(%
\begin{array}
[c]{cc}%
0 & 1\\
1 & 0
\end{array}
)$ is Pauli matrix and $\vec{1}=(%
\begin{array}
[c]{cc}%
1 & 0\\
0 & 1
\end{array}
)$. The leapfrogging process is characterized by $\frac{\hbar_{\mathrm{eff}%
}\omega^{\ast}}{2}\cdot(\tau_{x}-\vec{1})$. As a result, the pseudo-quantum
mechanics describes the dynamics of two entangled vortex-membranes with fixed
volume in the 5D fluid. Under the constraint, the total degree of freedom of a
vortex-membrane is reduced to $2$.

In pseudo-quantum mechanics, there are also three conserved physical
quantities for entangled vortex-membranes: the total energy $H(\vec
{p}_{\mathrm{Lamb}})$ that is proportional to the total volume of the two
vortex-membranes $V_{P}=\mathrm{volume}(P)$; the total momentum $\vec
{p}_{\mathrm{Lamb}}$ that is proportional to the linking number between two
vortex-membranes (see below discussion); and the total (Lamb impulse) angular
momentum (the effective Planck constant $\hbar_{\mathrm{eff}}$) that is
proportional to the total volume of the two vortex-membranes in the 5D fluid
$V_{P}\cdot r_{0}^{2}$.

\subsection{Linking number and linking-number density}

In above part, we derive the equation of motion of two entangled
vortex-membranes. However, due to the leapfrogging process, the winding number
and winding-number density are not well defined. Instead, it is the linking
number and linking-number density that characterize the entanglement between
two vortex-membranes.

Firstly, we introduce the linking number to characterize entanglement between
two vortex-membranes.

The (Gauss) linking-number $\zeta_{1D}$ for two 1D vortex-lines is a
topological invariable to characterize the entanglement that is defined
as\cite{gauss}
\begin{equation}
\zeta_{1D}=\frac{1}{4\pi}\oint_{C_{\mathrm{A}}}\oint_{C_{\mathrm{B}}}%
\frac{(\mathbf{s}_{\mathrm{A}}-\mathbf{s}_{\mathrm{B}})\cdot d\mathbf{s}%
_{\mathrm{A}}\times d\mathbf{s}_{\mathrm{B}}}{|\mathbf{s}_{\mathrm{A}%
}-\mathbf{s}_{\mathrm{B}}|^{3}}.
\end{equation}
The 1D linking number is also integer number. The system with different
linking numbers has different entanglement patterns between two vortex-lines.

To locally characterize the entanglement, we define the density of
linking-number $\rho_{\mathrm{link}},$
\begin{equation}
\rho_{\mathrm{link}}=\frac{\Delta \zeta_{1D}}{\Delta x}.
\end{equation}
For the entangled vortex-membranes, owing to the intrinsic reversal symmetry
between vortex-membrane-\textrm{A} and vortex-membrane-\textrm{B}, we can
obtain the linking-number density from the winding-number density at special
time of the leapfrogging motion, $t=0,$ i.e.,
\begin{equation}
\left(
\begin{array}
[c]{c}%
\mathrm{z}_{\mathrm{A}}(x,t)\\
\mathrm{z}_{\mathrm{B}}(x,t)
\end{array}
\right)  =\left(
\begin{array}
[c]{c}%
r_{0}e^{ik\cdot x-i\omega t}\\
0
\end{array}
\right)  .
\end{equation}
At this time, the winding-number density of vortex-membrane-\textrm{A} is
$\rho_{\mathrm{wind,A}}=\frac{k}{2\pi}$. The linking-number density is equal
to $\rho_{\mathrm{wind,A}},$
\begin{equation}
\rho_{\mathrm{link}}=\rho_{\mathrm{wind,A}}=\frac{\zeta_{1D}}{l}=\frac{k}%
{2\pi}.
\end{equation}
Because the linking number is a conserved quantity and doesn't change during
leapfrogging motion, the corresponding operator of linking-number density is
given by
\begin{equation}
\rho_{\mathrm{linking}}\rightarrow \hat{\rho}_{\mathrm{linking}}=\frac{\hat{k}%
}{2\pi}=-i\frac{1}{2\pi}\frac{d}{dx}.
\end{equation}
For two entangled 1D vortex-line, the linking-number is defined by
\begin{align}
\zeta_{1D}  &  =\left \langle \mathbf{Z}(x,t)\right]  (l\cdot \hat{\rho
}_{\mathrm{linking}})\left[  \mathbf{Z}(x,t)\right \rangle \nonumber \\
&  =\int \mathbf{Z}^{\ast}(x,t)(-i\frac{1}{2\pi}\frac{1}{r_{0}^{2}}\frac{d}%
{dx})\mathbf{Z}(x,t)dx\nonumber \\
&  =\int \mathrm{z}_{\mathrm{A}}^{\ast}(x,t)(-i\frac{1}{2\pi}\frac{1}{r_{0}%
^{2}}\frac{d}{dx})\mathrm{z}_{\mathrm{A}}(x,t)dx\nonumber \\
&  +\int \mathrm{z}_{\mathrm{B}}^{\ast}(x,t)(-i\frac{1}{2\pi}\frac{1}{r_{0}%
^{2}}\frac{d}{dx})\mathrm{z}_{\mathrm{B}}(x,t)dx
\end{align}
where $\mathbf{Z}(x,t)=\left(
\begin{array}
[c]{c}%
\mathrm{z}_{\mathrm{A}}(\vec{x},t)\\
\mathrm{z}_{\mathrm{B}}(\vec{x},t)
\end{array}
\right)  .$ Thus, the linking-number density that is proportional to the total
momentum $p_{\mathrm{Lamb}}$. is%
\begin{align}
\rho_{\mathrm{linking}}  &  =\frac{1}{r_{0}^{2}}\mathbf{Z}^{\ast}%
(x,t)\hat{\rho}_{\mathrm{linking}}\mathbf{Z}(x,t)\nonumber \\
&  =\mathrm{z}_{\mathrm{A}}^{\ast}(x,t)(-i\frac{1}{2\pi}\frac{1}{r_{0}^{2}%
}\frac{d}{dx})\mathrm{z}_{\mathrm{A}}(x,t)\nonumber \\
&  +\mathrm{z}_{\mathrm{B}}^{\ast}(x,t)(-i\frac{1}{2\pi}\frac{1}{r_{0}^{2}%
}\frac{d}{dx})\mathrm{z}_{\mathrm{B}}(x,t).
\end{align}

In 5D space, the 2D Gauss linking number can only be defined between two 2D
closed, oriented, submanifolds given by\cite{comm}
\begin{equation}
\zeta_{2D}=\frac{1}{\mathrm{vol}\,S^{n}}\int_{K\times L}\frac{\Omega_{k,\ell
}(\alpha)}{|x|^{k+1}|y|^{\ell+1}\sin^{n}\alpha}\,[x,dx,y,dy]
\end{equation}
where $\Omega_{k,\ell}(\alpha)=\int_{\theta=\alpha}^{\pi}\sin^{k}%
(\theta-\alpha)\sin^{\ell}\theta \,d\theta.$ Here $\alpha(x,y)$ is the angle
between $x\in X$ and $y\in Y$, thought of as vectors in 5D space. So there is
no 3D linking number to characterize the entanglement between two 3D
vortex-membranes in 5D fluid.

Instead, we discuss entanglement between two 3D vortex-membranes via
(projected) 1D linking number. There are three 1D linking-numbers $\zeta
_{1D}^{I}$ ($I=X,Y,Z$) along different directions that are
\begin{equation}
\zeta_{1D}^{I}=\frac{1}{4\pi}\oint_{C_{x^{I},\mathrm{A}}}\oint_{C_{x^{I}%
,\mathrm{B}}}\frac{(\mathbf{s}_{\mathrm{A}}^{I}-\mathbf{s}_{\mathrm{B}}%
^{I})\cdot d\mathbf{s}_{\mathrm{A}}^{I}\times d\mathbf{s}_{\mathrm{B}}^{I}%
}{|\mathbf{s}_{\mathrm{A}}^{I}-\mathbf{s}_{\mathrm{B}}^{I}|^{3}}%
\end{equation}
where $\mathbf{s}^{I}=\mathbf{r}\cdot \vec{e}^{I}$. The vector of linking
number is defined as $\vec{\zeta}_{1D}=(\zeta_{1D}^{X},\zeta_{1D}^{Y}%
,\zeta_{1D}^{Z})$. We define the vector of linking-number densities and
linking-number density operators for 3D vortex-membranes,
\begin{equation}
\vec{\rho}_{\mathrm{linking}}^{I}(x,t)=\frac{1}{r_{0}^{2}}\mathbf{Z}^{\ast
}(x,t)\mathbf{\hat{\rho}}_{\mathrm{linking}}^{I}\mathbf{Z}(x,t),
\end{equation}
and
\begin{equation}
\mathbf{\hat{\rho}}_{\mathrm{linking}}=(-\frac{i}{2\pi r_{0}^{2}}\frac
{d}{dx^{X}},-\frac{i}{2\pi r_{0}^{2}}\frac{d}{dx^{Y}},-\frac{i}{2\pi r_{0}%
^{2}}\frac{d}{dx^{Z}}),
\end{equation}
respectively. For a vortex-membrane described by a plane Kelvin wave
$\mathrm{z}_{\mathrm{A}}(x^{I},t)=r_{0}e^{-i\omega \cdot t+ik^{I}\cdot x^{I}}$
and a constant vortex-line described by $\mathrm{z}_{\mathrm{B}}%
(x^{I},t)\equiv0$, the 1D linking-number along direction $\vec{e}^{I}$ is
given by $\zeta_{1D}^{I}=\frac{\left \vert k^{I}\right \vert }{2\pi}L_{I}$ and
the density of linking-number $\rho_{\mathrm{link}}$ is $\rho_{\mathrm{link}%
}^{I}=\frac{\zeta_{1D}^{I}}{L_{I}}.$

As a result, we answer the question -- \emph{how to characterize the evolution
of non-uniform entangled vortex-membranes?} For a given initial state
$\mathbf{Z}(\vec{x})$, the linking-number density for the local deformation of
entangled vortex-membranes is obtained as $\vec{\rho}_{\mathrm{linking}}%
^{I}(\vec{x},t)=\frac{1}{r_{0}^{2}}\mathbf{Z}^{\ast}(\vec{x},t)\mathbf{\hat
{\rho}}_{\mathrm{linking}}^{I}\mathbf{Z}(\vec{x},t)$ where
\begin{align}
\mathbf{Z}(\vec{x})  &  =%
%TCIMACRO{\dsum \nolimits_{k}}%
%BeginExpansion
{\displaystyle \sum \nolimits_{k}}
%EndExpansion
c_{k}e^{i\vec{k}\cdot \vec{x}}\\
&  \rightarrow \mathbf{Z}(\vec{x},t)=%
%TCIMACRO{\dsum \nolimits_{k}}%
%BeginExpansion
{\displaystyle \sum \nolimits_{k}}
%EndExpansion
c_{k}e^{i\vec{k}\cdot \vec{x}}e^{-i\omega \cdot t}.\nonumber
\end{align}

From the point view of topology, the Sommerfeld quantization condition for
two-component particle is the topological condition of the linking number,
i.e.,%
\[%
%TCIMACRO{\doint }%
%BeginExpansion
{\displaystyle \oint}
%EndExpansion
p_{\mathrm{Lamb}}\cdot dx=h_{\mathrm{eff}}\cdot \zeta_{1D}.
\]
where $\zeta_{1D}$ is the linking number of the two entangled vortex-membranes.

\subsection{Tensor representation for entangled vortex-membranes}

\subsubsection{Tensor states for plane Kelvin waves}

In this part, we introduce the tensor representation for Kelvin waves by
separating the wave vector along a given direction into positive part and
negative part, $\pm \vec{k}$.

\begin{figure}[ptb]
\includegraphics[clip,width=0.65\textwidth]{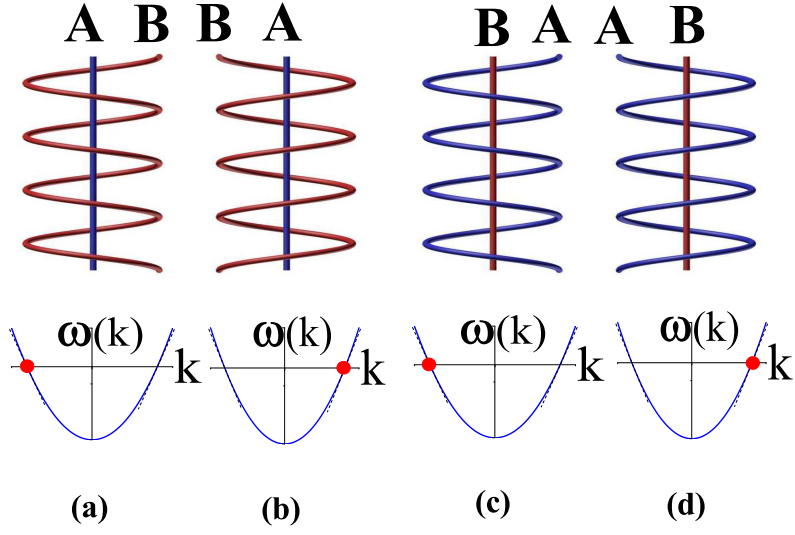}\caption{An illustration
of basis of four internal degrees of freedom for plane Kelvin waves in
pseudo-quantum mechanics, $\left[  \vec{k},\mathrm{B}\right \rangle $, $\left[
-\vec{k},\mathrm{B}\right \rangle ,$ $\left[  \vec{k},\mathrm{A}\right \rangle
,$ $\left[  -\vec{k},\mathrm{A}\right \rangle .$ The lower figures illustrate
linear dispersion for the four elementary states. The red dots denote the four
possible elementary states with opposite wave vectors of the Kelvin waves. }%
\end{figure}

For a plane Kelvin wave with fixed wavelength $2a=\frac{2\pi}{\left \vert
\vec{k}\right \vert }$ along $\vec{e}$-direction, there are two helical degrees
of freedom: $\mathrm{z}_{\vec{k}}(\vec{x},t)$ at $\vec{k}$ or $\mathrm{z}%
_{-\vec{k}}(\vec{x},t)$ at $-\vec{k}$. After considering the vortex degrees of
freedom $\mathrm{A}$ or $\mathrm{B}$ that characterize the Kelvin waves on
different vortex-membranes, there are four degenerate states to characterize a
plane Kelvin wave with fixed wave-length $2a$, i.e., the basis is
\begin{align}
\left[  \vec{k},\mathrm{A}\right \rangle  &  =\left(
\begin{array}
[c]{c}%
\mathrm{z}_{\vec{k},\mathrm{A}}(\vec{x})\\
0\\
0\\
0
\end{array}
\right)  ,\text{ }\left[  \vec{k},\mathrm{B}\right \rangle =\left(
\begin{array}
[c]{c}%
0\\
\mathrm{z}_{\vec{k},\mathrm{B}}(\vec{x})\\
0\\
0
\end{array}
\right)  ,\\
\left[  -\vec{k},\mathrm{A}\right \rangle  &  =\left(
\begin{array}
[c]{c}%
0\\
0\\
\mathrm{z}_{-\vec{k},\mathrm{A}}(\vec{x})\\
0
\end{array}
\right)  ,\text{ }\left[  -\vec{k},\mathrm{B}\right \rangle =\left(
\begin{array}
[c]{c}%
0\\
0\\
0\\
\mathrm{z}_{-\vec{k},\mathrm{B}}(\vec{x})
\end{array}
\right)  .\nonumber
\end{align}
Here, $\left[  \vec{k},\mathrm{A}\right \rangle $ denotes a plane Kelvin-wave
with wave vector $\vec{k}$ on vortex-membrane-\textrm{A, }%
\begin{equation}
\mathrm{z}_{\vec{k},\mathrm{A}}(\vec{x})=r_{0}e^{i\vec{k}\cdot \vec{x}};
\end{equation}
$\left[  \vec{k},\mathrm{B}\right \rangle $ denotes a plane Kelvin-wave with
wave vector $\vec{k}$ on vortex-membrane-\textrm{B, }%
\begin{equation}
\mathrm{z}_{\vec{k},\mathrm{B}}(\vec{x})=r_{0}e^{i\vec{k}\cdot \vec{x}};
\end{equation}
$\left[  -\vec{k},\mathrm{A}\right \rangle $ denotes a plane Kelvin-wave with
wave vector $-\vec{k}$ on vortex-membrane-\textrm{A, }%
\begin{equation}
\mathrm{z}_{-\vec{k},\mathrm{A}}(\vec{x})=r_{0}e^{-i\vec{k}\cdot \vec{x}};
\end{equation}
$\left[  -\vec{k},\mathrm{B}\right \rangle $ denotes a plane Kelvin-wave with
wave vector $-\vec{k}$ on vortex-membrane-\textrm{B, }%
\begin{equation}
\mathrm{z}_{-\vec{k},\mathrm{B}}(\vec{x})=r_{0}e^{-i\vec{k}\cdot \vec{x}}.
\end{equation}
See the illustration in Fig.7.

Thus, an arbitrary plane Kelvin wave can be characterized by a superposed
state of the basis along different spatial directions. We introduce a tensor
representation to define the plane Kelvin waves with fixed wave-length,%
\begin{equation}
\left[  \mathbf{Z}\right \rangle =%
%TCIMACRO{\dprod \limits_{I}}%
%BeginExpansion
{\displaystyle \prod \limits_{I}}
%EndExpansion
\left[  \mathbf{Z}^{I}\right \rangle
\end{equation}
with
\begin{equation}
\left[  \mathbf{Z}^{I}\right \rangle =\left(
\begin{array}
[c]{c}%
\alpha_{1,1}^{I}\\
\alpha_{1,-1}^{I}\\
\alpha_{-1,1}^{I}\\
\alpha_{-1,-1}^{I}%
\end{array}
\right)
\end{equation}
where $I$ denotes the spatial indices, $X,$ $Y$, $Z,$ ...; $i=1,$ $-1$ denotes
helical degrees of freedom $\vec{k}$ or $-\vec{k}$; $j=1,$ $-1$ denotes the
vortex-degrees of freedom \textrm{A} or \textrm{B}. In general, there are
$d\times2\times2$ elements $\alpha_{i,j}^{I}$. Each element $\alpha_{i,j}^{I}$
is proportional to the winding radius of the corresponding plane Kelvin waves.
According to the geometric constraint condition, we have
\begin{align}
\left \vert \alpha_{1,1}^{X}\right \vert ^{2}+\left \vert \alpha_{1,-1}%
^{X}\right \vert ^{2}+\left \vert \alpha_{-1,1}^{X}\right \vert ^{2}+\left \vert
\alpha_{-1,-1}^{X}\right \vert ^{2}  &  =1,\\
\left \vert \alpha_{1,1}^{Y}\right \vert ^{2}+\left \vert \alpha_{1,-1}%
^{Y}\right \vert ^{2}+\left \vert \alpha_{-1,1}^{Y}\right \vert ^{2}+\left \vert
\alpha_{-1,-1}^{Y}\right \vert ^{2}  &  =1,\nonumber \\
\left \vert \alpha_{1,1}^{Z}\right \vert ^{2}+\left \vert \alpha_{1,-1}%
^{Z}\right \vert ^{2}+\left \vert \alpha_{-1,1}^{Z}\right \vert ^{2}+\left \vert
\alpha_{-1,-1}^{Z}\right \vert ^{2}  &  =1,\nonumber \\
&  ...\nonumber
\end{align}
So the resulting function for a Kelvin wave with fixed wave-length is given
by
\begin{align}
\mathbf{Z}^{I}(\vec{x})  &  =\alpha_{1,1}^{I}\mathrm{z}_{\vec{k},\mathrm{A}%
}(\vec{x})+\alpha_{1,-1}^{I}z_{\vec{k},\mathrm{B}}(\vec{x})\nonumber \\
&  +\alpha_{-1,1}^{I}z_{-\vec{k},\mathrm{A}}(\vec{x})+\alpha_{-1,-1}%
^{I}z_{-\vec{k},\mathrm{B}}(\vec{x}).
\end{align}
Owing to the fact of the energy degeneracy, all superposed states that consist
of four degenerate states along different spatial directions denoted by
different weights $\alpha_{i,j}^{I}$ have the same energy.

\subsubsection{Tensor operators and tensor order}

The tensor states of a plane Kelvin wave can be classified by operator representation.

We define the $d\times3\times4$ tensor operators $\hat{\Gamma}_{i,j}%
^{I}=\sigma_{i}^{I}\otimes \tau_{j}^{I}$ ($I=X,Y,Z,...;$ $i=x,y,z$ for helical
degrees of freedom; $j=0,x,y,z$ for the vortex-degrees of freedom) to be%
\begin{align}
\hat{\Gamma}_{x,x}^{X}  &  =\sigma_{x}^{X}\otimes \tau_{x}^{X},\text{ }%
\hat{\Gamma}_{x,x}^{Y}=\sigma_{x}^{Y}\otimes \tau_{x}^{Y},\text{ }\hat{\Gamma
}_{x,x}^{Z}=\sigma_{x}^{Z}\otimes \tau_{x}^{Z},\text{ }\hat{\Gamma}_{x,0}%
^{X}=\sigma_{x}^{X}\otimes \tau_{0}^{X},\\
\hat{\Gamma}_{x,y}^{X}  &  =\sigma_{x}^{X}\otimes \tau_{y}^{X},\text{ }%
\hat{\Gamma}_{x,y}^{Y}=\sigma_{x}^{Y}\otimes \tau_{y}^{Y},\text{ }\hat{\Gamma
}_{x,y}^{Z}=\sigma_{x}^{Z}\otimes \tau_{y}^{Z},\text{ }\hat{\Gamma}_{x,0}%
^{Y}=\sigma_{x}^{Y}\otimes \tau_{0}^{Y},\nonumber \\
\hat{\Gamma}_{x,z}^{X}  &  =\sigma_{x}^{X}\otimes \tau_{z}^{X},\text{ }%
\hat{\Gamma}_{x,z}^{Y}=\sigma_{x}^{Y}\otimes \tau_{z}^{Y},\text{ }\hat{\Gamma
}_{x,z}^{Z}=\sigma_{x}^{Z}\otimes \tau_{z}^{Z},\text{ }\hat{\Gamma}_{x,0}%
^{Z}=\sigma_{x}^{Z}\otimes \tau_{0}^{Z},\nonumber \\
\hat{\Gamma}_{y,x}^{X}  &  =\sigma_{y}^{X}\otimes \tau_{x}^{X},\text{ }%
\hat{\Gamma}_{y,x}^{Y}=\sigma_{y}^{Y}\otimes \tau_{x}^{Y},\text{ }\hat{\Gamma
}_{y,x}^{Z}=\sigma_{y}^{Z}\otimes \tau_{x}^{Z},\text{ }\hat{\Gamma}_{y,0}%
^{X}=\sigma_{y}^{X}\otimes \tau_{0}^{X},\nonumber \\
\hat{\Gamma}_{y,y}^{X}  &  =\sigma_{y}^{X}\otimes \tau_{y}^{X},\text{ }%
\hat{\Gamma}_{y,y}^{Y}=\sigma_{y}^{Y}\otimes \tau_{y}^{Y},\text{ }\hat{\Gamma
}_{y,y}^{Z}=\sigma_{y}^{Z}\otimes \tau_{y}^{Z},\text{ }\hat{\Gamma}_{y,0}%
^{Y}=\sigma_{y}^{Y}\otimes \tau_{0}^{Y},\nonumber \\
\hat{\Gamma}_{y,z}^{X}  &  =\sigma_{y}^{X}\otimes \tau_{z}^{X},\text{ }%
\hat{\Gamma}_{y,z}^{Y}=\sigma_{y}^{Y}\otimes \tau_{z}^{Y},\text{ }\hat{\Gamma
}_{y,z}^{Z}=\sigma_{y}^{Z}\otimes \tau_{z}^{Z},\text{ }\hat{\Gamma}_{z,z}%
^{Z}=\sigma_{y}^{Z}\otimes \tau_{0}^{Z},\nonumber \\
\hat{\Gamma}_{z,x}^{X}  &  =\sigma_{z}^{X}\otimes \tau_{x}^{X},\text{ }%
\hat{\Gamma}_{z,x}^{Y}=\sigma_{z}^{Y}\otimes \tau_{x}^{Y},\text{ }\hat{\Gamma
}_{z,x}^{Z}=\sigma_{z}^{Z}\otimes \tau_{x}^{Z},\text{ }\hat{\Gamma}_{z,0}%
^{X}=\sigma_{z}^{X}\otimes \tau_{0}^{X},\nonumber \\
\hat{\Gamma}_{z,y}^{X}  &  =\sigma_{z}^{X}\otimes \tau_{y}^{X},\text{ }%
\hat{\Gamma}_{z,y}^{Y}=\sigma_{z}^{Y}\otimes \tau_{y}^{Y},\text{ }\hat{\Gamma
}_{z,y}^{Z}=\sigma_{z}^{Z}\otimes \tau_{y}^{Z},\text{ }\hat{\Gamma}_{z,0}%
^{Y}=\sigma_{z}^{Y}\otimes \tau_{0}^{Y},\nonumber \\
\hat{\Gamma}_{z,z}^{X}  &  =\sigma_{z}^{X}\otimes \tau_{z}^{X},\text{ }%
\hat{\Gamma}_{z,z}^{Y}=\sigma_{z}^{Y}\otimes \tau_{z}^{Y},\text{ }\hat{\Gamma
}_{z,z}^{Z}=\sigma_{z}^{Z}\otimes \tau_{z}^{Z},\text{ }\hat{\Gamma}_{z,0}%
^{Z}=\sigma_{z}^{Z}\otimes \tau_{0}^{Z},\nonumber \\
&  ...\nonumber
\end{align}
Here, $\tau_{0}^{X/Y/Z}=\vec{1}=\left(
\begin{array}
[c]{cc}%
1 & 0\\
0 & 1
\end{array}
\right)  $. Different tensor states have different tensor operators
\begin{align}
\left \langle \mathbf{\hat{\Gamma}}^{I}\right \rangle  &  =\left \langle
\mathbf{Z}_{\mathrm{knot-crystal}}\right]  \mathbf{\hat{\Gamma}}^{I}\left[
\mathbf{Z}_{\mathrm{knot-crystal}}\right \rangle \\
&  =\vec{\Gamma}^{I}=\vec{n}_{\sigma}^{I}\otimes(\vec{n}_{\tau}^{I}%
,\tau)\nonumber
\end{align}
where $\mathbf{\hat{\Gamma}}^{I}=\mathbf{\sigma}^{I}\otimes(\mathbf{\tau}%
^{I},\vec{1})$ and $\mathbf{\sigma}^{I}$, $\mathbf{\tau}^{I}$ are $2\times2$
Pauli matrices for helical and vortex degrees of freedom, respectively. On the
basis of pseudo-spin base $(\tau_{x}^{I},\tau_{y}^{I},\tau_{z}^{I},\vec{1})$,
the tensor states are determined by the "direction" of pseudo-spin order for
helical degrees of freedom,
\begin{align}
\left \langle \mathbf{\tau}^{I}\right \rangle  &  =\left \langle \mathbf{Z}%
\right]  (\vec{1}\otimes \tau_{i}^{I})\left[  \mathbf{Z}\right \rangle
\nonumber \\
&  =\left(  \left \langle \vec{1}\otimes \tau_{x}^{I}\right \rangle ,\left \langle
\vec{1}\otimes \tau_{y}^{I}\right \rangle ,\left \langle \vec{1}\otimes \tau
_{z}^{I}\right \rangle \right)  =\vec{n}_{\tau}^{I},\nonumber \\
\left \langle \vec{1}\right \rangle  &  =\left \langle \mathbf{Z}\right]
(\vec{1}\otimes \vec{1})\left[  \mathbf{Z}\right \rangle =\tau
\end{align}
In general, we have $\left \vert \vec{n}_{\tau}^{I}\right \vert ^{2}+\left \vert
\tau \right \vert ^{2}=1$. On the basis of pseudo-spin base $(\sigma_{x}%
^{I},\sigma_{y}^{I},\sigma_{z}^{I})$, the tensor states are determined by the
"direction" of pseudo-spin order for helical degrees of freedom,
\begin{align}
\left \langle \mathbf{\sigma}^{I}\right \rangle  &  =\left \langle \mathbf{Z}%
\right]  (\sigma_{i}^{I}\otimes \vec{1})\left[  \mathbf{Z}\right \rangle \\
&  =\left(  \left \langle \sigma_{x}^{I}\otimes \vec{1}\right \rangle
,\left \langle \sigma_{y}^{I}\otimes \vec{1}\right \rangle ,\left \langle
\sigma_{z}^{I}\otimes \vec{1}\right \rangle \right)  =\vec{n}_{\sigma}%
^{I}.\nonumber
\end{align}
In general, we have $\left \vert \vec{n}_{\sigma}^{I}\right \vert ^{2}=1$. For
example, $\vec{n}_{\sigma}^{X}=(1,0,0)$ is a tensor state along spatial
x-direction and helical x-direction, $\left \langle \sigma_{x}^{X}\otimes
\vec{1}\right \rangle =1,$ $\left \langle \sigma_{y}^{X}\otimes \vec
{1}\right \rangle =0,$ $\left \langle \sigma_{z}^{X}\otimes \vec{1}\right \rangle
=0$. Because the vortex-degrees of freedom are always trivial, the
vortex-vector $\vec{n}_{\tau}^{I}$ is a constant vector along different
directions, i.e., $\vec{n}_{\tau}^{X}=\vec{n}_{\tau}^{Y}=\vec{n}_{\tau}%
^{Z}=\vec{n}_{\tau}$ and $\tau=\tau_{0}$. Therefore, in the following parts,
we focus on the tensor states $\vec{n}_{\sigma}^{I}$ for helical degrees of freedom.

\subsubsection{Example}

For a 1D travelling Kelvin wave $\mathbf{Z}(\vec{x})=\left(
\begin{array}
[c]{c}%
\mathrm{z}_{\mathrm{A}}(\vec{x})\\
\mathrm{z}_{\mathrm{B}}(\vec{x})
\end{array}
\right)  =r_{0}e^{i\vec{k}\cdot \vec{x}}\left(
\begin{array}
[c]{c}%
1\\
1
\end{array}
\right)  ,$ or $\mathbf{Z}(\vec{x})=\left(
\begin{array}
[c]{c}%
\mathrm{z}_{\mathrm{A}}(\vec{x})\\
\mathrm{z}_{\mathrm{B}}(\vec{x})
\end{array}
\right)  =r_{0}e^{-i\vec{k}\cdot \vec{x}}\left(
\begin{array}
[c]{c}%
1\\
1
\end{array}
\right)  $, the tensor state is represented by
\begin{align}
\left[  \mathbf{Z}^{X}\right \rangle  &  =\left(
\begin{array}
[c]{c}%
\alpha_{1,1}^{I}\\
\alpha_{1,-1}^{I}\\
\alpha_{-1,1}^{I}\\
\alpha_{-1,-1}^{I}%
\end{array}
\right) \nonumber \\
&  =\left(
\begin{array}
[c]{c}%
1\\
1\\
0\\
0
\end{array}
\right)  ,
\end{align}
or
\begin{align}
\left[  \mathbf{Z}^{X}\right \rangle  &  =\left(
\begin{array}
[c]{c}%
\alpha_{1,1}^{I}\\
\alpha_{1,-1}^{I}\\
\alpha_{-1,1}^{I}\\
\alpha_{-1,-1}^{I}%
\end{array}
\right) \\
&  =\left(
\begin{array}
[c]{c}%
0\\
0\\
1\\
1
\end{array}
\right)  ,\nonumber
\end{align}
of which the tensor state is denoted by
\begin{equation}
\left \langle \mathbf{\sigma}^{X}(x)\otimes \vec{1}\right \rangle =\vec
{n}_{\sigma}^{X}=(0,0,\pm1).
\end{equation}
We call it plane $\sigma_{z}$-Kelvin wave.

For a standing Kelvin wave described by $\mathbf{Z}(\vec{x})=\left(
\begin{array}
[c]{c}%
\mathrm{z}_{\mathrm{A}}(\vec{x})\\
\mathrm{z}_{\mathrm{B}}(\vec{x})
\end{array}
\right)  =\sqrt{2}r_{0}\cos \left(  k\cdot x\right)  \left(
\begin{array}
[c]{c}%
1\\
1
\end{array}
\right)  $ or $i\sqrt{2}r_{0}\sin \left(  k\cdot x\right)  \left(
\begin{array}
[c]{c}%
1\\
1
\end{array}
\right)  $, the tensor state is represented by
\begin{equation}
\left[  \mathbf{Z}^{X}\right \rangle =\frac{1}{\sqrt{2}}\left(
\begin{array}
[c]{c}%
1\\
1\\
1\\
1
\end{array}
\right)  ,
\end{equation}
or
\begin{equation}
\left[  \mathbf{Z}^{X}\right \rangle =\frac{1}{\sqrt{2}}\left(
\begin{array}
[c]{c}%
1\\
1\\
-1\\
-1
\end{array}
\right)  ,
\end{equation}
of which the tensor state is denoted by
\begin{equation}
\left \langle \mathbf{\sigma}^{X}(x)\otimes \vec{1}\right \rangle =\vec
{n}_{\sigma}^{X}=(\pm1,0,0).
\end{equation}
We call it $\sigma_{x}$-Kelvin wave.

Next, we discuss the 2D plane Kelvin waves.

For a 2D $\sigma_{z}$-Kelvin wave described by $\mathbf{Z}(\vec{x})=\left(
\begin{array}
[c]{c}%
\mathrm{z}_{\mathrm{A}}(\vec{x})\\
\mathrm{z}_{\mathrm{B}}(\vec{x})
\end{array}
\right)  =r_{0}\left(
\begin{array}
[c]{c}%
1\\
1
\end{array}
\right)  e^{\pm i\vec{k}\cdot \vec{x}},$ the tensor state can be represented
by
\begin{equation}
\left[  \mathbf{Z}^{X}\right \rangle =\left[  \mathbf{Z}^{Y}\right \rangle
=\left(
\begin{array}
[c]{c}%
1\\
1\\
0\\
0
\end{array}
\right)
\end{equation}
or
\begin{equation}
\left[  \mathbf{Z}^{X}\right \rangle =\left[  \mathbf{Z}^{Y}\right \rangle
=\left(
\begin{array}
[c]{c}%
0\\
0\\
1\\
1
\end{array}
\right)  .
\end{equation}
The tensor state is denoted by
\begin{align}
\left \langle \mathbf{\sigma}^{X}\otimes \vec{1}\right \rangle  &  =\vec
{n}_{\sigma}^{X}=(0,0,\pm1),\\
\left \langle \mathbf{\sigma}^{Y}\otimes \vec{1}\right \rangle  &  =\vec
{n}_{\sigma}^{Y}=(0,0,\pm1).\nonumber
\end{align}
Another 2D plane Kelvin wave with fixed wave-length is standing Kelvin-wave.
Along x-direction, the function of plane Kelvin wave becomes
\begin{align}
\mathbf{Z}(x)  &  =\left(
\begin{array}
[c]{c}%
\mathrm{z}_{\mathrm{A}}(x)\\
\mathrm{z}_{\mathrm{B}}(x)
\end{array}
\right)  =\frac{1}{\sqrt{2}}r_{0}(e^{ik\cdot x}+e^{-ik\cdot x})\left(
\begin{array}
[c]{c}%
1\\
1
\end{array}
\right) \\
&  =\sqrt{2}r_{0}\cos(k\cdot x)\left(
\begin{array}
[c]{c}%
1\\
1
\end{array}
\right)  ;\nonumber
\end{align}
along y-direction, the function of the plane Kelvin wave becomes
\begin{equation}
\mathbf{Z}(y)=\left(
\begin{array}
[c]{c}%
\mathrm{z}_{\mathrm{A}}(y)\\
\mathrm{z}_{\mathrm{B}}(y)
\end{array}
\right)  =\frac{1}{\sqrt{2}}r_{0}(e^{ik\cdot y}+ie^{-ik\cdot y})\left(
\begin{array}
[c]{c}%
1\\
1
\end{array}
\right)  .
\end{equation}
The tensor state is represented by
\begin{equation}
\left[  \mathbf{Z}^{X}\right \rangle =\frac{1}{\sqrt{2}}\left(
\begin{array}
[c]{c}%
1\\
1\\
1\\
1
\end{array}
\right)
\end{equation}
or
\begin{equation}
\left[  \mathbf{Z}^{Y}\right \rangle =\frac{1}{\sqrt{2}}\left(
\begin{array}
[c]{c}%
1\\
1\\
i\\
i
\end{array}
\right)  .
\end{equation}
The tensor state is denoted by
\begin{align}
\left \langle \mathbf{\sigma}^{X}\otimes \vec{1}\right \rangle  &  =\vec
{n}_{\sigma}^{X}=(1,0,0),\\
\left \langle \mathbf{\sigma}^{Y}\otimes \vec{1}\right \rangle  &  =\vec
{n}_{\sigma}^{Y}=(0,1,0).\nonumber
\end{align}
For the plane Kelvin wave along x-direction, we have,
\begin{equation}
\left \langle \mathbf{\sigma}^{X}\otimes \vec{1}\right \rangle =(1,0,0);
\end{equation}
For the plane Kelvin wave along y-direction, we have,
\begin{equation}
\left \langle \mathbf{\sigma}^{Y}\otimes \vec{1}\right \rangle =(0,1,0).
\end{equation}
\ 

Thirdly, we discuss 3D plane Kelvin waves with fixed wave-length.

For a 3D plane $\sigma_{z}$-Kelvin waves with fixed wave-length described by
$\mathbf{Z}(\vec{x})=\left(
\begin{array}
[c]{c}%
\mathrm{z}_{\mathrm{A}}(\vec{x})\\
\mathrm{z}_{\mathrm{B}}(\vec{x})
\end{array}
\right)  =r_{0}e^{\pm i\vec{k}\cdot \vec{x}}\left(
\begin{array}
[c]{c}%
1\\
1
\end{array}
\right)  ,$ the tensor state is represented by
\begin{equation}
\left[  \mathbf{Z}^{X}\right \rangle =\left[  \mathbf{Z}^{Y}\right \rangle
=\left[  \mathbf{Z}^{Z}\right \rangle =\left(
\begin{array}
[c]{c}%
1\\
1\\
0\\
0
\end{array}
\right)  ,
\end{equation}
or%
\begin{equation}
\left[  \mathbf{Z}^{X}\right \rangle =\left[  \mathbf{Z}^{Y}\right \rangle
=\left[  \mathbf{Z}^{Z}\right \rangle =\left(
\begin{array}
[c]{c}%
0\\
0\\
1\\
1
\end{array}
\right)  ,
\end{equation}
of which the tensor state is also denoted by
\begin{equation}
\left \langle \mathbf{\sigma}^{X}\otimes \vec{1}\right \rangle =\vec{n}_{\sigma
}^{X}=\vec{n}_{\sigma}^{Y}=\vec{n}_{\sigma}^{Z}=(0,0,\pm1).
\end{equation}
Another 3D plane Kelvin waves with fixed wave-length is described by the
following tensor state,
\begin{align}
\left[  \mathbf{Z}^{X}\right \rangle  &  =\frac{1}{\sqrt{2}}\left(
\begin{array}
[c]{c}%
1\\
1\\
1\\
1
\end{array}
\right)  ,\\
\left[  \mathbf{Z}^{Y}\right \rangle  &  =\frac{1}{\sqrt{2}}\left(
\begin{array}
[c]{c}%
1\\
1\\
i\\
i
\end{array}
\right)  ,\nonumber \\
\left[  \mathbf{Z}^{Z}\right \rangle  &  =\frac{1}{\sqrt{2}}\left(
\begin{array}
[c]{c}%
1\\
1\\
0\\
0
\end{array}
\right)  .\nonumber
\end{align}
Along x-direction, the function of the plane Kelvin wave becomes
\begin{align}
\mathbf{Z}(x)  &  =\left(
\begin{array}
[c]{c}%
\mathrm{z}_{\mathrm{A}}(x)\\
\mathrm{z}_{\mathrm{B}}(x)
\end{array}
\right)  =\frac{1}{\sqrt{2}}r_{0}(e^{ik\cdot x}+e^{-ik\cdot x})\left(
\begin{array}
[c]{c}%
1\\
1
\end{array}
\right) \\
&  =\sqrt{2}r_{0}\cos(k\cdot x)\left(
\begin{array}
[c]{c}%
1\\
1
\end{array}
\right)  ;
\end{align}
along y-direction, the function of the plane Kelvin wave becomes%
\begin{equation}
\mathbf{Z}(y)=\left(
\begin{array}
[c]{c}%
\mathrm{z}_{\mathrm{A}}(y)\\
\mathrm{z}_{\mathrm{B}}(y)
\end{array}
\right)  =\frac{1}{\sqrt{2}}r_{0}(e^{ik\cdot y}+ie^{-ik\cdot y})\left(
\begin{array}
[c]{c}%
1\\
1
\end{array}
\right)  ;
\end{equation}
along z-direction, the function of the plane Kelvin wave becomes
\begin{equation}
\mathbf{Z}(z)=\left(
\begin{array}
[c]{c}%
\mathrm{z}_{\mathrm{A}}(z)\\
\mathrm{z}_{\mathrm{B}}(z)
\end{array}
\right)  =r_{0}e^{ik\cdot z}\left(
\begin{array}
[c]{c}%
1\\
1
\end{array}
\right)  .
\end{equation}
For the tensor state along x-direction, we have,
\begin{equation}
\left \langle \mathbf{\sigma}^{X}\otimes \vec{1}\right \rangle =\vec{n}_{\sigma
}^{X}=(1,0,0);
\end{equation}
For the tensor state along y-direction, we have,
\begin{equation}
\left \langle \mathbf{\sigma}^{Y}\otimes \vec{1}\right \rangle =\vec{n}_{\sigma
}^{Y}=(0,1,0);
\end{equation}
\ For the tensor state along z-direction, we have,
\begin{equation}
\left \langle \mathbf{\sigma}^{Z}\otimes \vec{1}\right \rangle =\vec{n}_{\sigma
}^{Z}=(0,0,1).
\end{equation}
\ 

\subsubsection{Generation operator}

In this part, we define generator operators for Kelvin waves of different
tensor states.

Firstly, we generate the four degenerate states along a given direction
$\left[  \vec{k},\mathrm{A}\right \rangle ,$ $\left[  \vec{k},\mathrm{B}%
\right \rangle ,$ $\left[  -\vec{k},\mathrm{A}\right \rangle ,$ $\left[
-\vec{k},\mathrm{B}\right \rangle $ by four operators%
\begin{align}
\left[  \vec{k},\mathrm{A}\right \rangle  &  =\mathrm{\hat{U}}(\vec
{k},\mathrm{A})\left[  \text{\textrm{0}}\right \rangle ,\\
\left[  \vec{k},\mathrm{B}\right \rangle  &  =\mathrm{\hat{U}}(\vec
{k},\mathrm{B})\left[  \text{\textrm{0}}\right \rangle ,\nonumber \\
\left[  -\vec{k},\mathrm{A}\right \rangle  &  =\mathrm{\hat{U}}(-\vec
{k},\mathrm{A})\left[  \text{\textrm{0}}\right \rangle ,\nonumber \\
\left[  -\vec{k},\mathrm{B}\right \rangle  &  =\mathrm{\hat{U}}(-\vec
{k},\mathrm{B})\left[  \text{\textrm{0}}\right \rangle \nonumber
\end{align}
with%
\begin{align}
\mathrm{\hat{U}}(\vec{k},\mathrm{A})  &  =e^{i\int[\phi_{\vec{k},\mathrm{A}%
}(x)\cdot \hat{K}]dx}\cdot \mathrm{\hat{\digamma}}(r_{0}),\\
\mathrm{\hat{U}}(\vec{k},\mathrm{B})  &  =e^{i\int[\phi_{\vec{k},\mathrm{B}%
}(x)\cdot \hat{K}]dx}\cdot \mathrm{\hat{\digamma}}(r_{0}),\nonumber \\
\mathrm{\hat{U}}(-\vec{k},\mathrm{A})  &  =e^{i\int[\phi_{-\vec{k},\mathrm{A}%
}(x)\cdot \hat{K}]dx}\cdot \mathrm{\hat{\digamma}}(r_{0}),\nonumber \\
\mathrm{\hat{U}}(-\vec{k},\mathrm{B})  &  =e^{i\int[\phi_{-\vec{k},\mathrm{B}%
}(x)\cdot \hat{K}]dx}\cdot \mathrm{\hat{\digamma}}(r_{0}),\nonumber
\end{align}
where $\mathrm{\hat{\digamma}}(r_{0})$ is an expanding operator by shifting
radius from $0$ to $r_{0}$ on the membrane, $\hat{K}=-i\frac{d}{d\phi}$,
$\phi(x)=\pm kx,$ and $x$ is coordinate along $\vec{x}$ direction. Here
$\left[  \text{\textrm{0}}\right \rangle $ denotes a vortex-membrane with
$r_{0}=0$.

Based on the basis $\left(
\begin{array}
[c]{c}%
\left[  \vec{k},\mathrm{A}\right \rangle \\
\left[  \vec{k},\mathrm{B}\right \rangle \\
\left[  -\vec{k},\mathrm{A}\right \rangle \\
\left[  -\vec{k},\mathrm{B}\right \rangle
\end{array}
\right)  ,$ the generator operator is simplified into
\begin{equation}
\mathrm{\hat{U}}(\vec{k}\mathrm{,}\mathbf{\vec{\Gamma}}^{I})=e^{i\int
\mathbf{\vec{\Gamma}}^{I}[\phi_{\vec{k}}(x)\cdot \hat{K}]dx}\cdot
\mathrm{\hat{\digamma}}(r_{0})
\end{equation}
with $\mathbf{\vec{\Gamma}}^{I}=(\vec{n}_{\sigma}^{I}\mathbf{\sigma}%
^{I})\otimes(\vec{n}_{\tau}\mathbf{\tau}+\vec{1}\tau_{0})$. Then we get
\begin{equation}
\left(
\begin{array}
[c]{c}%
\left[  \vec{k},\mathrm{A}\right \rangle \\
\left[  \vec{k},\mathrm{B}\right \rangle \\
\left[  -\vec{k},\mathrm{A}\right \rangle \\
\left[  -\vec{k},\mathrm{B}\right \rangle
\end{array}
\right)  =\mathrm{\hat{U}}(\vec{k}\mathrm{,}\mathbf{\vec{\Gamma}}^{I})\left(
\begin{array}
[c]{c}%
\left[  \text{\textrm{0}}\right \rangle \\
\left[  \text{\textrm{0}}\right \rangle \\
\left[  \text{\textrm{0}}\right \rangle \\
\left[  \text{\textrm{0}}\right \rangle
\end{array}
\right)  .
\end{equation}

For different Kelvin waves with fixed wave-length, the generation operators
can be changed from a basis $\left(
\begin{array}
[c]{c}%
\left[  \vec{k},\mathrm{A}\right \rangle \\
\left[  \vec{k},\mathrm{B}\right \rangle \\
\left[  -\vec{k},\mathrm{A}\right \rangle \\
\left[  -\vec{k},\mathrm{B}\right \rangle
\end{array}
\right)  $ to another $\left(
\begin{array}
[c]{c}%
\left[  \vec{k},\mathrm{A}\right \rangle ^{\prime}\\
\left[  \vec{k},\mathrm{B}\right \rangle ^{\prime}\\
\left[  -\vec{k},\mathrm{A}\right \rangle ^{\prime}\\
\left[  -\vec{k},\mathrm{B}\right \rangle ^{\prime}%
\end{array}
\right)  $ by an \textrm{SU(2)}$\otimes$\textrm{SU(2)} rotation operator
\begin{equation}
\mathcal{\hat{S}}=\hat{S}\otimes \hat{V},
\end{equation}
$\hat{S}=%
%TCIMACRO{\dprod \limits_{I}}%
%BeginExpansion
{\displaystyle \prod \limits_{I}}
%EndExpansion
\hat{S}^{I}$ with $\hat{S}^{I}$ the \textrm{SU(2)} operation on helical
degrees of freedom%
\begin{equation}
\hat{S}^{I}\left(
\begin{array}
[c]{c}%
\left[  \vec{k}\right \rangle \\
\left[  -\vec{k}\right \rangle
\end{array}
\right)  =\left(
\begin{array}
[c]{c}%
\left[  \vec{k}\right \rangle ^{\prime}\\
\left[  -\vec{k}\right \rangle ^{\prime}%
\end{array}
\right)
\end{equation}
and $\hat{V}=%
%TCIMACRO{\dprod \limits_{I}}%
%BeginExpansion
{\displaystyle \prod \limits_{I}}
%EndExpansion
\hat{V}^{I}$ with $\hat{V}^{I}$ the \textrm{SU(2)} operation on vortex degrees
of freedom%
\begin{equation}
\hat{V}^{I}\left(
\begin{array}
[c]{c}%
\left[  \mathrm{A}\right \rangle \\
\left[  \mathrm{B}\right \rangle
\end{array}
\right)  =\left(
\begin{array}
[c]{c}%
\left[  \mathrm{A}\right \rangle ^{\prime}\\
\left[  \mathrm{B}\right \rangle ^{\prime}%
\end{array}
\right)  .
\end{equation}
In this paper, we consider a trivial $\hat{V}$ operation, $\hat{V}\equiv1$ and
focus on the effect from $\hat{S}^{I}$, i.e.,
\begin{equation}
\mathcal{\hat{S}}=\hat{S}\otimes \mathbf{1}=%
%TCIMACRO{\dprod \limits_{I}}%
%BeginExpansion
{\displaystyle \prod \limits_{I}}
%EndExpansion
\hat{S}^{I}\otimes \mathbf{1}.
\end{equation}
By doing $\hat{S}^{I}$ operation, we have%
\begin{align}
\mathbf{\vec{\sigma}}^{I}\otimes(\vec{n}_{\tau}\mathbf{\tau}+\vec{1}\tau_{0})
&  \rightarrow \left(  \mathbf{\vec{\sigma}}^{I}\right)  ^{\prime}\otimes
(\vec{n}_{\tau}\mathbf{\tau}+\vec{1}\tau_{0})\nonumber \\
&  =\hat{S}[\mathbf{\vec{\sigma}}^{I}\otimes(\vec{n}_{\tau}\mathbf{\tau}%
+\vec{1}\tau_{0})](\hat{S}^{I})^{-1},
\end{align}
or
\begin{align}
\mathbf{\vec{\Gamma}}^{I}  &  =\mathbf{\vec{\sigma}}^{I}\otimes(\vec{n}_{\tau
}\mathbf{\tau}+\vec{1}\tau_{0})\rightarrow(\mathbf{\vec{\Gamma}}^{I})^{\prime
}=\mathcal{\hat{S}}\vec{\Gamma}^{I}\mathcal{\hat{S}}^{-1}\\
&  =(\mathbf{\vec{\sigma}}^{I})^{\prime}\otimes(\vec{n}_{\tau}\mathbf{\tau
}+\vec{1}\tau_{0}).\nonumber
\end{align}
By doing \textrm{SU(2)}$\otimes$\textrm{SU(2)} rotation operation
$\mathcal{\hat{S}},$ the generation operator rotates as%
\begin{align}
\mathrm{\hat{U}}(\vec{k},\mathbf{\vec{\Gamma}}^{I})  &  \rightarrow
\lbrack \mathrm{\hat{U}}(\vec{k},\mathbf{\vec{\Gamma}}^{I})]^{\prime}\\
&  =\mathcal{\hat{S}}\mathrm{\hat{U}}(\vec{k},\mathbf{\vec{\Gamma}}%
^{I})\mathcal{\hat{S}}^{-1}\nonumber \\
&  =\mathrm{\hat{U}}(\vec{k},(\mathbf{\vec{\Gamma}}^{I})^{\prime}).\nonumber
\end{align}
As a result, after \textrm{SU(2)}$\otimes$\textrm{SU(2)} rotation operation, a
tensor state described by $\vec{n}_{\sigma}^{I}$ changes to another described
by $\left(  \vec{n}_{\sigma}^{I}\right)  ^{\prime}$. Thus, a new basis
becomes
\begin{align}
\left(
\begin{array}
[c]{c}%
\left[  \vec{k},\mathrm{A}\right \rangle \\
\left[  \vec{k},\mathrm{B}\right \rangle \\
\left[  -\vec{k},\mathrm{A}\right \rangle \\
\left[  -\vec{k},\mathrm{B}\right \rangle
\end{array}
\right)   &  \rightarrow \left(
\begin{array}
[c]{c}%
\left[  \vec{k},\mathrm{A}\right \rangle ^{\prime}\\
\left[  \vec{k},\mathrm{B}\right \rangle ^{\prime}\\
\left[  -\vec{k},\mathrm{A}\right \rangle ^{\prime}\\
\left[  -\vec{k},\mathrm{B}\right \rangle ^{\prime}%
\end{array}
\right)  =[\mathrm{\hat{U}}^{I}(\vec{k},\mathbf{\vec{\Gamma}}^{I})]^{\prime
}\left(
\begin{array}
[c]{c}%
0\\
0\\
0\\
0
\end{array}
\right) \\
&  =\mathcal{\hat{S}}\mathrm{\hat{U}}^{I}(\vec{k},\mathbf{\vec{\Gamma}}%
^{I})\mathcal{\hat{S}}^{-1}\left(
\begin{array}
[c]{c}%
0\\
0\\
0\\
0
\end{array}
\right) \nonumber \\
&  =\mathcal{\hat{S}}\left(
\begin{array}
[c]{c}%
\left[  \vec{k},\mathrm{A}\right \rangle \\
\left[  \vec{k},\mathrm{B}\right \rangle \\
\left[  -\vec{k},\mathrm{A}\right \rangle \\
\left[  -\vec{k},\mathrm{B}\right \rangle
\end{array}
\right)  .\nonumber
\end{align}

For example, by using the \textrm{SU(2)}$\otimes$\textrm{SU(2)} rotation
operation $\mathcal{\hat{S}}$, we can change a 3D $\sigma_{z}$-Kelvin waves to
another standing Kelvin waves by the rotating generation operators
\begin{align}
\hat{S}^{X}  &  =e^{i\frac{\pi}{2}\sigma_{y}\otimes \vec{1}},\\
\hat{S}^{Y}  &  =e^{-i\frac{\pi}{2}\sigma_{x}\otimes \vec{1}},\nonumber \\
\hat{S}^{Z}  &  =1.\nonumber
\end{align}

\subsubsection{Generalized spatial translation symmetry}

We discuss the spatial translation symmetry of plane Kelvin waves with fixed wave-length.

Different plane Kelvin waves of different tensor states $\left \langle
\mathbf{\hat{\Gamma}}^{I}\right \rangle =\vec{\Gamma}^{I}$ have different
translation symmetries. We define translation operator $\mathcal{\hat{T}%
}(\Delta \vec{x})$\ as
\begin{equation}
\mathcal{\hat{T}}(\Delta \vec{x})\mathbf{Z}(\vec{x})=\mathbf{Z}^{\prime}%
(\vec{x}+\Delta \vec{x}).
\end{equation}
For the case of
\begin{equation}
\mathcal{T}(2a)\mathbf{Z}(\vec{x})=\mathbf{Z}(\vec{x}),
\end{equation}
the system has translation symmetry.

The generalized translation symmetry of the plane Kelvin waves is
characterized by three translation operators,
\begin{align}
\mathcal{\hat{T}}_{x}(2a)  &  =e^{i2a(\hat{k}^{X}\cdot \mathbf{\vec{\Gamma}%
}^{X})},\nonumber \\
\mathcal{\hat{T}}_{y}(2a)  &  =e^{i2a(\hat{k}^{Y}\cdot \mathbf{\vec{\Gamma}%
}^{Y})},\nonumber \\
\mathcal{\hat{T}}_{z}(2a)  &  =e^{i2a(\hat{k}^{Z}\cdot \mathbf{\vec{\Gamma}%
}^{Z})},
\end{align}
where $\mathbf{\vec{\Gamma}}^{I}$ is a four-by-four matrix,%
\begin{equation}
\mathbf{\vec{\Gamma}}^{I}=(\vec{n}_{\sigma}^{I}\mathbf{\sigma}^{I}%
)\otimes(\vec{n}_{\tau}\mathbf{\tau}+\vec{1}\tau_{0}).\nonumber
\end{equation}
$\hat{k}^{X}=-i\frac{d}{dx}$ (or $\hat{k}^{Y}=-i\frac{d}{dy},$ $\hat{k}%
^{Z}=-i\frac{d}{dz}$) is wave vector operator. For the plane Kelvin waves
described by tensor state $\vec{n}_{\sigma}^{I}$ and $\tau_{0}=1$, we have
\begin{equation}
\vec{\Gamma}^{I}\mathbf{Z}(\vec{x})=\pm \mathbf{I}\cdot \mathbf{Z}(\vec{x}),
\end{equation}
where $\mathbf{I}=\left(
\begin{array}
[c]{cccc}%
1 & 0 & 0 & 0\\
0 & 1 & 0 & 0\\
0 & 0 & -1 & 0\\
0 & 0 & 0 & -1
\end{array}
\right)  .$

It looks like all plane Kelvin waves with different tensor states $\vec
{\Gamma}^{I}$ have the same translation symmetries as
\begin{equation}
\mathcal{\hat{T}}(2a)\mathbf{Z}(\vec{x})=\mathbf{Z}(\vec{x}).
\end{equation}
However, in addition to the translation symmetry, there exists generalized
translation symmetry by doing a translation operation $\mathcal{T}(\Delta
x)=e^{i\Delta x(\hat{k}^{I}\cdot \mathbf{\vec{\Gamma}}^{I})}$
\begin{align}
\mathbf{Z}(\vec{x})  &  \rightarrow \mathbf{Z}^{\prime}(\vec{x})\\
&  =\mathcal{T}(\Delta x)\mathbf{Z}(\vec{x})=e^{\pm i\cdot \mathbf{I\cdot
}k\cdot \Delta x}\mathbf{Z}(\vec{x}).\nonumber
\end{align}
Thus, an arbitrary continuous spatial translation operation is combination of
a discrete spatial translation operation
\begin{align}
\left \vert \mathbf{Z}(\vec{x},t)\right \rangle  &  \rightarrow \mathcal{T}%
(\Delta \vec{x})\mathbf{Z}(\vec{x})\\
&  =\mathcal{T}(\left \vert \Delta \vec{x}\right \vert =2a)\mathbf{Z}(\vec
{x})=\mathbf{Z}(\vec{x})\nonumber
\end{align}
and a global gauge transformation operation
\begin{equation}
\mathbf{Z}(\vec{x})\rightarrow \mathbf{Z}(\vec{x})e^{i\cdot \mathbf{I\cdot
}\Delta \phi}%
\end{equation}
where $\Delta \phi=ik\cdot \lbrack2a[(\Delta x)\operatorname{mod}2a]$.

\subsubsection{Biot-Savart equation in tensor representation}

Then we derive the Biot-Savart equation in tensor representation.

For perturbative Kelvin waves, the Biot-Savart equation $i\frac{d\mathrm{z}%
_{\mathrm{A/B}}(\vec{x},t)}{dt}=\frac{\delta \mathrm{\hat{H}}(\mathrm{z}%
_{\mathrm{A/B}}(\vec{x},t))}{\delta z_{\mathrm{A/B}}^{\ast}}$ turns into the
pseudo-Schr\"{o}dinger equation
\begin{equation}
i\hbar_{\mathrm{eff}}\frac{d\mathbf{Z}(\vec{x},t)}{dt}=\mathrm{\hat{H}}%
\cdot \mathbf{Z}(\vec{x},t)
\end{equation}
where $\mathbf{Z}(\vec{x},t)=\left(
\begin{array}
[c]{c}%
\alpha_{1,1}^{I}(\vec{x},t)\\
\alpha_{1,-1}^{I}(\vec{x},t)\\
\alpha_{-1,1}^{I}(\vec{x},t)\\
\alpha_{-1,-1}^{I}(\vec{x},t)
\end{array}
\right)  $ denotes perturbative Kelvin waves around the four degenerate Kelvin
states and $\hbar_{\mathrm{eff}}$ is the effective Planck constant in
pseudo-quantum mechanics. Based on the basis $\left(
\begin{array}
[c]{c}%
\left[  \vec{k}\right \rangle \\
\left[  -\vec{k}\right \rangle
\end{array}
\right)  \otimes \left(
\begin{array}
[c]{c}%
\left[  \mathrm{A}\right \rangle \\
\left[  \mathrm{B}\right \rangle
\end{array}
\right)  $ ($k>0$), we obtain the total Hamiltonian as%
\begin{align}
\mathrm{\hat{H}}  &  =\hat{T}+\hat{V}\nonumber \\
&  =%
%TCIMACRO{\dsum \limits_{I}}%
%BeginExpansion
{\displaystyle \sum \limits_{I}}
%EndExpansion
[\frac{(\hat{p}_{\mathrm{Lamb}}^{I}(\mathbf{\vec{\sigma}}^{I}\otimes \vec
{1}))^{2}}{2m_{\mathrm{pseudo}}}]+\frac{\hbar_{\mathrm{eff}}\omega^{\ast}}%
{2}(\vec{1}\otimes(\tau_{x}-\vec{1}))
\end{align}
Thus, in tensor representation, the Biot-Savart equation is changed into
\begin{align}
i\hbar_{\mathrm{eff}}\frac{d\mathbf{Z}(\vec{x},t)}{dt}  &  =\{%
%TCIMACRO{\dsum \limits_{I}}%
%BeginExpansion
{\displaystyle \sum \limits_{I}}
%EndExpansion
[\frac{(\hat{p}_{\mathrm{Lamb}}^{I}(\mathbf{\vec{\sigma}}^{I}\otimes \vec
{1}))^{2}}{2m_{\mathrm{pseudo}}}]\\
&  +\frac{\hbar_{\mathrm{eff}}\omega^{\ast}}{2}(\vec{1}\otimes(\tau_{x}%
-\vec{1}))\} \cdot \mathbf{Z}(\vec{x},t)\nonumber \\
&  =\{%
%TCIMACRO{\dsum \limits_{I}}%
%BeginExpansion
{\displaystyle \sum \limits_{I}}
%EndExpansion
[\frac{(\hat{p}_{\mathrm{Lamb}}^{I})^{2}}{2m_{\mathrm{pseudo}}}]+\frac
{\hbar_{\mathrm{eff}}\omega^{\ast}}{2}(\vec{1}\otimes(\tau_{x}-\vec
{1}))\} \nonumber \\
&  \cdot \mathbf{Z}(\vec{x},t)\nonumber
\end{align}
where $(\hat{p}_{\mathrm{Lamb}}^{I}(\mathbf{\vec{\sigma}}^{I}\otimes \vec
{1}))^{2}=(\hat{p}_{\mathrm{Lamb}}^{I})^{2}$ ($p_{\mathrm{Lamb}}^{I}>0$).

\subsubsection{Zeros and zero-density}

In the previous section, we have defined linking number and linking-number
density to characterize the deformed two vortex-membranes. However, the
winding number and winding-number density are not well defined for standing
waves. To locally characterize different Kelvin waves, we introduce the zeros
between two projected entangled vortex-membranes. Thus, it is the zero number
and zero-number density that characterize the deformation of Kelvin waves in
tensor representation.

\paragraph{Projection}

A d-dimensional vortex-membrane\ in d+2D space $\{ \vec{x},\xi(\vec{x}%
),\eta(\vec{x})\}$ can be described by the two-component function $\left(
\begin{array}
[c]{c}%
\xi(\vec{x})\\
\eta(\vec{x})
\end{array}
\right)  $. We then introduce the concept of \emph{projection}: a projection
of a vortex-membrane along a given direction $\theta$ on $\{ \xi,\eta \}$
space. In mathematics, the projection is defined by
\begin{equation}
\hat{P}_{\theta}\left(
\begin{array}
[c]{c}%
\xi(\vec{x})\\
\eta(\vec{x})
\end{array}
\right)  =\left(
\begin{array}
[c]{c}%
\xi_{\theta}(\vec{x})\\
\left[  \eta_{\theta}(\vec{x})\right]  _{0}%
\end{array}
\right)
\end{equation}
where $\xi_{\theta}(\vec{x})$ is variable and $\left[  \eta_{\theta}(\vec
{x})\right]  _{0}$ is constant. In the following parts we use $\hat{P}$ to
denote the projection operators. Because the projection direction out of
vortex-membrane is characterized by an angle $\theta$ in $\{ \xi,\eta \}$
space, we have
\begin{equation}
\left(
\begin{array}
[c]{c}%
\xi_{\theta}\\
\eta_{\theta}%
\end{array}
\right)  =\left(
\begin{array}
[c]{cc}%
\cos \theta & \sin \theta \\
\sin \theta & -\cos \theta
\end{array}
\right)  \left(
\begin{array}
[c]{c}%
\xi \\
\eta
\end{array}
\right)
\end{equation}
where $\theta$ is angle, i.e. $\theta \operatorname{mod}2\pi=0.$ So the
projected vortex-membrane is described by the function
\begin{equation}
\xi_{\theta}(\vec{x})=\xi(\vec{x})\cos \theta+\eta(\vec{x})\sin \theta.
\end{equation}

\paragraph{Zeros between two projected entangled vortex-lines}

We then define the projection between two entangled vortex-lines $\{
\xi_{\mathrm{A/B}}(x,t),\eta_{\mathrm{A/B}}(x,t)\}$ along a given direction
$\theta$ in 3D space by
\begin{equation}
\hat{P}_{\theta}\left(
\begin{array}
[c]{c}%
\xi_{\mathrm{A/B}}(x,t)\\
\eta_{\mathrm{A/B}}(x,t)
\end{array}
\right)  =\left(
\begin{array}
[c]{c}%
\xi_{\mathrm{A/B},\theta}(x,t)\\
\left[  \eta_{\mathrm{A/B},\theta}(x,t)\right]  _{0}%
\end{array}
\right)
\end{equation}
where $\xi_{\mathrm{A/B},\theta}(x,t)=\xi_{\mathrm{A/B}}(x,t)\cos \theta
+\eta_{\mathrm{A/B}}(x,t)\sin \theta$ is variable and $\left[  \eta
_{\mathrm{A/B},\theta}(x,t)\right]  _{0}=\xi_{\mathrm{A/B}}(x,t)\sin
\theta-\eta_{\mathrm{A/B}}(x,t)\cos \theta$ is constant. So the projected
vortex-line is described by the function $\xi_{\mathrm{A/B},\theta}(x,t).$ For
two projected vortex-lines described by $\xi_{\mathrm{A},\theta}(x,t)$ and
$\xi_{\mathrm{B},\theta}(x,t),$ a zero is solution of the equation
\begin{align}
\hat{P}_{\theta}[\mathrm{z}_{\mathrm{A}}(x,t)]  &  \equiv \xi_{\mathrm{A}%
,\theta}(x,t)\\
&  =\hat{P}_{\theta}[\mathrm{z}_{\mathrm{B}}(x,t)]\equiv \xi_{\mathrm{B}%
,\theta}(x,t).\nonumber
\end{align}
We call the equation to be zero-equation and its solutions to be zero-solution
(See the below discussion).

For vortex-lines described by plane wave $\mathbf{Z}(x,t)=\left(
\begin{array}
[c]{c}%
r_{0}e^{-i\omega \cdot t+ik\cdot x}\\
0
\end{array}
\right)  $, there exist periodic zeros. From the zero-equation $\xi
_{0,\mathrm{A},\theta}(x,t)=\xi_{0,\mathrm{B},\theta}(x,t)$ or $\cos(kx-\omega
t-\theta)=0,$ we get the zero-solutions to be
\begin{equation}
x(t)=a\cdot n+\frac{a}{\pi}\omega t+\frac{a}{\pi}(\theta+\frac{\pi}{2})
\end{equation}
where $n$ is an integer number and $\theta=-\frac{\pi}{2}$. From the
projection, we have a 1D crystal of zeros, of which the zero density
$\rho_{\text{\textrm{zero}}}$ is
\begin{equation}
\rho_{\text{\textrm{zero}}}=\frac{k}{\pi}=\frac{1}{a}.
\end{equation}
Fig.8 shows zeros between two projected vortex-lines. In addition, for this
case with $\mathrm{z}_{\mathrm{B}}=0$, the zero-number density is twice as the
linking-number density between two entangled vortex-membranes that is also
equal to the winding-number density for vortex-membrane-\textrm{A}, i.e.,
\begin{equation}
\rho_{\text{\textrm{zero}}}\equiv2\rho_{\text{\textrm{linking}}}\equiv
2\rho_{\text{\textrm{wind}}}.
\end{equation}

The zeros from the zero-solution don't change for a plane Kelvin wave with
different tensor states. For a plane Kelvin wave with clockwise winding
characterized by $\left[  \vec{k}\right \rangle $ and another plane Kelvin wave
with counterclockwise winding characterized by $\left[  -\vec{k}\right \rangle
,$ the functions for plane Kelvin waves are $r_{0}e^{-i\omega \cdot t+ik\cdot
x}$ and $r_{0}e^{-i\omega \cdot t-ik\cdot x},$ respectively. For both cases
with fixed time $t=0$, up to a constant phase factor, we have the same
zero-equation $\cos(kx)=0$ and the same zero-solution $x(t)=a\cdot n$,
respectively. For different plane Kelvin waves with different helical degrees
of freedom, we have $\left[  \mathbf{Z}\right \rangle =\alpha \left[  \vec
{k}\right \rangle +\beta \left[  -\vec{k}\right \rangle $ with $\left \vert
\alpha \right \vert ^{2}+\left \vert \beta \right \vert ^{2}=1.$ As a result, up to
a constant phase $\theta_{0}$, a plane Kelvin wave with the same wave-length
but different tensor orders has the same zero-equation
\begin{equation}
\cos(kx-\theta_{0})=0
\end{equation}
and the same zero-solution
\begin{equation}
x(t)=a\cdot n+\frac{a}{\pi}(\theta_{0}+\frac{\pi}{2}),
\end{equation}
respectively.

Next, we define the zeros from the projection of 3D vortex-membranes in 5D
space. a 3D vortex-membrane\ in 5D space $\{ \vec{x},\xi(\vec{x}),\eta(\vec
{x})\}$ ($\vec{x}=(x,y,z)$) can be described by the two-component function
$\left(
\begin{array}
[c]{c}%
\xi(\vec{x})\\
\eta(\vec{x})
\end{array}
\right)  $. Along $\vec{e}$-direction, for the helical vortex-membrane
described by $\mathrm{z}(x,t)=r_{0}e^{-i\omega \cdot t+i\vec{k}\cdot \vec{x}}$,
there exist periodic zeros between them. From the zero-equation $\xi
_{0,\mathrm{A},\theta}(\vec{x},t)=\xi_{0,\mathrm{B},\theta}(\vec{x},t)$ or
$\cos(\vec{k}\cdot \vec{x}-\omega t-\theta)=0,$ we get the zero-solutions to
be
\begin{equation}
\vec{x}(t)=(a\cdot n+\frac{a}{\pi}\omega t)\vec{e}%
\end{equation}
where $n$ is an integer number and $\theta=-\frac{\pi}{2}$. So there exist the
vector of zero density $\vec{\rho}_{\text{\textrm{zero}}}=(\rho
_{x,\text{\textrm{zero}}},$ $\rho_{y,\text{\textrm{zero}}},$ $\rho
_{z,\text{\textrm{zero}}})$ along three different directions $(\vec{e}_{x}$
$\vec{e}_{y}$, $\vec{e}_{z}).$

\paragraph{Zero-density operator}

We define zero-density operator to characterize two vortex-membranes.

According to pseudo-quantum mechanics, for each plane Kelvin wave
$e^{-i\omega \cdot t+ik\cdot x}$, the linking-number density is given by
$\rho_{\mathrm{zero}}=2\rho_{\mathrm{linking}}=\frac{k}{\pi}.$ Thus, the
corresponding zero-density operator is given by
\begin{equation}
\rho_{\mathrm{zero}}\rightarrow \hat{\rho}_{\mathrm{zero}}=\frac{\hat{k}}{\pi
}=-i\frac{1}{\pi}\frac{d}{dx}.
\end{equation}
In tensor representation, we can classify different types of zeros. For two
entangled 1D vortex-line, the zero-number along $\sigma_{z}$-internal
direction is given by
\begin{align}
N_{\mathrm{zero}}^{\sigma_{z}}  &  =\int \mathbf{\tilde{Z}}^{\ast}%
(x,t)[(\sigma^{z}\otimes \vec{1})\cdot(-i\frac{1}{\pi r_{0}^{2}}\frac{d}%
{dx})]\mathbf{\tilde{Z}}(x,t)dx\nonumber \\
&  =\int \mathrm{z}_{\uparrow,\mathrm{A}}^{\ast}(x,t)(-i\frac{1}{\pi r_{0}^{2}%
}\frac{d}{dx})\mathrm{z}_{\uparrow,\mathrm{A}}(x,t)dx\nonumber \\
&  +\int \mathrm{z}_{\uparrow,\mathrm{B}}^{\ast}(x,t)(-i\frac{1}{\pi r_{0}^{2}%
}\frac{d}{dx})\mathrm{z}_{\uparrow,\mathrm{B}}(x,t)dx\nonumber \\
&  +\int \mathrm{z}_{\downarrow,\mathrm{A}}^{\ast}(x,t)(-i\frac{1}{\pi
r_{0}^{2}}\frac{d}{dx})\mathrm{z}_{\downarrow,\mathrm{A}}(x,t)dx\nonumber \\
&  +\int \mathrm{z}_{\downarrow,\mathrm{B}}^{\ast}(x,t)(-i\frac{1}{\pi
r_{0}^{2}}\frac{d}{dx})\mathrm{z}_{\downarrow,\mathrm{B}}(x,t)dx
\end{align}
where $\mathbf{\tilde{Z}}(x,t)=\left(
\begin{array}
[c]{c}%
\mathrm{z}_{\uparrow \mathrm{A}}(x,t)\\
\mathrm{z}_{\uparrow \mathrm{B}}(x,t)\\
\mathrm{z}_{\downarrow \mathrm{A}}(x,t)\\
\mathrm{z}_{\downarrow \mathrm{B}}(\vec{x},t)
\end{array}
\right)  .$ The zero-number density is given by
\begin{equation}
\frac{1}{r_{0}^{2}}\mathbf{\tilde{Z}}^{\ast}(x,t)[(\sigma^{z}\otimes \vec
{1})\cdot \hat{\rho}_{\mathrm{zero}}]\mathbf{\tilde{Z}}(x,t).
\end{equation}
Using similar approach, we define the zero density for standing wave along
$\sigma_{x}$-internal direction
\begin{equation}
N_{\mathrm{zero}}^{\sigma_{x}}=\frac{1}{r_{0}^{2}}\int \mathbf{\tilde{Z}}%
^{\ast}(x,t)[(\sigma^{x}\otimes \vec{1})\cdot \hat{\rho}_{\mathrm{zero}%
}]\mathbf{\tilde{Z}}(x,t)dx.
\end{equation}
The corresponding zero density along $\sigma_{x}$-internal direction is given
by
\begin{equation}
\rho_{\mathrm{zero}}^{\sigma_{x}}=\frac{1}{r_{0}^{2}}\mathbf{\tilde{Z}}^{\ast
}(x,t)[(\sigma^{x}\otimes \vec{1})\cdot \hat{\rho}_{\mathrm{zero}}%
]\mathbf{\tilde{Z}}(x,t).
\end{equation}
In general, the zero density and zero-density operator are given by
\begin{equation}
\rho_{\mathrm{zero}}^{\vec{\sigma}}=\frac{1}{r_{0}^{2}}\mathbf{\tilde{Z}%
}^{\ast}(x,t)[(\vec{\sigma}\otimes \vec{1})\cdot \hat{\rho}_{\mathrm{zero}%
}]\mathbf{\tilde{Z}}(x,t).
\end{equation}

We also define the tensor of zero densities $\rho_{\mathrm{zero}}%
^{I,\vec{\sigma}}(x^{I},t)$ and its operators $\hat{\rho}_{\mathrm{zero}%
}^{I,\vec{\sigma}}$ for 3D vortex-membranes, i.e.,
\begin{align}
\rho_{\mathrm{zero}}^{I,\vec{\sigma}}(x^{I},t)  &  =\mathbf{\tilde{Z}}^{\ast
}(x^{I},t)[(\vec{\sigma}\otimes \vec{1})\\
&  \cdot(-i\frac{1}{\pi r_{0}^{2}}\frac{d}{dx^{I}})]\mathbf{\tilde{Z}}%
(x^{I},t),\nonumber
\end{align}
and
\begin{equation}
\hat{\rho}_{\mathrm{zero}}^{I,\vec{\sigma}}=(\vec{\sigma}\otimes \vec{1}%
)\cdot(-\frac{i}{\pi r_{0}^{2}}\frac{d}{dx^{I}}),
\end{equation}
respectively.

Finally, we could use the zero-density tensor $\rho_{\mathrm{zero}}%
^{I,\vec{\sigma}}(x^{I},t)$ to describe the time-evolution of two non-uniform
entangled vortex-membranes: for two entangled vortex-membranes with given
initial state $\mathbf{\tilde{Z}}(\vec{x})$, the time-dependent zero-density
tensor $\rho_{\mathrm{zero}}^{I,\vec{\sigma}}(x^{I},t)$ is obtained as
$\rho_{\mathrm{zero}}^{I,\vec{\sigma}}(x^{I},t)=\mathbf{\tilde{Z}}^{\ast
}(x^{I},t)[(\vec{\sigma}\otimes \vec{1})\cdot(-i\frac{1}{\pi r_{0}^{2}}\frac
{d}{dx^{I}})]\mathbf{\tilde{Z}}(x^{I},t)$ where
\begin{align}
\mathbf{\tilde{Z}}(\vec{x})  &  =%
%TCIMACRO{\dsum \nolimits_{k}}%
%BeginExpansion
{\displaystyle \sum \nolimits_{k}}
%EndExpansion
c_{k}e^{i\vec{k}\cdot \vec{x}}\nonumber \\
&  \rightarrow \mathbf{\tilde{Z}}(\vec{x},t)=%
%TCIMACRO{\dsum \nolimits_{k}}%
%BeginExpansion
{\displaystyle \sum \nolimits_{k}}
%EndExpansion
c_{k}e^{i\vec{k}\cdot \vec{x}}e^{-i\omega \cdot t}.
\end{align}

From the point view of information (see below discussion), the Sommerfeld
quantization condition for four-component particle is the condition of the
zero number, i.e.,%
\[%
%TCIMACRO{\doint }%
%BeginExpansion
{\displaystyle \oint}
%EndExpansion
p_{\mathrm{Lamb}}\cdot dx=h_{\mathrm{eff}}\cdot2N_{\mathrm{zero}}^{\sigma_{z}%
}.
\]

\section{Emergent quantum mechanics in knot physics}

The issue of emergent quantum mechanics is about the dynamic evolution for an
object of a zero (a knot) with volume-changing, another types of local
deformation of two entangled vortex-membranes. Here, we ask another question
-- \emph{how to characterize the evolution of the zeros with volume-changing?}

In last section, we found that the deformed knot-crystal is described by the
pseudo-quantum mechanics, of which two entangled vortex-membranes with fixed
volume are mapped onto a particle with two internal degrees of freedom in
tensor representation. It is obvious that the physical degrees of freedom for
a vortex-membrane (the size $V_{P}\rightarrow \infty$) are seriously reduced.
Because a vortex-membrane is an extended object, in general, each vortex-piece
(vortex filament) with infinitesimal size ($\Delta V_{P}\rightarrow0$) has a
degree of freedom. So the pseudo-quantum mechanics cannot characterize the
zero fluctuations from volume-changing on entangled vortex-membranes. In the
following parts, from point view of information, we consider the smallest
volume-changing on entangled vortex-membranes as an object with a single zero
($\Delta V_{\mathrm{knot}}\equiv a^{3}r_{0}^{2}$). We call such object a knot.
To characterize knot dynamics, the Biot-Savart mechanics becomes emergent
quantum mechanics, the true quantum mechanics.

\subsection{Basic theory for knots}

Before developing emergent quantum mechanics for knots, we review the concept
of "information".

\begin{figure}[ptb]
\includegraphics[clip,width=0.65\textwidth]{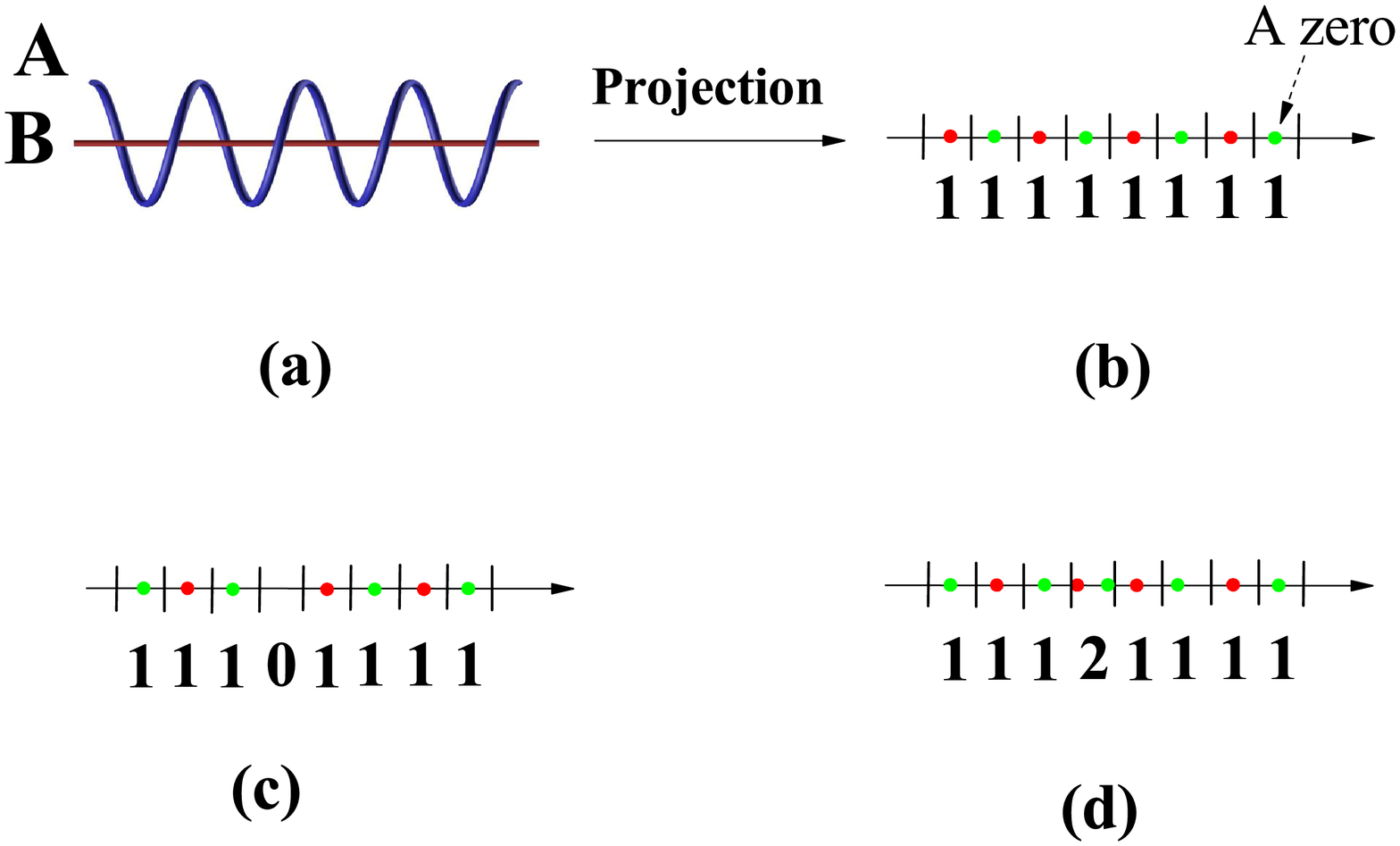}\caption{(a) Perfect
entangled vortex-membranes; (b) The information of perfect entangled
vortex-membranes by projection ...$11111111$...; (c) The information from
projected entangled vortex-membranes with a hole-like knot that is described
by the information, ...$11101111$...; (d) The information from projected
entangled vortex-membranes with a particle-like knot that is described by the
information, ...$11121111$...}%
\end{figure}

A generalized definition of the concept "information" is "\emph{An answer to a
specific question}". In general, the answer is related to data that represents
values attributed to parameters. In other words, information can be
interpreted as a data pattern or an ordered sequence of symbols from an
alphabet. Take DNA as an example. The sequence of nucleotides is a pattern
that influences the formation and development of an organism. In addition,
information processing consists of an input-output function that maps any
input sequence into an output sequence. Therefore, information reduces
uncertainty. The uncertainty of an event is measured by its probability of
occurrence. The more uncertain an event, the more information is required to
resolve uncertainty of that event.

To represent information, information unit must be defined. The bit is a
typical unit of information. Example: information in one "fair" coin flip:
$\log_{2}(\frac{2}{1})=1$ bit, and in two fair coin flips is $\log_{2}%
(\frac{4}{1})=2$ bits. In summary, the information is defined by the following
equation%
\[
\text{Information = a data pattern of information unit.}%
\]

The use of knotted strings for record-keeping purposes in China dates back to
ancient times. In 2003 J. D. Bekenstein claimed that a growing trend in
physics was to define the physical world as being made up of information
itself (and thus information is defined in this way).

By using projected representation, we show that different entanglement
patterns between two vortex-membranes correspond to different patterns of
zeros that are obtained from different zero-solutions, i.e.,%
\begin{align}
\text{Information }  &  =\text{ a pattern of zeros between }\nonumber \\
&  \text{projected vortex-membranes.}%
\end{align}
In physics, the information is characterized by the distribution of zeros.
Here, the information unit is a zero between projected vortex-membranes,
\begin{align}
\text{Information unit}  &  =\text{ A zero between two }\nonumber \\
&  \text{projected vortex-membranes.}%
\end{align}
For the case of an extra zero, we have a knot; for the case of missing zero,
we have an anti-knot. See the illustration in Fig.8. The colors of the dots in
Fig.8 denotes $\mathrm{L}$ and $\mathrm{R}$ types of zeros.

\subsubsection{Definition of a unified knot}

Each zero corresponds to a piece of vortex-membranes. We call the object to be
a \emph{knot}, a piece of plane Kelvin-wave with fixed wave-length. As a
result, the system (plane Kelvin-wave with fixed wave-length) can be regarded
as a crystal of knots. I call the system \emph{knot-crystal}. In this section,
we will study knot and show its properties on a knot-crystal.

We give the mathematic definition of knot.

From point view of information, a knot is an information unit with fixed
geometric properties. For example, for 1D knot, people divide the knot-crystal
into $N$ identical pieces, each of which is just a knot. As a result, some
properties of knot can be obtained by scaling the properties of a
knot-crystal; From point view of a classical field theory, a knot is
elementary topological defect of a knot-crystal that is always anti-phase
changing along arbitrary direction $\vec{e}$, $\mathbf{Z}(\vec{x}%
,t)\rightarrow \mathbf{Z}_{\mathrm{knot}}(\vec{x},t),$ i.e.,
\begin{equation}
\frac{\mathbf{Z}_{\mathrm{knot}}(\vec{x}\rightarrow-\infty,t)}{\mathbf{Z}%
_{\mathrm{knot}}(\vec{x}\rightarrow \infty,t)}=-\frac{\mathbf{Z}(\vec
{x}\rightarrow-\infty,t)}{\mathbf{Z}(\vec{x}\rightarrow \infty,t)}%
\end{equation}
Let us show the fact that an information unit must be an anti-phase changing
on Kelvin waves.

Based on the projected vortex-membranes, we define a knot by a monotonic
function $F_{\theta}(x)=\xi_{\mathrm{A},\theta}(x)-\xi_{\mathrm{B},\theta}(x)$
with
\begin{equation}
\mathrm{sgn}\text{ }[F_{\theta}(x\rightarrow-\infty)\cdot F_{\theta
}(x\rightarrow \infty)]=-1
\end{equation}
where $x$ denotes the position along the given direction $\vec{e}$. A knot can
be regarded as domain wall of $\pi$-phase shifting. When there exists a knot,
the periodic boundary condition of Kelvin waves along arbitrary direction is
changed into anti-periodic boundary condition. So the sign-switching character
can be labeled by winding number $w_{1D}$. The winding number $w_{1D}$ for a
knot along given direction is $\pm \frac{1}{2},$ i.e.,%
\begin{align}
w_{1D}^{I}  &  =\frac{-i}{2\pi \left(  r_{0}\right)  ^{2}}\int_{-\infty
}^{\infty}\mathbf{Z}_{\mathrm{knot}}^{\ast}(x^{I})d\mathbf{Z}_{\mathrm{knot}%
}(x^{I})\\
&  =\frac{1}{2\pi}\int_{-\infty}^{\infty}\frac{d\phi(x^{I})}{dx^{I}}%
dx^{I}\nonumber \\
&  =\frac{1}{2\pi}\int d\phi(x^{I})=\pm \frac{1}{2}\nonumber
\end{align}
where $x$ is the coordinate along the given direction. Because the zero number
is twice of winding-number, each zero $\Delta N_{\mathrm{zero}}=\pm1$
corresponds to a knot with half winding-number, $\Delta w_{1D}=\pm \frac{1}{2}%
$. A knot is an object with a fractional winding-number. As a result, the
quantum number for a 3D knot can be denoted by
\begin{equation}
N_{\mathrm{knot}}=8\Delta w_{1D}^{X}\cdot \Delta w_{1D}^{Y}\cdot \Delta
w_{1D}^{Z}=\pm1.
\end{equation}

On the other hand, from the topological character of a knot, there must exist
a point, at which $\xi_{\mathrm{A}}(x)$ is equal to $\xi_{\mathrm{B}}(x)$. The
position of the point $x$ is determined by a local solution of the
\emph{zero-equation}
\begin{equation}
F_{\theta}(x)=0
\end{equation}
or%
\begin{equation}
\xi_{\mathrm{A},\theta}(x)=\xi_{\mathrm{B},\theta}(x).
\end{equation}
So each knot corresponds to a zero between two vortex-membranes along the
given direction.

In summary, for a knot, there are three properties:

1) Conservation condition: The conservation condition indicates that as an
information unit, the existence of a zero for a knot is independent on the
directions of projection angle $\theta$. When one gets a zero-solution along a
given direction $\theta$, it will never split or disappear whatever changing
the projection direction, $\theta \rightarrow \theta^{\prime}$. The type of
zeros is switched $\mathrm{L}\leftrightarrow \mathrm{R}$ by rotating the
projection angle $\theta \rightarrow \theta+\pi$ (see below discussion);

2) Isotropic property:\emph{ }According to isotropic condition, there exists a
single knot solution along arbitrary direction $\vec{e}$. During rotation
operation $\vec{e}\rightarrow \vec{e}^{\prime}=R\vec{e}$, the zero of the knot
still exists but its spin degrees of freedom is rotating synchronously,
$\sigma \rightarrow \sigma^{\prime}=R\sigma$;

3) Z2 topological property -- each knot is $\pi$-phase changing along
arbitrary direction $\vec{e}$. When there exists a knot, for all Kelvin waves
the periodic boundary condition along the given direction is changed into
anti-periodic boundary condition.

For a knot, there are three conserved physical quantities: the energy of a
(static) knot that is proportional to the volume of the knot on
vortex-membrane $V_{P}=a^{3}$ where $a$ is a characteristic winding length
along different directions; the (Lamb impulse) angular momentum (the effective
Planck constant $\hbar_{\mathrm{knot}}$) that is proportional to the volume of
the knot in the 5D fluid $V_{\mathrm{knot}}=V_{P}\cdot r_{0}^{2}=a^{3}\cdot
r_{0}^{2}$ where $r_{0}$ is a characteristic winding radius; the winding
number $\left \vert \Delta w_{1D}\right \vert =\frac{1}{2}$ or the zero number
$\left \vert \Delta N_{\mathrm{zero}}\right \vert =1$. According to the
geometric character of the three conserved physical quantities, the shape of
knot will never be changed. However, the knot can split and the three physical
quantities are conserved for all knot-pieces. Quantum mechanics is a mechanics
to determine the distribution of knot-pieces.

\subsubsection{Degrees of freedom of a knot}

For a knot, there exists a zero satisfying conservation condition and Z2
topological property. A knot (a zero) has four degrees of freedom: two spin
degrees of freedom $\uparrow$ or $\downarrow$ from the helicity degrees of
freedom, the other two vortex degrees of freedom from the vortex degrees of
freedom that characterize the vortex-membranes, $\mathrm{A}$ or $\mathrm{B}$.
The basis to define the microscopic structure of a knot is given by
\begin{align}
\left \vert \uparrow,\mathrm{A}\right \rangle  &  =\left(
\begin{array}
[c]{c}%
\mathrm{z}_{\mathrm{knot},\uparrow,\mathrm{A}}(x,t)\\
0\\
0\\
0
\end{array}
\right)  ,\text{ }\left \vert \uparrow,\mathrm{B}\right \rangle =\left(
\begin{array}
[c]{c}%
0\\
\mathrm{z}_{\mathrm{knot},\uparrow,\mathrm{B}}(x,t)\\
0\\
0
\end{array}
\right)  ,\\
\left \vert \downarrow,\mathrm{A}\right \rangle  &  =\left(
\begin{array}
[c]{c}%
0\\
0\\
\mathrm{z}_{\mathrm{knot},\downarrow,\mathrm{A}}(x,t)\\
0
\end{array}
\right)  ,\text{ }\left \vert \downarrow,\mathrm{B}\right \rangle =\left(
\begin{array}
[c]{c}%
0\\
0\\
0\\
\mathrm{z}_{\mathrm{knot},\downarrow,\mathrm{B}}(x,t)
\end{array}
\right) \nonumber
\end{align}
where \textrm{A/B} denotes the degrees of freedom of vortex-membranes;
$\uparrow$/$\downarrow$ denotes the degrees of freedom of (pseudo-) spin.
$\mathrm{z}_{\mathrm{knot},\uparrow,\mathrm{A/B}}(x,t)$ is the knot-function
for a knot on vortex-membrane \textrm{A/B} with clockwise winding;
$\mathrm{z}_{\mathrm{knot},\downarrow,\mathrm{A/B}}(x,t)$ is the knot-function
for a knot on vortex-membrane \textrm{A/B} with counterclockwise winding.

Firstly, we discuss spin degrees of freedom of knots. We have shown that there
exist two types of winding directions for a knot%
\begin{equation}
\mathbf{Z}_{\mathrm{knot}}(x,t)=(%
\begin{array}
[c]{c}%
\mathrm{z}_{\mathrm{knot},\uparrow}(x,t)\\
\mathrm{z}_{\mathrm{knot},\downarrow}(x,t)
\end{array}
)\rightarrow(%
\begin{array}
[c]{c}%
\left \vert \uparrow \right \rangle \\
\left \vert \downarrow \right \rangle
\end{array}
).
\end{equation}
A knot with the eigenstates of $\sigma_{z}$ described by $\left \vert
\uparrow \right \rangle $ or $\left \vert \downarrow \right \rangle $ is really a
piece of traveling Kelvin waves; A knot with the eigenstates of $\sigma_{x}$
described by $\left \vert \rightarrow \right \rangle =\frac{1}{\sqrt{2}}\left(
\left \vert \uparrow \right \rangle +\left \vert \downarrow \right \rangle \right)
$ or $\left \vert \leftarrow \right \rangle =\frac{1}{\sqrt{2}}\left(  \left \vert
\uparrow \right \rangle -\left \vert \downarrow \right \rangle \right)  $ is a
piece of standing Kelvin waves; A knot with the eigenstates of $\sigma_{y}$
described by $\frac{1}{\sqrt{2}}\left(  \left \vert \uparrow \right \rangle
+i\left \vert \downarrow \right \rangle \right)  $ or $\frac{1}{\sqrt{2}}\left(
\left \vert \uparrow \right \rangle -i\left \vert \downarrow \right \rangle \right)
$ is a piece of another type of standing Kelvin waves. An important property
of spin degrees of freedom is time-reversal symmetry. In physics, under
time-reversal operation, a clockwise winding knot (spin-$\uparrow$ particle)
turns into a counterclockwise winding knot (spin-$\downarrow$ particle). An
important property of spin degrees of freedom is time-reversal symmetry. In
physics, under time-reversal operation, a clockwise winding knot
(spin-$\uparrow$ particle) turns into a counterclockwise winding knot
(spin-$\downarrow$ particle). In addition, in next sections we will show that
owing to the existence of spin-orbital coupling (SOC) in the Dirac model, the
degrees of freedom $\uparrow$/$\downarrow$ indeed play the role of spin
degrees of freedom as those in quantum mechanics.

Next, we discuss the vortex degrees of freedom of knots. The vortex-degrees of
freedom of knots is denoted by
\begin{equation}
\mathbf{Z}_{\mathrm{knot}}(x,t)=(%
\begin{array}
[c]{c}%
\mathrm{z}_{\mathrm{knot,A}}(x,t)\\
\mathrm{z}_{\mathrm{knot,B}}(x,t)
\end{array}
)\rightarrow(%
\begin{array}
[c]{c}%
\left \vert \mathrm{A}\right \rangle \\
\left \vert \mathrm{B}\right \rangle
\end{array}
).
\end{equation}
For knots, we identify $\mathrm{z}_{\mathrm{knot,A}}(x,t)$ the function of
A-vortex-membrane to be the wave-function of $\left \vert \mathrm{A}%
\right \rangle $, $\mathrm{z}_{\mathrm{knot,B}}(x,t)$ the function of an
B-vortex-membrane to be the wave-function of $\left \vert \mathrm{B}%
\right \rangle $.

Let us give the functions of four 1D knot states:

1) The up-spin knot on vortex-membrane-\textrm{A,} $\left \vert \uparrow
,\mathrm{A}\right \rangle $ is defined by
\begin{equation}
\mathrm{z}_{\mathrm{knot},\uparrow,\mathrm{A}}(x,t)=r_{\uparrow,\mathrm{A}%
}(x)\exp[i\phi_{\mathrm{knot},\uparrow,\mathrm{A}}(x,t)]
\end{equation}
where $\phi_{\mathrm{knot},\uparrow,\mathrm{A}}(x)$ is a monotonic function
changing $\pi$-phase. For unified knot, we have%
\begin{equation}
r_{\uparrow,\mathrm{A}}(x)\rightarrow \left \{
\begin{array}
[c]{c}%
0,\text{ }x\in(-\infty,x_{0}]\\
r_{0},\text{ }x\in(x_{0},x_{0}+a]\\
0,\text{ }x\in(x_{0}+a,\infty)
\end{array}
\right \}
\end{equation}
and
\begin{equation}
\phi_{\mathrm{knot},\uparrow,\mathrm{A}}(x)\rightarrow \phi
_{\mathrm{unified-knot},\uparrow,\mathrm{A}}(x)=\left \{
\begin{array}
[c]{c}%
\phi_{0}-\frac{\pi}{2},\text{ }x\in(-\infty,x_{0}]\\
\phi_{0}-\frac{\pi}{2}+k_{0}(x-x_{0}),\text{ }x\in(x_{0},x_{0}+a]\\
\phi_{0}+\frac{\pi}{2},\text{ }x\in(x_{0}+a,\infty)
\end{array}
\right \}
\end{equation}
where $x$ is the coordination on the axis and $\phi_{0}$ is an arbitrary
constant angle, $k_{0}=\frac{\pi}{a}.$ The center of the unified knot is at
$x=x_{0}+\frac{a}{2}$. For fragmentized knot, we have
\begin{equation}
\phi_{\mathrm{knot},\uparrow,\mathrm{A}}(x)\rightarrow \phi
_{\mathrm{fragmentized},\uparrow,\mathrm{A}}(x).
\end{equation}
We will give the definition of $\phi_{\mathrm{fragmentized},\uparrow
,\mathrm{A}}(x)$ in the following section. For different types of knots, there
exists a linear relationship between $\phi(x)$ and $x$, i.e., $\frac{d\phi
(x)}{dx}\propto k_{0}$ in the phase-changing region of $x_{0}<x\leq x_{0}+a$.
The linear character comes from the minimization of the volume
$\mathrm{volume}(P)$ of vortex-membrane. See the illustration of unified knots
in Fig.9. In the limit of $x\rightarrow-\infty$, we have $\phi(x)=\phi
_{0}-\frac{\pi}{2};$ in the limit of $x\rightarrow \infty$, we have
$\phi(x)=\phi_{0}+\frac{\pi}{2}$. So we obtain
\begin{equation}
\phi_{\mathrm{knot},\uparrow,\mathrm{A}}(x\rightarrow \infty)-\phi
_{\mathrm{knot},\uparrow,\mathrm{A}}(x\rightarrow-\infty)=\pi
\end{equation}
and
\begin{equation}
\frac{\mathrm{z}_{\mathrm{knot},\uparrow,\mathrm{A}}(x\rightarrow-\infty
,t)}{\mathrm{z}_{\mathrm{knot},\uparrow,\mathrm{A}}(x\rightarrow \infty,t)}=-1;
\end{equation}

\begin{figure}[ptb]
\includegraphics[clip,width=0.63\textwidth]{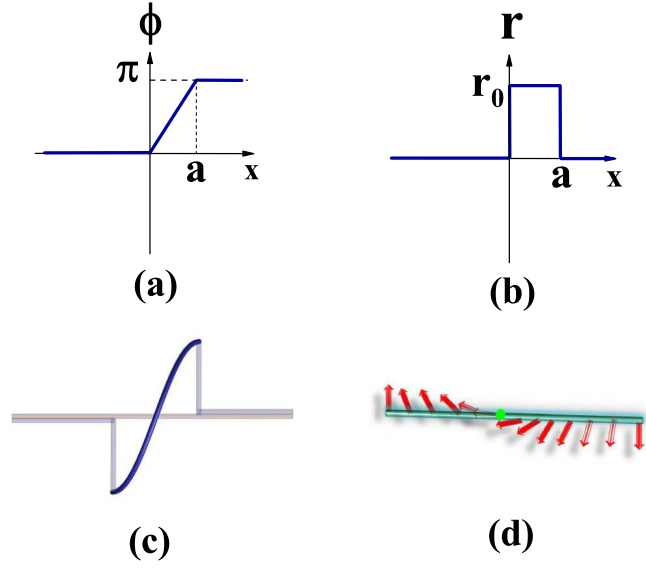}\caption{(a)The relation
between phase and coordination of a unified knot; (b)The relation between
radius and coordination of a unified knot; (c) The illustration of a unified
knot; (d) In complex field representation, a unified knot is an anti-phase
domain wall (the red arrows denote the directions of phase angles).}%
\end{figure}

2) The up-spin knot on vortex-membrane-\textrm{B, }$\left \vert \uparrow
,\mathrm{B}\right \rangle $ is defined by
\begin{equation}
\mathrm{z}_{\mathrm{knot},\uparrow,\mathrm{B}}(x,t)=r_{\uparrow,\mathrm{B}%
}\exp[i\phi_{\mathrm{knot},\uparrow,\mathrm{B}}(x,t)]
\end{equation}
where
\begin{equation}
r_{\uparrow,\mathrm{B}}(x)\rightarrow \left \{
\begin{array}
[c]{c}%
0,\text{ }x\in(-\infty,x_{0}]\\
r_{0},\text{ }x\in(x_{0},x_{0}+a]\\
0,\text{ }x\in(x_{0}+a,\infty)
\end{array}
\right \}
\end{equation}
and%
\begin{equation}
\phi_{\mathrm{knot},\uparrow,\mathrm{B}}(x)\rightarrow \phi
_{\mathrm{unified-knot},\uparrow,\mathrm{B}}(x)=\left \{
\begin{array}
[c]{c}%
\phi_{0}-\frac{\pi}{2},\text{ }x\in(-\infty,x_{0}]\\
\phi_{0}-\frac{\pi}{2}+k_{0}(x-x_{0}),\text{ }x\in(x_{0},x_{0}+a]\\
\phi_{0}+\frac{\pi}{2},\text{ }x\in(x_{0}+a,\infty)
\end{array}
\right \}  .
\end{equation}
Thus, we have
\begin{equation}
\phi_{\mathrm{knot},\uparrow,\mathrm{B}}(x\rightarrow \infty)-\phi
_{\mathrm{knot},\uparrow,\mathrm{B}}(x\rightarrow-\infty)=\pi
\end{equation}
and
\begin{equation}
\frac{\mathrm{z}_{\mathrm{knot},\uparrow,\mathrm{B}}(x\rightarrow-\infty
,t)}{\mathrm{z}_{\mathrm{knot},\uparrow,\mathrm{B}}(x\rightarrow \infty,t)}=-1;
\end{equation}

3) The down-spin knot on vortex-membrane-\textrm{A,} $\left \vert
\downarrow,\mathrm{A}\right \rangle $ is defined by
\begin{equation}
\mathrm{z}_{\mathrm{knot},\downarrow,\mathrm{A}}(x,t)=r_{\downarrow
,\mathrm{A}}\exp[i\phi_{\mathrm{knot},\downarrow,\mathrm{A}}(x,t)]
\end{equation}
where%
\begin{equation}
r_{\downarrow,\mathrm{A}}(x)\rightarrow \left \{
\begin{array}
[c]{c}%
0,\text{ }x\in(-\infty,x_{0}]\\
r_{0},\text{ }x\in(x_{0},x_{0}+a]\\
0,\text{ }x\in(x_{0}+a,\infty)
\end{array}
\right \}
\end{equation}
and
\begin{equation}
\phi_{\mathrm{knot},\downarrow,\mathrm{A}}(x)\rightarrow \phi
_{\mathrm{unified-knot},\downarrow,\mathrm{A}}(x)=\left \{
\begin{array}
[c]{c}%
\phi_{0}+\frac{\pi}{2},\text{ }x\in(-\infty,x_{0}]\\
\phi_{0}+\frac{\pi}{2}-k_{0}(x-x_{0}),\text{ }x\in(x_{0},x_{0}+a]\\
\phi_{0}-\frac{\pi}{2},\text{ }x\in(x_{0}+a,\infty)
\end{array}
\right \}  .
\end{equation}
Thus, we have
\begin{equation}
\phi_{\mathrm{knot},\downarrow,\mathrm{A}}(x\rightarrow \infty)-\phi
_{\mathrm{knot},\downarrow,\mathrm{A}}(x\rightarrow-\infty)=-\pi
\end{equation}
and
\begin{equation}
\frac{\mathrm{z}_{\mathrm{knot},\downarrow,\mathrm{A}}(x\rightarrow-\infty
,t)}{\mathrm{z}_{\mathrm{knot},\downarrow,\mathrm{A}}(x\rightarrow \infty
,t)}=-1;
\end{equation}

4) The down-spin knot on vortex-membrane-\textrm{B,} $\left \vert
\downarrow,\mathrm{B}\right \rangle $ is defined by
\begin{equation}
\mathrm{z}_{\mathrm{knot},\downarrow,\mathrm{B}}(x,t)=r_{\downarrow
,\mathrm{B}}\exp[i\phi_{\mathrm{knot},\downarrow,\mathrm{B}}(x,t)]
\end{equation}
where%
\begin{equation}
r_{\downarrow,\mathrm{B}}(x)\rightarrow \left \{
\begin{array}
[c]{c}%
0,\text{ }x\in(-\infty,x_{0}]\\
r_{0},\text{ }x\in(x_{0},x_{0}+a]\\
0,\text{ }x\in(x_{0}+a,\infty)
\end{array}
\right \}
\end{equation}
and%
\begin{equation}
\phi_{\mathrm{knot},\downarrow,\mathrm{B}}(x)\rightarrow \phi
_{\mathrm{unified-knot},\downarrow,\mathrm{B}}(x)=\left \{
\begin{array}
[c]{c}%
\phi_{0}+\frac{\pi}{2},\text{ }x\in(-\infty,x_{0}]\\
\phi_{0}+\frac{\pi}{2}-k_{0}(x-x_{0}),\text{ }x\in(x_{0},x_{0}+a]\\
\phi_{0}-\frac{\pi}{2},\text{ }x\in(x_{0}+a,\infty)
\end{array}
\right \}  .
\end{equation}
Thus, we have
\begin{equation}
\phi_{\mathrm{knot},\downarrow,\mathrm{B}}(x\rightarrow \infty)-\phi
_{\mathrm{knot},\downarrow,\mathrm{B}}(x\rightarrow-\infty)=-\pi
\end{equation}
and
\begin{equation}
\frac{\mathrm{z}_{\mathrm{knot},\downarrow,\mathrm{B}}(x\rightarrow-\infty
,t)}{\mathrm{z}_{\mathrm{knot},\downarrow,\mathrm{B}}(x\rightarrow \infty
,t)}=-1.
\end{equation}

The knots with different internal degrees of freedom are described by
different superposed states of the basis
\begin{equation}
\left(
\begin{array}
[c]{c}%
\left \vert \uparrow,\mathrm{A}\right \rangle \\
\left \vert \uparrow,\mathrm{B}\right \rangle \\
\left \vert \downarrow,\mathrm{A}\right \rangle \\
\left \vert \downarrow,\mathrm{B}\right \rangle
\end{array}
\right)  =\left(
\begin{array}
[c]{c}%
\left \vert \uparrow \right \rangle \\
\left \vert \downarrow \right \rangle
\end{array}
\right)  \otimes \left(
\begin{array}
[c]{c}%
\left \vert \mathrm{A}\right \rangle \\
\left \vert \mathrm{B}\right \rangle
\end{array}
\right)  .
\end{equation}
Along arbitrary direction $\vec{e}$, due to sign-switching character for each
of element states $\left \vert \uparrow,\mathrm{A}\right \rangle ,$ $\left \vert
\uparrow,\mathrm{B}\right \rangle ,$\ $\left \vert \downarrow,\mathrm{A}%
\right \rangle ,$\ $\left \vert \downarrow,\mathrm{B}\right \rangle ,$\ all
superposed states $\left \vert \psi \right \rangle =\alpha \left \vert
\uparrow,\mathrm{A}\right \rangle +\beta \left \vert \uparrow,\mathrm{B}%
\right \rangle +\gamma \left \vert \downarrow,\mathrm{A}\right \rangle
+\kappa \left \vert \downarrow,\mathrm{B}\right \rangle $ change sign as
\[
\frac{\psi(\vec{x}\rightarrow-\infty,t)}{\psi(\vec{x}\rightarrow \infty
,t)}=-1.
\]

On the other hand, for each of element states $\left \vert \uparrow
,\mathrm{A}\right \rangle ,$ $\left \vert \uparrow,\mathrm{B}\right \rangle
,$\ $\left \vert \downarrow,\mathrm{A}\right \rangle ,$\ $\left \vert
\downarrow,\mathrm{B}\right \rangle $ along a given projection direction
$\vec{e}$ and a fixed $\theta,$ we get a zero-equation%
\begin{align}
\phi(x)  &  =\phi_{0}\mp \frac{\pi}{2}\pm k_{0}(x-x_{0})\\
&  =\pm \frac{\pi}{2}+\theta.\nonumber
\end{align}
Due to the existence of single zero between two projected vortex-membranes for
each state $\left \vert \uparrow,\mathrm{A}\right \rangle ,$ $\left \vert
\uparrow,\mathrm{B}\right \rangle ,$\ $\left \vert \downarrow,\mathrm{A}%
\right \rangle ,$\ $\left \vert \downarrow,\mathrm{B}\right \rangle ,$ all
superposed states $\left \vert \psi \right \rangle =\alpha \left \vert
\uparrow,\mathrm{A}\right \rangle +\beta \left \vert \uparrow,\mathrm{B}%
\right \rangle +\gamma \left \vert \downarrow,\mathrm{A}\right \rangle
+\kappa \left \vert \downarrow,\mathrm{B}\right \rangle $ have a single zero.
According to the conservation of the information unit, there exists a
constraint,%
\begin{equation}
\int \psi^{\dagger}(x,t)\cdot \psi(x,t)dx\equiv1
\end{equation}
that gives a normalization condition as
\begin{equation}
\left \vert \alpha \right \vert ^{2}+\left \vert \beta \right \vert ^{2}+\left \vert
\gamma \right \vert ^{2}+\left \vert \kappa \right \vert ^{2}=1.
\end{equation}
The normalization implies a normalized volume condition,
\begin{align}
&
%TCIMACRO{\dint }%
%BeginExpansion
{\displaystyle \int}
%EndExpansion
z^{\ast}(x,t)z(x,t)dx\nonumber \\
&  =l\cdot \lbrack \left \vert r_{\uparrow,\mathrm{A}}\right \vert ^{2}+\left \vert
r_{\downarrow,\mathrm{A}}\right \vert ^{2}+\left \vert r_{\uparrow,\mathrm{B}%
}\right \vert ^{2}+\left \vert r_{\downarrow,\mathrm{B}}\right \vert
^{2}]\nonumber \\
&  =V_{\mathrm{knot}}%
\end{align}
or
\begin{equation}
\left \vert r_{\uparrow,\mathrm{A}}\right \vert ^{2}+\left \vert r_{\uparrow
,\mathrm{B}}\right \vert ^{2}+\left \vert r_{\downarrow,\mathrm{A}}\right \vert
^{2}+\left \vert r_{\downarrow,\mathrm{B}}\right \vert ^{2}=r_{0}^{2}.
\end{equation}
The total volume of the knot (volume-changing of two entangled
vortex-membranes) $V_{\mathrm{knot}}=l\cdot r_{0}^{2}$ is always fixed.

In addition to the four degrees of freedom, we have another degrees of freedom
-- chiral degrees of freedom by projecting a knot. Under projection with given
projected angle $\theta$, each projected knot state corresponds to two
possible zeros -- one zero with \textrm{sgn}$[\Upsilon]=1$ is denoted by
$\left \vert \mathrm{L}\right \rangle $; the other with \textrm{sgn}%
$[\Upsilon]=-1$ is denoted by $\left \vert \mathrm{R}\right \rangle $
($\Upsilon=\eta_{\mathrm{A},\theta}(x,t)-\eta_{\mathrm{B},\theta}(x,t)$).

\subsubsection{Tensor states for a knot}

For a 3D knot with four internal degrees of freedom $\left(
\begin{array}
[c]{c}%
\left \vert \uparrow,\mathrm{A},x_{0}\right \rangle \\
\left \vert \uparrow,\mathrm{B},x_{0}\right \rangle \\
\left \vert \downarrow,\mathrm{A},x_{0}\right \rangle \\
\left \vert \downarrow,\mathrm{B},x_{0}\right \rangle
\end{array}
\right)  ,$ we introduce a tensor representation to define a knot,%
\begin{equation}
\left \vert \mathbf{Z}_{\mathrm{knot}}\right \rangle =%
%TCIMACRO{\dprod \limits_{I}}%
%BeginExpansion
{\displaystyle \prod \limits_{I}}
%EndExpansion
\left \vert \mathbf{Z}_{\mathrm{knot}}^{I}\right \rangle
\end{equation}
with%
\begin{equation}
\left \vert \mathbf{Z}_{\mathrm{knot}}^{I}\right \rangle =\left(
\begin{array}
[c]{c}%
\beta_{1,1}^{I}\\
\beta_{1,-1}^{I}\\
\beta_{-1,1}^{I}\\
\beta_{-1,-1}^{I}%
\end{array}
\right)  .
\end{equation}
For $\beta_{i,j}^{I},$ $I$ denotes the spatial indices $X,$ $Y$, $Z;$ $i=1,$
$-1$ denotes helical degrees of freedom; $j=1,$ $-1$ denotes the
vortex-degrees of freedom. According to the constraint condition, we have
\begin{equation}
\left \langle \mathbf{Z}_{\mathrm{knot}}^{I}\right \vert \left \vert
\mathbf{Z}_{\mathrm{knot}}^{I}\right \rangle =1,
\end{equation}
So the function for a 3D knot is given by
\begin{equation}
\mathbf{Z}_{\mathrm{knot}}(\vec{x}-\vec{x}_{0},t)=%
%TCIMACRO{\dprod \limits_{I,\vec{x}_{0}}}%
%BeginExpansion
{\displaystyle \prod \limits_{I,\vec{x}_{0}}}
%EndExpansion
\mathbf{Z}_{\mathrm{knot}}^{I}(\vec{x}-\vec{x}_{0},t).\nonumber
\end{equation}

The tensor states for a knot can also be classified by operator
representation. We define the $3\times3\times3$ tensor operators $\hat{\Gamma
}_{\mathrm{knot,}i,j}^{I}=\sigma_{\mathrm{knot,}i}^{I}\otimes(\tau
_{\mathrm{knot,}j}^{I},\vec{1})$ ($I=X,Y,Z$ for different spatial directions;
$i=x,y,z$ for spin degrees of freedom; $j=x,y,z,0$ for the vortex-degrees of
freedom). Different tensor states are described by
\begin{align}
\left \langle \mathbf{\hat{\Gamma}}_{\mathrm{knot}}^{I}\right \rangle  &
=\left \langle \mathbf{Z}_{\mathrm{knot}}\right \vert \mathbf{\hat{\Gamma}}%
^{I}\left \vert \mathbf{Z}_{\mathrm{knot}}\right \rangle \\
&  =\vec{\Gamma}_{\mathrm{knot}}^{I}=\vec{n}_{\mathrm{knot,}\sigma}^{I}%
\otimes(\vec{n}_{\tau}\mathbf{\tau}+\vec{1}\tau_{0}).\nonumber
\end{align}
One type of 3D knot is a $\sigma_{z}$-type knot that is characterized by
\begin{equation}
\left \langle \mathbf{\sigma}_{\mathrm{knot}}^{X}\otimes \vec{1}\right \rangle
=\left \langle \mathbf{\sigma}_{\mathrm{knot}}^{Y}\otimes \vec{1}\right \rangle
=\left \langle \mathbf{\sigma}_{\mathrm{knot}}^{Z}\otimes \vec{1}\right \rangle
=(0,0,\pm1).
\end{equation}
Another 3D knot is called SOC knot that is described
\begin{align}
\left \langle \mathbf{\sigma}_{\mathrm{knot}}^{X}\otimes \vec{1}\right \rangle
&  =(1,0,0),\\
\left \langle \mathbf{\sigma}_{\mathrm{knot}}^{Y}\otimes \vec{1}\right \rangle
&  =(0,1,0),\nonumber \\
\left \langle \mathbf{\sigma}_{\mathrm{knot}}^{Z}\otimes \vec{1}\right \rangle
&  =(0,0,1).\nonumber
\end{align}

\subsubsection{Tensor-network state for a knot-crystal}

Periodic entangled vortex-membranes form knot-crystal. In pseudo-quantum
mechanics, a knot-crystal corresponds to two entangled vortex-membranes
described by a special pure state of Kelvin waves with fixed wave length. In
pseudo-quantum mechanics, we consider flat vortex-membrane as the ground
state, i.e.,
\begin{equation}
\mathbf{Z}(\vec{x},t)=\left(
\begin{array}
[c]{c}%
\mathrm{z}_{\mathrm{A}}(\vec{x},t)\\
\mathrm{z}_{\mathrm{B}}(\vec{x},t)
\end{array}
\right)  \equiv \left(
\begin{array}
[c]{c}%
0\\
0
\end{array}
\right)  \rightarrow \left[  \text{\textrm{vacuum}}\right \rangle .
\end{equation}
A Kelvin wave becomes excited state around $\left[  \text{\textrm{vacuum}%
}\right \rangle $; in emergent quantum mechanics, we consider knot-crystal as a
ground state,
\begin{align}
\mathbf{Z}_{\mathrm{knot-crystal}}(\vec{x},t)  &  =\left(
\begin{array}
[c]{c}%
\mathrm{z}_{\mathrm{A}}(\vec{x},t)\\
\mathrm{z}_{\mathrm{B}}(\vec{x},t)
\end{array}
\right) \\
&  =\left(
\begin{array}
[c]{c}%
r_{\mathrm{A}}\\
r_{\mathrm{B}}%
\end{array}
\right)  e^{i\vec{k}_{0}\cdot \vec{x}-i\omega_{0}t+i\omega^{\ast}%
t/2}\nonumber \\
&  \rightarrow \left \vert \text{\textrm{vacuum}}\right \rangle ,\nonumber
\end{align}
the knots become topological excitations on it.

Because a knot-crystal is a plane Kelvin wave with fixed wave vector $k_{0}$,
we can use the tensor representation to characterize knot-crystals,
$\mathbf{\vec{\Gamma}}_{\mathrm{knot-crystal}}^{I}=(\vec{n}_{\sigma}%
^{I}\mathbf{\sigma}^{I})\otimes(\vec{n}_{\tau}\mathbf{\tau}+\vec{1}\tau_{0})$
and define corresponding generation operator $\mathrm{\hat{U}}%
_{\mathrm{knot-crystal}}(\vec{k}_{0}\mathrm{,}\mathbf{\vec{\Gamma}%
}_{\mathrm{knot-crystal}}^{I})$ for it.

Firstly, we define generate operator by knot-crystal%
\begin{equation}
\mathrm{\hat{U}}_{\mathrm{knot-crystal}}(\vec{k}\mathrm{,}\mathbf{\vec{\Gamma
}}_{\mathrm{knot-crystal}}^{I})=e^{i\int \mathbf{\vec{\Gamma}}^{I}[\phi
_{\vec{k}_{0}}(x)\cdot \hat{K}]dx}\cdot \mathrm{\hat{\digamma}}(r_{0}),
\end{equation}
where $\mathrm{\hat{\digamma}}(r_{0})$ is an expanding operator by shifting
radius from $0$ to $r_{0}$ on the membrane, $\hat{K}=-i\frac{d}{d\phi}$,
$\phi(x)=\pm kx,$ and $x$ is coordinate along $\vec{x}$ direction. Here
$\left[  \text{\textrm{0}}\right \rangle $ denotes a vortex-membrane with
$r_{0}=0$. The tensor state of the knot-crystal is determined by
$\mathbf{\vec{\Gamma}}_{\mathrm{knot-crystal}}^{I}=(\vec{n}_{\sigma}%
^{I}\mathbf{\sigma}^{I})\otimes \vec{1}$.

For example, a particular knot-crystal is called SOC knot-crystal that is
described by a 3D plane Kelvin waves with fixed wave-length $\mathbf{Z}%
_{\mathrm{knot-crystal}}(\vec{x})$. The tensor state of it is
\begin{align}
\left \langle \mathbf{\sigma}^{X}\otimes \vec{1}\right \rangle  &  =\vec
{n}_{\sigma}^{X}=(1,0,0),\\
\left \langle \mathbf{\sigma}^{Y}\otimes \vec{1}\right \rangle  &  =\vec
{n}_{\sigma}^{Y}=(0,1,0),\nonumber \\
\left \langle \mathbf{\sigma}^{Z}\otimes \vec{1}\right \rangle  &  =\vec
{n}_{\sigma}^{Z}=(0,0,1).\nonumber
\end{align}

On the one hand, a knot is a piece of knot-crystal; On the other hand, a
knot-crystal can be regarded as a composite system with multi-knot, each of
which is described by same tensor state. After projection, the knot-crystal
becomes a zero lattice, of which there are two sublattices $R/L$ and a
tensor-network state $\vec{\Gamma}_{\mathrm{knot-crystal}}%
^{\mathrm{Tensor-network}}$ describes the entanglement relationship between
two nearest-neighbor zeros. The tensor-network state comes from the product of
the basis of all knots
\begin{align}
&
%TCIMACRO{\dprod \limits_{\vec{x}_{0}}}%
%BeginExpansion
{\displaystyle \prod \limits_{\vec{x}_{0}}}
%EndExpansion
\left(
\begin{array}
[c]{c}%
\left \vert L,\vec{x}_{0}\right \rangle
\end{array}
\right)  \otimes \left(
\begin{array}
[c]{c}%
\left \vert R,\vec{x}_{0}-a\cdot \vec{e}_{x}\right \rangle
\end{array}
\right) \\
&  \otimes \left(
\begin{array}
[c]{c}%
\left \vert R,\vec{x}_{0}-a\cdot \vec{e}_{y}\right \rangle
\end{array}
\right)  \otimes \left(
\begin{array}
[c]{c}%
\left \vert R,\vec{x}_{0}-a\cdot \vec{e}_{z}\right \rangle
\end{array}
\right) \nonumber \\
&  ...\nonumber
\end{align}
The summation of $\vec{x}_{0}$ is obtained by summing the zeros between
projected vortex-membranes. The generalized translation symmetry for a
knot-crystal becomes
\begin{equation}
\left(
\begin{array}
[c]{c}%
\left \vert R,\vec{x}_{0}-a\cdot \vec{e}\right \rangle
\end{array}
\right)  =\mathcal{T}(\Delta \vec{x}=-a\cdot \vec{e})\iota^{-}\cdot \left(
\begin{array}
[c]{c}%
\left \vert L,\vec{x}_{0}\right \rangle
\end{array}
\right)
\end{equation}
where $\iota^{\pm}$ is an operator denoting switching of the chiral
(sub-lattice) degrees of freedom as
\begin{equation}
\iota^{+}=(R\rightarrow L),\text{ }\iota^{-}=(L\rightarrow R).
\end{equation}
There exist topological defects of the tensor-network states that are just
extra zeros (the knots). The chiral degrees of freedom become the sublattice
degrees of freedom for knots on knot-crystal! In particular, the emergent
Lorentz invariance for knots comes from the two-sublattice structure.

By considering the local degrees of freedom, the state space (the Hilbert
space) becomes suddenly enlarged from $2$ in pseudo-quantum mechanics to
$4^{N_{\mathrm{knot}}}$ in emergent quantum mechanics. The chiral degrees of
freedom become the freedom of "lattice" and label the position of the sublattices.

\subsubsection{Knot-operation and knot-number operator}

To characterize knots (the elementary volume-changing of two entangled
vortex-membranes), we introduce its operator-representation.

For 1D unified knot in Dirac representation, we define it by
\begin{align}
\mathbf{Z}_{\mathrm{knot}}(x,t)  &  =\mathrm{\hat{U}}_{\mathrm{knot}%
}\mathbf{Z}_{\mathrm{knot-crystal}}(x,t)\\
&  =\mathrm{\hat{U}}_{\mathrm{knot}}\mathrm{\hat{U}}_{\mathrm{knot-crystal}%
}\left(
\begin{array}
[c]{c}%
\left[  \text{\textrm{0}}\right \rangle \\
\left[  \text{\textrm{0}}\right \rangle \\
\left[  \text{\textrm{0}}\right \rangle \\
\left[  \text{\textrm{0}}\right \rangle
\end{array}
\right) \nonumber
\end{align}
where $\mathrm{\hat{U}}_{\mathrm{knot}}$ is a generation operator for a knot
that is an eigenstate of spin operator $\sigma_{z}$,
\begin{equation}
\mathrm{\hat{U}}_{\mathrm{knot}}=e^{i%
%TCIMACRO{\dint }%
%BeginExpansion
{\displaystyle \int}
%EndExpansion
(\sigma_{z}\otimes \vec{1})[\phi_{\mathrm{knot}}(x)\cdot \hat{K}]dx}%
\cdot \mathrm{\hat{\digamma}}_{\mathrm{knot}}(r_{0})
\end{equation}
where $\mathrm{\hat{\digamma}}_{\mathrm{knot}}(r_{0})$ is an expanding
operator by shifting $0$ to $r_{0}$ in the winding region of a knot (for
example, $x\in(x_{0},x_{0}+a]$). $\mathrm{\hat{U}}_{\mathrm{knot-crystal}}$
denotes the generation operator for a 1D knot-crystal that is the plane Kelvin
wave with fixed wave-length and becomes the same tensor state for a knot.

We define the generation operator for a 3D knot by doing a knot-operation
\begin{equation}
\mathrm{\hat{U}}_{\mathrm{knot}}=\mathrm{\hat{U}}_{\mathrm{knot}}%
^{X}\mathrm{\hat{U}}_{\mathrm{knot}}^{Y}\mathrm{\hat{U}}_{\mathrm{knot}}^{Z}%
\end{equation}
where
\begin{align}
\mathrm{\hat{U}}_{\mathrm{knot}}^{X}  &  =e^{i\int(\sigma^{X}\otimes \vec
{1})[\phi_{\mathrm{knot}}(x)\cdot \hat{K}]dx}\cdot \mathrm{\hat{\digamma}%
}_{\mathrm{knot}}(r_{0}),\\
\mathrm{\hat{U}}_{\mathrm{knot}}^{Y}  &  =e^{i\int(\sigma^{Y}\otimes \vec
{1})[\phi_{\mathrm{knot}}(y)\cdot \hat{K}]dy}\cdot \mathrm{\hat{\digamma}%
}_{\mathrm{knot}}(r_{0}),\nonumber \\
\mathrm{\hat{U}}_{\mathrm{knot}}^{Z}  &  =e^{i\int(\sigma^{Z}\otimes \vec
{1})[\phi_{\mathrm{knot}}(z)\cdot \hat{K}]dz}\cdot \mathrm{\hat{\digamma}%
}_{\mathrm{knot}}(r_{0}).\nonumber
\end{align}
For example, 3D SOC knot is defined by%
\begin{align}
\mathrm{\hat{U}}_{\mathrm{knot}}^{X}  &  =e^{\int i(\sigma_{x}\otimes \vec
{1})[\phi_{\mathrm{knot}}(x)\cdot \hat{K}]dx}\cdot \mathrm{\hat{\digamma}%
}_{\mathrm{knot}}(r_{0}),\\
\mathrm{\hat{U}}_{\mathrm{knot}}^{Y}  &  =e^{\int i(\sigma_{y}\otimes \vec
{1})[\phi_{\mathrm{knot}}(y)\cdot \hat{K}]dy}\cdot \mathrm{\hat{\digamma}%
}_{\mathrm{knot}}(r_{0}),\nonumber \\
\mathrm{\hat{U}}_{\mathrm{knot}}^{Z}  &  =e^{\int i(\sigma_{z}\otimes \vec
{1})[\phi_{\mathrm{knot}}(z)\cdot \hat{K}]dz}\cdot \mathrm{\hat{\digamma}%
}_{\mathrm{knot}}(r_{0});\nonumber
\end{align}
3D $\sigma_{z}$-knot is defined by%
\begin{align}
\mathrm{\hat{U}}_{\mathrm{knot}}^{X}  &  =e^{\int i(\sigma_{z}\otimes \vec
{1})[\phi_{\mathrm{knot}}(x)\cdot \hat{K}]dx}\cdot \mathrm{\hat{\digamma}%
}_{\mathrm{knot}}(r_{0}),\\
\mathrm{\hat{U}}_{\mathrm{knot}}^{Y}  &  =e^{\int i(\sigma_{z}\otimes \vec
{1})[\phi_{\mathrm{knot}}(y)\cdot \hat{K}]dy}\cdot \mathrm{\hat{\digamma}%
}_{\mathrm{knot}}(r_{0}),\nonumber \\
\mathrm{\hat{U}}_{\mathrm{knot}}^{Z}  &  =e^{\int i(\sigma_{z}\otimes \vec
{1})[\phi_{\mathrm{knot}}(z)\cdot \hat{K}]dz}\cdot \mathrm{\hat{\digamma}%
}_{\mathrm{knot}}(r_{0}).\nonumber
\end{align}

In addition, the knot-number of a 3D knot is defined by%
\begin{align}
\left \langle \hat{K}\right \rangle  &  =\left \langle \hat{K}_{x}\hat{K}_{y}%
\hat{K}_{z}\right \rangle \nonumber \\
&  =\frac{1}{\pi^{3}r_{0}^{2}}[\int \mathbf{Z}_{\mathrm{knot}}^{\ast}(\vec
{x},t)\cdot \hat{K}_{x}\hat{K}_{y}\hat{K}_{z}\nonumber \\
&  \cdot \mathbf{Z}_{\mathrm{knot}}(\vec{x},t)]d\phi(x)d\phi(y)d\phi(z).
\end{align}

\subsection{Emergent quantum mechanics}

In above section, we show that a knot (the elementary volume-changing of two
entangled vortex-membranes) is an elementary topological defect of
knot-crystal and also fundamental information carrier. In this section, we
will show that knots obey the emergent quantum mechanics.

\subsubsection{Information scaling between knot and knot-crystal}

From the point view of pseudo-quantum mechanics, knot-crystal is a special
entangled vortex-membranes that is described by a tensor state. The total
degrees of freedom for a knot-crystal is $2$. From the point view of emergent
quantum mechanics, a knot-crystal is periodic zeros between projected
vortex-membranes that is described by tensor-network state for knots. The
total degrees of freedom of a knot-crystal becomes $4^{N_{\mathrm{knot}}}$. In
particular, there exists a holographic principle: the pseudo-spin-tensor state
for a knot on a knot-crystal is same to the tensor state for the knot-crystal
in emergent quantum mechanics.

On the other hand, a knot corresponds to the changing of one zero on a
knot-crystal. When a knot is generated on a knot-crystal, the boundary of all
Kelvin waves changes -- periodic boundary condition changes into anti-periodic
boundary condition, vice versa. Because the Hilbert space is never changed
during the changing the boundary condition, the Hilbert space of knots
$\left(
\begin{array}
[c]{c}%
\left \vert \psi_{\uparrow,\mathrm{A}}\right \rangle \\
\left \vert \psi_{\uparrow,\mathrm{B}}\right \rangle \\
\left \vert \psi_{\downarrow,\mathrm{A}}\right \rangle \\
\left \vert \psi_{\downarrow,\mathrm{B}}\right \rangle
\end{array}
\right)  $ corresponds to the Hilbert space of Kelvin waves $\left(
\begin{array}
[c]{c}%
\left \vert \psi_{\uparrow,\mathrm{A}}\right \rangle \\
\left \vert \psi_{\uparrow,\mathrm{B}}\right \rangle \\
\left \vert \psi_{\downarrow,\mathrm{A}}\right \rangle \\
\left \vert \psi_{\downarrow,\mathrm{B}}\right \rangle
\end{array}
\right)  $ by perturbating the knot-crystal, i.e.,
\begin{equation}
\left(
\begin{array}
[c]{c}%
\left \vert \psi_{\uparrow,\mathrm{A}}\right \rangle \\
\left \vert \psi_{\uparrow,\mathrm{B}}\right \rangle \\
\left \vert \psi_{\downarrow,\mathrm{A}}\right \rangle \\
\left \vert \psi_{\downarrow,\mathrm{B}}\right \rangle
\end{array}
\right)  \Longleftrightarrow \left(
\begin{array}
[c]{c}%
\left[  \psi_{\uparrow,\mathrm{A}}\right \rangle \\
\left[  \psi_{\uparrow,\mathrm{B}}\right \rangle \\
\left[  \psi_{\downarrow,\mathrm{A}}\right \rangle \\
\left[  \psi_{\downarrow,\mathrm{B}}\right \rangle
\end{array}
\right)  .
\end{equation}
According to the correspondence, several quantum quantities of knots (momentum
and angular momentum) can be easily obtained by using the information
scaling,
\begin{equation}
\hbar_{\mathrm{eff}}\longrightarrow \hbar_{\mathrm{knot}}=\frac{\hbar
_{\mathrm{eff}}}{N_{\mathrm{knot}}}%
\end{equation}
where $\hbar_{\mathrm{knot}}$ is the Planck constant for knots in emergent
quantum mechanics and $\hbar_{\mathrm{eff}}$ is the Planck constant for plane
Kelvin wave with fixed wave-length in pseudo-quantum mechanics.
$N_{\mathrm{knot}}$ is the total knot number in the knot-crystal.

\subsubsection{Fragmentized knot}

According to the geometric character of the three conserved physical
quantities, the shape of knot will never be changed. However, the knot can
split and the three physical quantities are conserved for all knot-pieces. To
characterize this property, we introduce the concept of "fragmentized knot".
Quantum mechanics is a mechanics to determine the distribution of knot-pieces.

We take 1D knot with fixed spin and vortex degrees of freedom as an example to
show the concept of "fragmentized knot" ($\sigma \rightarrow \uparrow,$
$\tau \rightarrow \mathrm{B}$). Before introducing fragmentized knot, we give
the definition of unified knot that is described by
\begin{equation}
\mathrm{z}_{\mathrm{knot}}(\phi(x))=r_{\mathrm{knot}}\exp[i\phi_{\mathrm{knot}%
}(x)]
\end{equation}
where
\begin{equation}
r_{\mathrm{knot}}(x)\rightarrow \left(
\begin{array}
[c]{c}%
0,\text{ }x\in(-\infty,x_{0}]\\
r_{0},\text{ }x\in(x_{0},x_{0}+a]\\
0,\text{ }x\in(x_{0}+a,\infty]
\end{array}
\right)  .
\end{equation}
$x$ is coordinate along an arbitrary direction $\vec{e}$ and $\phi
_{\mathrm{knot}}(x)$ is the corresponding phase angle.

\begin{figure}[ptb]
\includegraphics[clip,width=0.65\textwidth]{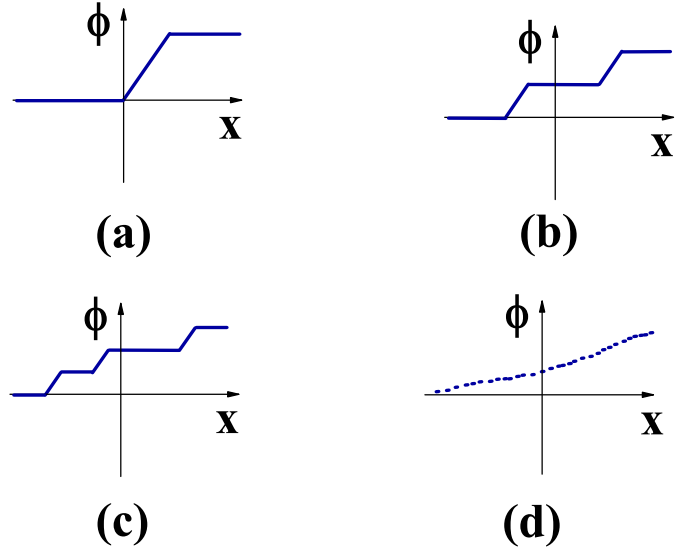}\caption{(a) A knot. The
inset illustrates the unified knot; (b) A fragmentized knot that is split into
two pieces; (c) A fragmentized knot that is split three pieces; (d) A
fragmentized knot that is split into infinite pieces. The blue spots denote
the fragmentized knot with $N\rightarrow \infty.$ For fragmentized knots with
$N=1,$ $2,$ $3,$ $\infty$, there always exists a single knot solution and the
knot number is $1$.}%
\end{figure}

We split a unified knot into two pieces, each of them is a half-knot with
$\frac{\pi}{2}$ phase-changing. As shown in Fig.10, we split the knot into two
pieces.\ The knot-function of two-piece fragmentized knot is $\mathrm{z}%
(\phi_{\mathrm{knot}}(x))$ where%
\begin{equation}
r_{\mathrm{knot}}(x)\rightarrow \left(
\begin{array}
[c]{c}%
0,\text{ }x\in(-\infty,x_{0}]\\
r_{0},\text{ }x\in(x_{0},x_{0}+\frac{a}{2}]\\
0,\text{ }x\in(x_{0}+\frac{a}{2},x_{0}+x_{0}^{\prime}]\\
r_{0},\text{ }x\in(x_{0}+x_{0}^{\prime},x_{0}+x_{0}^{\prime}+\frac{a}{2}]\\
0,\text{ }x\in(x_{0}+x_{0}^{\prime}+\frac{a}{2},\infty]
\end{array}
\right)
\end{equation}
and
\begin{equation}
\phi_{\mathrm{knot}}(x)=\left(
\begin{array}
[c]{c}%
-\frac{\pi}{2},\text{ }x\in(-\infty,x_{0}]\\
-\frac{\pi}{2}+\frac{\pi}{a}(x-x_{0}),\text{ }x\in(x_{0},x_{0}+\frac{a}{2}]\\
0,\text{ }x\in(x_{0}+\frac{a}{2},x_{0}+x_{0}^{\prime}]\\
\frac{\pi}{a}(x-x_{0}-x_{0}^{\prime}),\text{ }x\in(x_{0}+x_{0}^{\prime}%
,x_{0}+x_{0}^{\prime}+\frac{a}{2}]\\
\frac{\pi}{2},\text{ }x\in(x_{0}+x_{0}^{\prime}+\frac{a}{2},\infty]
\end{array}
\right)
\end{equation}
with a condition of $x_{0}^{\prime}>\frac{a}{2}.$ It is obvious that for the
fragmentized knot, there exists only single zero-solution due to topological
condition. And we have $50\%$ probability to find a zero-solution in the
region of $x_{0}<x\leq x_{0}+\frac{a}{2}$ and $50\%$ probability to find a
zero-solution in the region of $x_{0}+x_{0}^{\prime}<x\leq x_{0}+x_{0}%
^{\prime}+\frac{a}{2}$. We had defined a knot with half-winding to be
\begin{equation}
\mathrm{z}_{\mathrm{N}=\mathrm{1}}(\phi(x),x_{0})=\mathrm{\hat{U}}%
(\phi_{\mathrm{knot}},x_{0})\mathrm{z},
\end{equation}
of which the knot is at $x_{0}$. As a result, The knot-function of two-piece
fragmentized knot is defined by
\begin{align}
\left[  \mathrm{z}(\phi)\right]  _{\mathrm{fragment,N}=\mathrm{2}}  &
=\mathrm{\hat{U}}(\Delta \phi_{\frac{1}{2}\text{-}\mathrm{knot}},x_{0})\\
&  \cdot \mathrm{\hat{U}}(\Delta \phi_{\frac{1}{2}\text{-}\mathrm{knot}}%
,x_{0}+x_{0}^{\prime})\mathrm{z}_{0},\nonumber
\end{align}
of which the two half-knots are at $x_{0}$ and $x_{0}+x_{0}^{\prime},$ respectively.

Similarly, we may split a knot into $N$ pieces, each of which is an identical
knot-piece with $\frac{\pi}{N}$ phase-changing. The knot-function of $N$
knot-pieces is
\begin{align}
\left[  \mathrm{z}(\phi)\right]  _{\mathrm{fragment,N}}  &  =\mathrm{\hat{U}%
}(\Delta \phi_{\frac{1}{N}\text{-}\mathrm{knot}},\left(  x_{0}^{\prime}\right)
_{1})\\
&  \cdot \mathrm{\hat{U}}(\Delta \phi_{\frac{1}{N}\text{-}\mathrm{knot}},\left(
x_{0}^{\prime}\right)  _{2})...\nonumber \\
&  \cdot \mathrm{\hat{U}}(\Delta \phi_{\frac{1}{N}\text{-}\mathrm{knot}},\left(
x_{0}^{\prime}\right)  _{N})\mathrm{z}_{0}\nonumber \\
&  =%
%TCIMACRO{\dprod \limits_{i=1}^{N}}%
%BeginExpansion
{\displaystyle \prod \limits_{i=1}^{N}}
%EndExpansion
\mathrm{\hat{U}}(\Delta \phi_{\frac{1}{N}\text{-}\mathrm{knot}},\left(
x_{0}^{\prime}\right)  _{i})\mathrm{z}_{0}\nonumber
\end{align}
where $N$ identical knot-pieces are at $\left(  x_{0}^{\prime}\right)  _{1},$
$\left(  x_{0}^{\prime}\right)  _{2},$..., $\left(  x_{0}^{\prime}\right)
_{N}$, respectively. For $N$ knot-pieces, there also exists a zero and the
knot-number is conserved. We have $\frac{1}{N}$ probability to find a zero in
a knot-piece. By using similar argument, an arbitrary superposed states of
knots can be fragmentized. Fig.10.(d) is an illustration of a fragmentized knot.

The generation operator for a fragmentized knot with uniform distribution of
knot-pieces is defined as
\begin{equation}
\mathbf{\tilde{Z}}_{\mathrm{fragment}}=\mathrm{\hat{U}}_{\mathrm{fragment}%
}\left \vert 0\right \rangle
\end{equation}
where $\mathrm{\hat{U}}_{\mathrm{fragment}}=\mathrm{\hat{U}}%
_{\mathrm{fragment}}^{X}\mathrm{\hat{U}}_{\mathrm{fragment}}^{Y}%
\mathrm{\hat{U}}_{\mathrm{fragment}}^{Z}$ denotes a unform distributed
knot-pieces as%
\begin{align}
\mathrm{\hat{U}}_{\mathrm{fragment}}^{X}  &  =%
%TCIMACRO{\dprod \limits_{i=1}^{N\rightarrow\infty}}%
%BeginExpansion
{\displaystyle \prod \limits_{i=1}^{N\rightarrow \infty}}
%EndExpansion
\mathrm{\hat{U}}_{\mathrm{fragment}}^{X}(\Delta \phi_{\frac{1}{N}%
\text{-}\mathrm{knot}}\left(  x_{0}\right)  _{i}),\\
\mathrm{\hat{U}}_{\mathrm{fragment}}^{Y}  &  =%
%TCIMACRO{\dprod \limits_{i=1}^{N\rightarrow\infty}}%
%BeginExpansion
{\displaystyle \prod \limits_{i=1}^{N\rightarrow \infty}}
%EndExpansion
\mathrm{\hat{U}}_{\mathrm{fragment}}^{Y}(\Delta \phi_{\frac{1}{N}%
\text{-}\mathrm{knot}}\left(  y_{0}\right)  _{i}),\nonumber \\
\mathrm{\hat{U}}_{\mathrm{fragment}}^{Z}  &  =%
%TCIMACRO{\dprod \limits_{i=1}^{N\rightarrow\infty}}%
%BeginExpansion
{\displaystyle \prod \limits_{i=1}^{N\rightarrow \infty}}
%EndExpansion
\mathrm{\hat{U}}_{\mathrm{fragment}}^{Z}(\Delta \phi_{\frac{1}{N}%
\text{-}\mathrm{knot}}\left(  z_{0}\right)  _{i}).\nonumber
\end{align}
with
\begin{align}
\mathrm{\hat{U}}_{\mathrm{fragment}}^{X}(\Delta \phi_{\frac{1}{N}%
\text{-}\mathrm{knot}}\left(  x_{0}\right)  _{i})  &  =e^{\int i(\sigma
_{X}\otimes{\large \vec{1}})[\Delta \phi_{\frac{1}{N}\text{-}\mathrm{knot}%
}(\left(  x_{0}\right)  _{i})\cdot \hat{K}]dx}\cdot \hat{\digamma}_{\frac{1}%
{N}\text{-}\mathrm{knot}}(r_{0},\left(  x_{0}\right)  _{i}),\\
\mathrm{\hat{U}}_{\mathrm{fragment}}^{Y}(\Delta \phi_{\frac{1}{N}%
\text{-}\mathrm{knot}}\left(  y_{0}\right)  _{i})  &  =e^{\int i(\sigma
_{Y}\otimes{\large \vec{1}})[\Delta \phi_{\frac{1}{N}\text{-}\mathrm{knot}%
}(\left(  y_{0}\right)  _{i})\cdot \hat{K}]dy}\cdot \hat{\digamma}_{\frac{1}%
{N}\text{-}\mathrm{knot}}(r_{0},\left(  y_{0}\right)  _{i}),\nonumber \\
\mathrm{\hat{U}}_{\mathrm{fragment}}^{Z}(\Delta \phi_{\frac{1}{N}%
\text{-}\mathrm{knot}}\left(  z_{0}\right)  _{i})  &  =e^{\int i(\sigma
_{Z}\otimes{\large \vec{1}})[\Delta \phi_{\frac{1}{N}\text{-}\mathrm{knot}%
}(\left(  z_{0}\right)  _{i})\cdot \hat{K}]dz}\cdot \hat{\digamma}_{\frac{1}%
{N}\text{-}\mathrm{knot}}(r_{0},\left(  z_{0}\right)  _{i}),\nonumber
\end{align}
where $\hat{\digamma}_{\frac{1}{N}\text{-}\mathrm{knot}}(r_{0},\left(
x_{0}\right)  _{i})$ is an operator by shifting $0$ to $r_{0}$ in the winding
region of $x\in(\left(  x_{0}\right)  _{i},\left(  x_{0}\right)  _{i}+\frac
{a}{N}]$, $\hat{\digamma}_{\frac{1}{N}\text{-}\mathrm{knot}}(r_{0},\left(
y_{0}\right)  _{i})$ is an operator by shifting $0$ to $r_{0}$ in the winding
region of $y\in(\left(  y_{0}\right)  _{i},\left(  y_{0}\right)  _{i}+\frac
{a}{N}]$, $\hat{\digamma}_{\frac{1}{N}\text{-}\mathrm{knot}}(r_{0},\left(
z_{0}\right)  _{i})$ is an operator by shifting $0$ to $r_{0}$ in the winding
region of $z\in(\left(  z_{0}\right)  _{i},\left(  z_{0}\right)  _{i}+\frac
{a}{N}]$, and
\begin{align}
\Delta \phi_{\frac{1}{N}\text{-}\mathrm{knot}}(\left(  x_{0}\right)  _{i})  &
=\left \{
\begin{array}
[c]{c}%
-\phi_{0}-\frac{\pi}{2N},\text{ }x\in(-\infty,x_{0}]\\
-\phi_{0}-\frac{\pi}{2N}+k_{0}(x-\left(  x_{0}\right)  _{i}),\text{ }%
x\in(\left(  x_{0}\right)  _{i},\left(  x_{0}\right)  _{i}+\frac{a}{N}]\\
-\phi_{0}+\frac{\pi}{2N},\text{ }x\in(\left(  x_{0}\right)  _{i}+\frac{a}%
{N},\infty)
\end{array}
\right \}  ,\\
\Delta \phi_{\frac{1}{N}\text{-}\mathrm{knot}}(\left(  y_{0}\right)  _{i})  &
=\left \{
\begin{array}
[c]{c}%
-\phi_{0}-\frac{\pi}{2N},\text{ }y\in(-\infty,y_{0}]\\
-\phi_{0}-\frac{\pi}{2N}+k_{0}(y-\left(  y_{0}\right)  _{i}),\text{ }%
y\in(\left(  y_{0}\right)  _{i},\left(  y_{0}\right)  _{i}+\frac{a}{N}]\\
-\phi_{0}+\frac{\pi}{2N},\text{ }y\in(\left(  y_{0}\right)  _{i}+\frac{a}%
{N},\infty)
\end{array}
\right \}  ,\nonumber \\
\Delta \phi_{\frac{1}{N}\text{-}\mathrm{knot}}(\left(  z_{0}\right)  _{i})  &
=\left \{
\begin{array}
[c]{c}%
-\phi_{0}-\frac{\pi}{2N},\text{ }z\in(-\infty,x_{0}]\\
-\phi_{0}-\frac{\pi}{2N}+k_{0}(z-\left(  z_{0}\right)  _{i}),\text{ }%
z\in(\left(  z_{0}\right)  _{i},\left(  z_{0}\right)  _{i}+\frac{a}{N}]\\
-\phi_{0}+\frac{\pi}{2N},\text{ }\mathrm{z}\in(\left(  z_{0}\right)
_{i}+\frac{a}{N},\infty)
\end{array}
\right \}  ,
\end{align}
where $k_{0}=\frac{\pi}{a}.$ The center of the knot-piece with $\frac{\pi}{N}$
phase-changing is at $\left(  x_{0},y_{0},z_{0}\right)  _{i}$. For this case,
we assume the distribution of the knot-pieces is same as the zeros of the
knot-crystal,
\begin{align}
\left(  x_{0}\right)  _{i}  &  =a\cdot i,\text{ }i=0,1,2,...N_{x}=\frac{L_{x}%
}{a}\\
\left(  y_{0}\right)  _{i}  &  =a\cdot i,\text{ }i=0,1,2,...N_{y}=\frac{L_{y}%
}{a}\nonumber \\
\left(  z_{0}\right)  _{i}  &  =a\cdot i,\text{ }i=0,1,2,...N_{z}=\frac{L_{z}%
}{a}.\nonumber
\end{align}

An important issue is to obtain the spatial and temporal distribution of the
knot-pieces that has the information of a knot. In next part, we will show
that the quantum states of a knot -- the wave-functions describe the spatial
and temporal distribution of the knot-pieces determined by the perturbative
Kelvin waves on a knot-crystal.

\subsubsection{Quantum states of fragmentized knots}

Firstly, we consider a knot with uniform knot-pieces, of which the state is
denoted by
\begin{equation}
\mathrm{\hat{U}}_{\mathrm{fragment}}\left \vert \text{\textrm{0}}\right \rangle
\rightarrow \left \vert \psi \right \rangle =c_{\vec{k}=0}^{\dagger}\left \vert
\text{\textrm{vacuum}}\right \rangle
\end{equation}
where $c_{\vec{k}=0}^{\dagger}$ means that the knot has the same wave vectors
to those of its background. The density of knot-pieces is given by
\begin{equation}
\rho_{\mathrm{knot}}=\frac{1}{V_{P}}%
\end{equation}
where $V_{P}$ is volume of the system,
\begin{equation}
V_{P}=\left(
\begin{array}
[c]{c}%
L_{x},\text{ }d=1\\
L_{x}L_{y},\text{ }d=2\\
L_{x}L_{y}L_{z},\text{ }d=3
\end{array}
\right)  .
\end{equation}
$L_{x}/L_{y}/L_{z}$ denotes the size of vortex-membrane along $X/Y/X$
direction. It is obvious that the function for a knot with $\vec{k}=0$ is same
as that of a knot-crystal by scaling the linking number along given direction.

We take 1D knot with fixed spin and vortex degrees of freedom (for example,
$\sigma=\uparrow,$ $\tau=\mathrm{B}$) as an example to prove the knot density
$\rho_{\mathrm{knot}}=\frac{1}{L_{x}}$ for a uniform distribution of
knot-pieces. For 1D knot, the knot-number is defined as%
\begin{equation}
N_{\mathrm{knot}}=\left \langle \hat{K}\right \rangle =\frac{1}{\pi r_{0}^{2}%
}\int \mathrm{z}_{\mathrm{fragment}}^{\ast}\text{ }\hat{K}\text{ }%
\mathrm{z}_{\mathrm{fragment}}\text{ }d\phi(x)
\end{equation}
where $\hat{K}=-i\frac{d}{d\phi}$. To calculate the knot-number, we introduce
two types of integrals: one is \emph{on-vortex-membrane integral }with a
sorting of knot-positions in on vortex-membrane, the other is
\emph{out-vortex-membrane integral} with a sorting of knot-phases in extra
space $(\xi,\eta)$.

We can label a knot-piece by its position $\left(  x_{0}\right)  _{i_{x}}$ on
vortex-membrane space or label a knot-piece by its position $\left(
x_{0}\right)  _{i_{\phi}}$ in $(\xi,\eta)$ space. Here $i_{\phi}$ denotes a
sorting of ordering of phase $\phi$ from small to bigger and $i_{x}$ denotes a
sorting of coordination $x$ with a given order. Each $i_{\phi}$ corresponds to
an $i_{x}.$ As a result, there exists an \emph{integral-transformation }by
reorganizing all knot-pieces with different rules as%
\begin{align}
N_{\mathrm{knot}}  &  =\frac{1}{r_{0}^{2}\pi}\int \mathrm{z}^{\ast}\text{ }%
\hat{K}\text{ }\mathrm{z}_{\mathrm{fragment}}d\phi(x)\\
&  =\frac{1}{r_{0}^{2}}%
%TCIMACRO{\dsum \nolimits_{i_{\phi}=1}^{i_{\phi}=N}}%
%BeginExpansion
{\displaystyle \sum \nolimits_{i_{\phi}=1}^{i_{\phi}=N}}
%EndExpansion
\mathrm{z}_{\mathrm{fragment,N}}^{\ast}\  \hat{K}\text{ }\mathrm{z}%
_{\mathrm{fragment,N}}\nonumber \\
&  =\frac{1}{r_{0}^{2}}%
%TCIMACRO{\dsum \nolimits_{i_{x}=1}^{i_{x}=N}}%
%BeginExpansion
{\displaystyle \sum \nolimits_{i_{x}=1}^{i_{x}=N}}
%EndExpansion
\mathrm{z}_{\mathrm{fragment,N}}^{\ast}\text{ }\hat{K}\text{ }\mathrm{z}%
_{\mathrm{fragment,N}}\nonumber \\
&  \cdot%
%TCIMACRO{\dsum \nolimits_{i_{y}=1}^{i_{y}=N}}%
%BeginExpansion
{\displaystyle \sum \nolimits_{i_{y}=1}^{i_{y}=N}}
%EndExpansion
\mathrm{z}_{\mathrm{fragment,N}}^{\ast}\text{ }\hat{K}\text{ }\mathrm{z}%
_{\mathrm{fragment,N}}\nonumber \\
&  \cdot%
%TCIMACRO{\dsum \nolimits_{i_{z}=1}^{i_{z}=N}}%
%BeginExpansion
{\displaystyle \sum \nolimits_{i_{z}=1}^{i_{z}=N}}
%EndExpansion
\mathrm{z}_{\mathrm{fragment,N}}^{\ast}\text{ }\hat{K}\text{ }\mathrm{z}%
_{\mathrm{fragment,N}}\nonumber \\
&  =\frac{1}{r_{0}^{2}}\frac{1}{L_{x}}\int \text{ }\mathrm{z}%
_{\mathrm{fragment}}^{\ast}\text{ }\hat{K}\text{ }\mathrm{z}%
_{\mathrm{fragment}}\text{ }dx\nonumber
\end{align}
By transforming an out-vortex-membrane integral to on-vortex-membrane
integral, the knot-number is obtained as
\begin{equation}
N_{\mathrm{knot}}=\left \langle \hat{K}\right \rangle =\frac{1}{\pi r_{0}^{2}%
}\int \mathrm{z}_{\mathrm{fragment}}^{\ast}\text{ }\hat{K}\text{ }%
\mathrm{z}_{\mathrm{fragment}}\text{ }dx\nonumber
\end{equation}
where
\begin{equation}
\mathrm{z}_{\mathrm{fragment}}=\hat{U}_{\mathrm{fragment}}\mathrm{z}%
_{\mathrm{knot-crystal}}.
\end{equation}
As a result, we get
\begin{align}
N_{\mathrm{knot}}  &  =\left \langle \hat{K}\right \rangle \nonumber \\
&  =\frac{1}{\pi r_{0}^{2}}\int \mathrm{z}_{\mathrm{knot-crystal}}^{\ast}\{
\hat{U}_{\mathrm{fragment}}^{-1}\hat{K}\hat{U}_{\mathrm{fragment}})\}
\mathrm{z}_{\mathrm{knot-crystal}}dx\\
&  =\frac{1}{L_{x}}%
%TCIMACRO{\dint }%
%BeginExpansion
{\displaystyle \int}
%EndExpansion
dx=1.\nonumber
\end{align}
In the limit of $N\rightarrow \infty,$ we indeed have a uniform distribution of
the $N$ identical knot-pieces on vortex-membrane,%
\begin{equation}
\rho_{\mathrm{knot}}=\frac{1}{V_{P}}.
\end{equation}

Next, we consider a plane Kelvin wave with an extra fragmentized knot. Now,
the system is described by
\begin{align}
\mathbf{Z}_{\mathrm{knot-crystal}}(\vec{x},t)  &  \longrightarrow
\mathbf{Z}_{\mathrm{knot-crystal}}^{\prime}(\vec{x},t)\nonumber \\
&  =e^{i\cdot \mathbf{1}\cdot(\vec{k}\cdot \vec{x})-i\omega \cdot t}%
\mathbf{Z}_{\mathrm{knot-crystal}}(\vec{x},t).
\end{align}
Here, for simplify, we take a fragmentized knot with fixed spin (for example,
$\uparrow$) and vortex degrees of freedom (for example, \textrm{B}) as an
example. Then we consider the effect of an extra knot,
\begin{equation}
\mathrm{z}_{\mathrm{fragment}}=\frac{1}{\sqrt{V_{P}}}\exp[i\Delta \phi(\vec
{x},t)]
\end{equation}
where $\Delta \phi(\vec{x},t)=\vec{k}\cdot \vec{x}-\omega \cdot t$. We may also
denote the state by
\begin{equation}
\left \vert \psi_{\vec{k}}(\vec{x},t)\right \rangle =\mathrm{\hat{U}%
}_{\mathrm{fragment}}\left[  0\right \rangle =c_{\vec{k}}^{\dagger}\left \vert
0\right \rangle .
\end{equation}
The wave-function $\left \vert \psi_{\vec{k}}(\vec{x},t)\right \rangle $ and the
function $\left[  \psi_{\vec{k}}(\vec{x},t)\right \rangle $ correspond to
anti-periodic boundary condition and periodic boundary condition for the plane
Kelvin waves. $\hat{U}_{\vec{k}}$ is the operator changing the boundary
condition. So the number of zeros is changed by $\pm1.$ We say that
$\left \vert \psi_{\vec{k}}(\vec{x},t)\right \rangle $ becomes plane waves for knots.

From the superposition principle of Kelvin waves, an arbitrary quantum state
for a knot is given by%
\begin{equation}
\left \vert \mathbf{Z}_{\mathrm{knot}}(\vec{x},t)\right \rangle =\left(
\begin{array}
[c]{c}%
\mathrm{z}_{\uparrow,\mathrm{A}}(\vec{x},t)\\
\mathrm{z}_{\uparrow,\mathrm{B}}(\vec{x},t)\\
\mathrm{z}_{\downarrow,\mathrm{A}}(\vec{x},t)\\
\mathrm{z}_{\downarrow,\mathrm{B}}(\vec{x},t)
\end{array}
\right)
\end{equation}
where
\begin{align}
\mathrm{z}_{\sigma,\tau}(\vec{x},t)  &  =r_{\sigma,\tau}(\vec{x},t)\cdot
e^{i\phi_{\sigma,\tau}(\vec{x},t)}\\
&  =%
%TCIMACRO{\dsum \limits_{\vec{k}}}%
%BeginExpansion
{\displaystyle \sum \limits_{\vec{k}}}
%EndExpansion
(r_{\sigma,\tau,\vec{k}})\cdot e^{\pm i(\vec{k}\cdot \vec{x})\pm i(\omega \cdot
t)}\nonumber
\end{align}
where $r_{\sigma,\tau,\vec{k}}$ is the amplitude of given plane wave $\vec{k}$
and $\sigma=\uparrow$/$\downarrow$ denotes spin degrees of freedom,
$\tau=\mathrm{A}/\mathrm{B}$ denotes vortex degrees of freedom. The results
illustrate superposition principle for fragmentized knot. We may denote the
equation by
\[
\left \vert \psi(\vec{x},t)\right \rangle =c^{\dagger}(\vec{x},t)\left \vert
\text{\textrm{vacuum}}\right \rangle .
\]
For simplify, in this part we discuss the properties of quantum states for
knots by fixing the spin and vortex degrees of freedom, (for example,
$\sigma \rightarrow \uparrow,$ $\tau \rightarrow \mathrm{B}$).

We point out that the function of a Kelvin wave with an extra fragmentized
knot describes the distribution of the $N$ identical knot-pieces and plays the
role of the wave-function in quantum mechanics as
\begin{equation}
\frac{1}{\sqrt{V_{P}}}\frac{\mathrm{z}(\vec{x},t)}{r_{0}}=\sqrt{\rho
_{\mathrm{knot}}(x,t)}e^{i\Delta \phi(\vec{x},t)}\Longleftrightarrow \psi
(\vec{x},t),
\end{equation}
and
\begin{equation}
\rho_{\mathrm{knot}}(\vec{x},t)\Longleftrightarrow n_{\mathrm{knot}}(\vec
{x},t)
\end{equation}
where the function of Kelvin wave with a fragmentized knot $\frac
{\mathrm{z}(\vec{x},t)}{\sqrt{V_{P}}r_{0}}$ becomes the wave-function
$\psi(\vec{x},t)$ in emergent quantum mechanics, the angle $\Delta \phi(\vec
{x},t)$ becomes the quantum phase angle of wave-function, the knot density
$\rho_{\mathrm{knot}}=\left \langle \frac{\Delta \hat{K}}{\Delta V_{P}%
}\right \rangle $ becomes the density of knot-pieces $n_{\mathrm{knot}}(\vec
{x})$. Thus, the density of knot-pieces in a given region is obtained as
$\psi^{\ast}(\vec{x},t)\psi(\vec{x},t)$.

We then prove that the wave-function $\psi(\vec{x},t)=\sqrt{\rho
_{\mathrm{knot}}(x,t)}e^{i\phi(\vec{x},t)}$ is the function of Kelvin wave of
knots (elementary particles). The knot density $\rho_{\mathrm{knot}%
}=\left \langle \frac{\Delta \hat{K}}{\Delta V_{P}}\right \rangle $ is equal to
\begin{align}
\rho_{\mathrm{knot}}  &  =\frac{1}{r_{0}^{2}\pi^{3}}\mathrm{z}%
_{\mathrm{fragment,N}}^{\ast}\text{ }\hat{K}\text{ }\mathrm{z}%
_{\mathrm{fragment,N}}\text{ }\cdot d\phi(x)d\phi(y)d\phi(z)\nonumber \\
&  =\frac{1}{r_{0}^{2}}\sum_{i_{\phi(x)},i_{\phi(y)},i_{\phi(z)}}%
\mathrm{z}_{\mathrm{fragment,N}}^{\ast}(t,\left(  x_{0}\right)  _{i_{\phi}%
})\text{ }\hat{K}\text{ }\mathrm{z}_{\mathrm{fragment,N}}(t,\left(
x_{0}\right)  _{i_{\phi}})\nonumber \\
&  =\frac{1}{r_{0}^{2}}\sum_{i_{x}}\mathrm{z}_{\mathrm{fragment,N}}^{\ast
}(t,\left(  x_{0}\right)  _{i_{x}})\hat{K}\text{ }\mathrm{z}%
_{\mathrm{fragment,N}}(t,\left(  x_{0}\right)  _{i_{x}})\nonumber \\
&  \cdot \sum_{i_{y}}\mathrm{z}_{\mathrm{fragment,N}}^{\ast}(t,\left(
x_{0}\right)  _{i_{y}})\hat{K}\text{ }\mathrm{z}_{\mathrm{fragment,N}%
}(t,\left(  x_{0}\right)  _{i_{y}})\nonumber \\
&  \cdot \sum_{i_{z}}^{\ast}\mathrm{z}_{\mathrm{fragment,N}}^{\ast}(t,\left(
x_{0}\right)  _{i_{z}})\hat{K}\text{ }\mathrm{z}_{\mathrm{fragment,N}%
}(t,\left(  x_{0}\right)  _{i_{z}})\\
&  =\frac{1}{\Delta V_{P}\cdot r_{0}^{2}}\mathrm{z}_{\mathrm{fragment,N}%
}^{\ast}(\vec{x},t)\text{ }\hat{K}\text{ }\mathrm{z}_{\mathrm{fragment,N}%
}(\vec{x},t)dV_{P}\nonumber \\
&  =\frac{1}{\Delta V_{P}}\psi^{\ast}(\vec{x},t)(-i\frac{d}{d\phi})\psi
(\vec{x},t)dV_{P}\nonumber \\
&  =\psi^{\ast}(\vec{x},t)\psi(\vec{x},t).\nonumber
\end{align}
For this case, we have set the bare density of finding a knot to be zero
$\rho_{\mathrm{knot}}(\vec{x},t)\equiv0$. As a result, $\psi(\vec{x},t)$
indeed plays the role of wave-function in quantum mechanics, $\rho
_{\mathrm{knot}}(x,t)$ is the knot density and $\phi(\vec{x},t)$ is the phase
(the angle in extra space). For a single knot, we have (volume) normalization
condition $\int \rho_{\mathrm{knot}}(\vec{x},t)dV_{P}=1$. In the following
part, we will show the probability interpretation for wave-functions according
to the dynamic projection with fast-clock effect.

Thus, the quantum state is invariant under a global gauge transformation,
i.e.,
\begin{align}
\psi(\vec{x},t)  &  \rightarrow \psi^{\prime}(\vec{x},t)=\psi(\vec
{x},t)e^{i\phi_{0}}\\
&  =\sqrt{\rho_{\mathrm{knot}}(\vec{x},t)}e^{i\phi(\vec{x},t)}e^{i\phi_{0}%
}\nonumber
\end{align}
where $\phi_{0}$ is constant. Such invariant comes from a global rotation
symmetry in $(\xi,\eta)$ space.

\subsubsection{Momentum operator and energy operator for knots}

For a plane wave, $\psi(\vec{x},t)=\frac{1}{\sqrt{V_{P}}}e^{-i\omega \cdot
t+i\vec{k}\cdot \vec{x}}$ (for simplify, we have fixed the spin and vortex
degrees of freedom, $\sigma \rightarrow \uparrow,$ $\tau \rightarrow \mathrm{B}$),
the projected (Lamb impulse) energy of a knot is
\begin{equation}
E_{\mathrm{knot}}=\hbar_{\mathrm{knot}}\omega
\end{equation}
and the projected (Lamb impulse) momentum of a knot is
\begin{equation}
\vec{p}_{\mathrm{knot}}=\hbar_{\mathrm{knot}}\vec{k}%
\end{equation}
where the effective Planck constant $\hbar_{\mathrm{knot}}$ is obtained as
projected (Lamb impulse) angular momentum of a knot (the elementary
volume-changing of two entangled vortex-membranes)
\[
\hbar_{\mathrm{knot}}=J_{\mathrm{knot}}=\frac{1}{2}\rho_{0}\kappa
V_{\mathrm{knot}}.
\]

As a result, we have
\begin{equation}
\psi(\vec{x},t)=\frac{1}{\sqrt{V_{P}}}\exp(\frac{-iE_{\mathrm{knot}}t+i\vec
{p}_{\mathrm{knot}}\cdot \vec{x}}{\hbar_{\mathrm{knot}}}).
\end{equation}
From the fact of superposition principle of Kelvin waves, a generalized
wave-function can be
\begin{align}
\psi(\vec{x},t)  &  =\sqrt{\rho_{\mathrm{knot}}(x,t)}e^{i\phi(\vec{x},t)}\\
&  =%
%TCIMACRO{\dsum \nolimits_{\vec{p}}}%
%BeginExpansion
{\displaystyle \sum \nolimits_{\vec{p}}}
%EndExpansion
c_{\vec{p}}\exp(\frac{-iE_{\mathrm{knot}}t+i\vec{p}_{\mathrm{knot}}\cdot
\vec{x}}{\hbar_{\mathrm{knot}}}).\nonumber
\end{align}

Next, we calculate the expect values of the projected (Lamb impulse) energy
$\left \langle E_{\mathrm{knot}}\right \rangle $ and the projected (Lamb
impulse) momentum $\left \langle \vec{p}_{\mathrm{knot}}\right \rangle ,$ respectively.

From the wave-function $\psi(\vec{x},t)=%
%TCIMACRO{\dsum \nolimits_{p}}%
%BeginExpansion
{\displaystyle \sum \nolimits_{p}}
%EndExpansion
c_{p}\exp(\frac{-i(E_{\mathrm{knot}}t-\vec{p}_{\mathrm{knot}}\cdot \vec{x}%
)}{\hbar_{\mathrm{knot}}})$, we have
\begin{align}
\left \langle E\right \rangle  &  =\int E_{\mathrm{knot}}\rho_{\mathrm{knot}%
}(x)dV_{P}=\int \psi^{\ast}(x,t)E_{\mathrm{knot}}\psi(x,t)dV_{P}\\
&  =\int[%
%TCIMACRO{\dsum \nolimits_{p}}%
%BeginExpansion
{\displaystyle \sum \nolimits_{p}}
%EndExpansion
c_{p}^{\ast}\exp(\frac{i(E_{\mathrm{knot}}t-\vec{p}_{\mathrm{knot}}\cdot
\vec{x})}{\hbar_{\mathrm{knot}}})]\nonumber \\
&  (i\hbar_{\mathrm{knot}}\frac{d}{dt})[%
%TCIMACRO{\dsum \nolimits_{p^{\prime}}}%
%BeginExpansion
{\displaystyle \sum \nolimits_{p^{\prime}}}
%EndExpansion
c_{p^{\prime}}\exp(\frac{-i(E_{\mathrm{knot}}t-\vec{p}_{\mathrm{knot}}%
^{\prime}\cdot \vec{x})}{\hbar_{\mathrm{knot}}})]dV_{P}\nonumber \\
&  =\int \psi^{\ast}(\vec{x},t)(i\hbar_{\mathrm{knot}}\frac{d}{dt})\psi(\vec
{x},t)dV_{P}.\nonumber
\end{align}
This result indicates that the projected (Lamb impulse) energy becomes
operator%
\begin{equation}
E_{\mathrm{knot}}\rightarrow \hat{E}_{\mathrm{knot}}=i\hbar_{\mathrm{knot}%
}\frac{d}{dt}.
\end{equation}

Using the similar approach, we derive
\begin{align}
\left \langle \vec{p}_{\mathrm{knot}}\right \rangle  &  =\int \vec{p}%
_{\mathrm{knot}}\rho_{\mathrm{knot}}(\vec{x},t)dV_{P}=\int \psi^{\ast}%
(x,t)\vec{p}_{\mathrm{knot}}\psi(\vec{x},t)dV_{P}\\
&  =\int \psi^{\ast}(\vec{x},t)(-i\hbar_{\mathrm{knot}}\frac{d}{d\vec{x}}%
)\psi(\vec{x},t)dV_{P}.\nonumber
\end{align}
As a result, the projected (Lamb impulse) momentum also becomes operators%
\begin{equation}
\vec{p}_{\mathrm{knot}}\rightarrow \hat{p}_{\mathrm{knot}}=-i\hbar
_{\mathrm{knot}}\frac{d}{d\vec{x}}.
\end{equation}

\subsubsection{The Schr\"{o}dinger equation}

For deriving the Hamiltonian of knots, we treat the Hamiltonian of
perturbative Kelvin waves around the knot-crystal. For knot-crystal of two
entangled vortex-membranes, the Biot-Savart equation $i\frac{d\mathrm{z}%
_{\mathrm{A/B}}}{dt}=\frac{\delta \mathrm{\hat{H}}(\mathrm{z}_{\mathrm{A/B}}%
)}{\delta \mathrm{z}_{\mathrm{A/B}}^{\ast}}$ turns into the Schr\"{o}dinger
equation for constraint Kelvin waves
\begin{equation}
i\hbar_{\mathrm{eff}}\frac{d\mathbf{Z}_{\mathrm{knot-crystal}}(\vec{x},t)}%
{dt}=\mathrm{\hat{H}}\cdot \mathbf{Z}_{\mathrm{knot-crystal}}(\vec{x},t)
\end{equation}
where $\mathbf{Z}_{\mathrm{knot-crystal}}(\vec{x},t)$ denote perturbative
Kelvin waves around the knot-crystal (a special plane Kelvin wave) and the
Hamiltonian density is given by
\begin{equation}
\mathrm{\hat{H}}_{\mathrm{knot-crystal}}=%
%TCIMACRO{\dsum \limits_{I}}%
%BeginExpansion
{\displaystyle \sum \limits_{I}}
%EndExpansion
[\frac{(\hat{p}_{\mathrm{Lamb}}^{I})^{2}}{2m_{\mathrm{pseudo}}}]+\frac
{\hbar_{\mathrm{eff}}\omega^{\ast}}{2}(\vec{1}\otimes(\tau_{x}-\vec{1}))
\end{equation}
where $m_{\mathrm{pseudo}}=\frac{\hbar_{\mathrm{eff}}}{\alpha \kappa \ln
\epsilon}$ and $\hat{p}_{\mathrm{lamb}}^{I}=-i\hbar_{\mathrm{eff}}\frac
{d}{dx^{I}}$.

For a knot, by scaling the Planck constant, we derive the Schr\"{o}dinger
equation for knots as%
\begin{equation}
i\hbar_{\mathrm{knot}}\frac{d\psi(\vec{x},t)}{dt}=\mathrm{\hat{H}%
}_{\mathrm{knot}}\psi(\vec{x},t)
\end{equation}
where $\psi(\vec{x},t)=\left(
\begin{array}
[c]{c}%
\psi_{\uparrow,\mathrm{A}}(\vec{x},t)\\
\psi_{\uparrow,\mathrm{B}}(\vec{x},t)\\
\psi_{\downarrow,\mathrm{A}}(\vec{x},t)\\
\psi_{\downarrow,\mathrm{B}}(\vec{x},t)
\end{array}
\right)  $ denotes quantum states of a knot with four degrees of freedom and
$\mathrm{\hat{H}}_{\mathrm{knot}}$ is the Hamiltonian of knots. On flat
vortex-membranes, the Hamiltonian of SOC knots is obtained by scaling that of
plane Kelvin wave with same tensor state as
\begin{equation}
\mathrm{\hat{H}}_{\mathrm{knot}}=%
%TCIMACRO{\dsum \limits_{I}}%
%BeginExpansion
{\displaystyle \sum \limits_{I}}
%EndExpansion
[\frac{(\hat{p}_{\mathrm{knot}}^{I})^{2}}{2m_{\mathrm{pseudo}}}]+\frac
{\hbar_{\mathrm{knot}}\omega^{\ast}}{2}(\vec{1}\otimes(\tau_{x}-\vec{1}))
\end{equation}
where $m_{\mathrm{pseudo}}=\frac{\hbar_{\mathrm{knot}}}{\alpha \kappa
\ln \epsilon}$ and $\hat{p}_{\mathrm{knot}}^{I}=-i\hbar_{\mathrm{knot}}\frac
{d}{dx^{I}}$. In next section, we will derive the knot Hamiltonian
$\mathrm{\hat{H}}_{\mathrm{knot}}$ on the SOC knot-crystal that is different
from above Hamiltonian.

For the eigenstate with eigenvalue $E_{\mathrm{knot}},$%
\begin{equation}
\mathrm{\hat{H}}_{\mathrm{knot}}\psi(\vec{x},t)=E_{\mathrm{knot}}\psi(\vec
{x},t)
\end{equation}
the wave-function becomes
\begin{equation}
\psi(\vec{x},t)=f(\vec{x})\exp(\frac{iE_{\mathrm{knot}}t}{\hbar})
\end{equation}
where $f(\vec{x})$ is spatial function. This state corresponds to a
vortex-membrane with fixed angular velocity, $\omega=\frac{E_{\mathrm{knot}}%
}{\hbar_{\mathrm{knot}}}$. As a result, the quantized condition for knot is
due to the conservation of angular momentum in extra space (the volume of the
knot in 5D space) for information-unit, $\hbar_{\mathrm{knot}}$.

In particular, we have the topological representation on action $S$ for knots,
i.e.,
\begin{align}
S  &  =%
%TCIMACRO{\dint }%
%BeginExpansion
{\displaystyle \int}
%EndExpansion
p_{\mathrm{knot}}\cdot dx=\hbar_{\mathrm{knot}}%
%TCIMACRO{\dint }%
%BeginExpansion
{\displaystyle \int}
%EndExpansion
k\cdot dx\\
&  =\hbar_{\mathrm{knot}}%
%TCIMACRO{\dint }%
%BeginExpansion
{\displaystyle \int}
%EndExpansion
2\pi \rho_{\mathrm{wind}}\cdot dx=2\pi \hbar_{\mathrm{knot}}\cdot w_{1D}%
\nonumber \\
&  =h_{\mathrm{knot}}\cdot w_{1D}.\nonumber
\end{align}
From the point view of topology, we explain the Sommerfeld quantization
condition for knot-pieces
\begin{equation}%
%TCIMACRO{\doint }%
%BeginExpansion
{\displaystyle \oint}
%EndExpansion
p_{\mathrm{knot}}\cdot dx=nh_{\mathrm{knot}}%
\end{equation}
where $h_{\mathrm{knot}}=2\pi \hbar_{\mathrm{knot}}$ and $n$ is an integer
number. The Sommerfeld quantization condition is just the topological
condition of the winding number for a process of "static" knot-pieces, i.e.,%
\[%
%TCIMACRO{\doint }%
%BeginExpansion
{\displaystyle \oint}
%EndExpansion
p_{\mathrm{knot}}\cdot dx=h_{\mathrm{knot}}\cdot w_{1D}%
\]
where $w_{1D}$ is the winding number of the process that can be regarded as
the winding number for corresponding Kelvin waves.

\subsection{Emergent quantum field theory}

In this part, we regard knot-crystal as a multi-knot system that is described
by quantum field theory and a knot (the elementary volume-changing of two
entangled vortex-membranes) becomes elementary fermionic particle.

\subsubsection{Fermionic statistics}

To derive the formulation of quantum field theory for knot-crystal, we
consider the knot-crystal as a many-knot system. We may introduce
$N$-wave-function to describe the motions of $N$-knots, $\Psi(\vec{x}_{1}%
,\vec{x}_{2},...,\vec{x}_{N})$. For this many-body system, an important
feature is the statistics. In quantum mechanics, an assumption is
spin-statistics of fermions. An electron is a fermion with half spin. The
wave-function of a system of identical spin-1/2 particles changes sign when
two particles exchange. Particles with wave-functions antisymmetric under
exchange are called fermions.

To distinguish the statistics for the knots, we consider two knots. We show
that the wave-functions antisymmetric by exchanging knots is due to $\pi
$-phase changing nature. When we exchange two 1D knots, the angle in extra
space (that is just the phase of wave-function) is $\pi$. According to the
definition of knots, we have two static unified knots and get knot-functions
\begin{equation}
\Psi(\vec{x},\vec{x}^{\prime})=\mathrm{\hat{U}}_{\mathrm{knot}}(\vec
{x}^{\prime})\cdot \mathrm{\hat{U}}_{\mathrm{knot}}(\vec{x})\mathbf{Z}%
_{\mathrm{knot-crystal}}(\vec{x},t).
\end{equation}
Due to
\begin{equation}
\mathrm{\hat{U}}_{\mathrm{knot}}^{\ast}(\vec{x})\cdot \mathrm{\hat{U}%
}_{\mathrm{knot}}(\vec{x}^{\prime})\cdot \mathrm{\hat{U}}_{\mathrm{knot}}%
(\vec{x})=-\mathrm{\hat{U}}_{\mathrm{knot}}(\vec{x}^{\prime}),
\end{equation}
after exchanging two knots, we get
\begin{align}
\Psi(\vec{x}^{\prime},\vec{x})  &  =[\mathrm{\hat{U}}_{\mathrm{knot}}(\vec
{x}^{\prime})\cdot \mathrm{\hat{U}}_{\mathrm{knot}}(\vec{x})]\mathbf{Z}%
_{\mathrm{knot-crystal}}(\vec{x},t)\\
&  \rightarrow \Psi(\vec{x},\vec{x}^{\prime})=[\mathrm{\hat{U}}_{\mathrm{knot}%
}(\vec{x})\cdot \mathrm{\hat{U}}_{\mathrm{knot}}(\vec{x}^{\prime}%
)]\mathbf{Z}_{\mathrm{knot-crystal}}(\vec{x},t)\nonumber \\
&  =[\mathrm{\hat{U}}_{\mathrm{knot}}(\vec{x}^{\prime})\cdot \mathrm{\hat{U}%
}_{\mathrm{knot}}(\vec{x})]\mathbf{Z}_{\mathrm{knot-crystal}}^{\prime}(\vec
{x},t)\nonumber \\
&  =-\Psi(\vec{x},\vec{x}^{\prime})\nonumber
\end{align}
where
\begin{equation}
\mathbf{Z}_{\mathrm{knot-crystal}}^{\prime}(\vec{x},t)=\mathbf{Z}%
_{\mathrm{knot-crystal}}(\vec{x},t)e^{i\pi}.
\end{equation}
The results are independent of the dimensions. As a result, the knot obeys
fermionic statistics in different dimensions.

We then introduce second quantization representation for knots by defining
fermionic operator $c_{1}^{\dagger}(\vec{x})$, $c_{2}^{\dagger}(\vec{x})$ as
\begin{equation}
\mathrm{\hat{U}}_{\mathrm{knot}}(\vec{x})\Longrightarrow c_{1}^{\dagger}%
(\vec{x}),\text{ }\mathrm{\hat{U}}_{\mathrm{knot}}(\vec{x}^{\prime
})\Longrightarrow c_{2}^{\dagger}(\vec{x}).
\end{equation}
The knot-crystal without extra knots corresponds to the vacuum state
$\left \vert \text{\textrm{vacuum}}\right \rangle $
\begin{equation}
\mathbf{Z}_{\mathrm{knot-crystal}}(\vec{x},t)\Longrightarrow \left \vert
\text{\textrm{vacuum}}\right \rangle .
\end{equation}
Then after exchanging two identical particles, the many-body wave-function
\begin{equation}
\left \vert \Psi_{\mathrm{initial}}(\vec{x}^{\prime},\vec{x})\right \rangle
=c^{\dagger}(\vec{x})c^{\dagger}(\vec{x}^{\prime})\left \vert
\text{\textrm{vacuum}}\right \rangle
\end{equation}
changes by a phase to be
\begin{align}
\left \vert \Psi_{\mathrm{initial}}(\vec{x}^{\prime},\vec{x})\right \rangle  &
\rightarrow \Psi_{\mathrm{final}}(\vec{x}^{\prime},\vec{x})\nonumber \\
&  =e^{i\pi}c^{\dagger}(\vec{x}^{\prime})c^{\dagger}(\vec{x})\left \vert
\text{\textrm{vacuum}}\right \rangle .
\end{align}

After considering the four internal degrees of freedom, we have
\begin{equation}
c(\vec{x},t)=\left(
\begin{array}
[c]{c}%
c_{\uparrow,\mathrm{A}}(\vec{x},t)\\
c_{\uparrow,\mathrm{B}}(\vec{x},t)\\
c_{\downarrow,\mathrm{A}}(\vec{x},t)\\
c_{\downarrow,\mathrm{B}}(\vec{x},t)
\end{array}
\right)  .
\end{equation}
We identify $c_{\uparrow,\mathrm{A/B}}^{\dagger}(\vec{x},t)$ ($c_{\uparrow
,\mathrm{A/B}}(\vec{x},t)$) the creation (or annihilation) operator of a
spin-$\uparrow$ knot on vortex-membrane-\textrm{A/B}; $c_{\downarrow
,\mathrm{A/B}}^{\dagger}(\vec{x},t)$ ($c_{\downarrow,\mathrm{A/B}}(\vec{x}%
,t)$) the creation (or annihilation) operator of a spin-$\downarrow$ knot on
vortex-membrane-\textrm{A/B}. According to the fermionic statistics, there
exists anti-commutation relation
\begin{equation}
\{c(\vec{x},t),\text{ }c(\vec{x}^{\prime},t)\}=0.
\end{equation}
In addition, we can introduce the field representation via Grassmann number
\begin{equation}
c(\vec{x},t)\Longrightarrow \psi(\vec{x},t)=\left(
\begin{array}
[c]{c}%
\psi_{\uparrow,\mathrm{A}}(\vec{x},t)\\
\psi_{\uparrow,\mathrm{B}}(\vec{x},t)\\
\psi_{\downarrow,\mathrm{A}}(\vec{x},t)\\
\psi_{\downarrow,\mathrm{B}}(\vec{x},t)
\end{array}
\right)  .
\end{equation}

\subsubsection{Path integral formulation for quantum mechanics}

Path integral formulation for quantum mechanics is another formulation
describing the dynamic evolution of the distribution of knot-pieces. By
considering information scaling the effective Planck constant of knot-crystal
to that of knots, the probability amplitude $K_{\mathrm{knot}}(\vec{x}%
^{\prime},t_{f};\vec{x},t_{i})$ for a knot (a knot-piece) from an initial
position $\vec{x}$ at time $t=t_{i}$ (that is described by a state $\left \vert
t_{i},\vec{x}\right \rangle $) to position $\vec{x}^{\prime}$ at a later time
$t=t_{f}$ ($\left \vert t_{f},\vec{x}^{\prime}\right \rangle $) is obtained as,
\begin{align}
K_{\mathrm{knot}}(\vec{x}^{\prime},t_{f};\vec{x},t_{i})  &  =\langle
t_{f},\vec{x}^{\prime}\left \vert t_{i},\vec{x}\right \rangle =%
%TCIMACRO{\dsum \limits_{n}}%
%BeginExpansion
{\displaystyle \sum \limits_{n}}
%EndExpansion
e^{iS_{n}/\hbar_{\mathrm{knot}}}\nonumber \\
&  =\int \mathcal{D}\vec{p}_{\mathrm{knot}}(t)\mathcal{D}\vec{x}(t)e^{iS/\hbar
_{\mathrm{knot}}}%
\end{align}
where
\begin{equation}
S=%
%TCIMACRO{\dint }%
%BeginExpansion
{\displaystyle \int}
%EndExpansion
[\vec{p}_{\mathrm{knot}}\frac{d\vec{x}}{dt}-\mathrm{\hat{H}}_{\mathrm{knot}%
}(\vec{p}_{\mathrm{knot}},\vec{x})]dt.
\end{equation}
Each knot-piece's path contributes $e^{iS_{n}/\hbar_{\mathrm{knot}}}$ where
$S_{n}$ is the $n$-th classical action i.e.,%
\begin{equation}%
%TCIMACRO{\dsum \limits_{n}}%
%BeginExpansion
{\displaystyle \sum \limits_{n}}
%EndExpansion
e^{iS_{n}/\hbar_{\mathrm{knot}}}.
\end{equation}

With the help of particle operators, we define the many-body Hamiltonian of
the system with multi-knot as
\begin{equation}
\mathcal{\hat{H}}=%
%TCIMACRO{\dint }%
%BeginExpansion
{\displaystyle \int}
%EndExpansion
d^{3}x[\psi^{\dagger}(\vec{x})\mathrm{\hat{H}}_{\mathrm{knot}}\psi(\vec{x})]
\end{equation}
where $\mathrm{\hat{H}}_{\mathrm{knot}}$ is the Hamiltonian operator of single
knot and $\psi(\vec{x})=\frac{1}{\sqrt{V}}\sum_{p}c_{p}\exp(i\vec{p}\cdot
\vec{x}/\hbar_{\mathrm{knot}})$ ($c_{p}=\left(
\begin{array}
[c]{c}%
c_{\uparrow,p}\\
c_{\downarrow,p}%
\end{array}
\right)  $). As a result, we have
\begin{equation}
\mathcal{\hat{H}}=\sum_{k}\mathrm{\hat{H}}_{\mathrm{knot}}(\vec{p}%
)c_{p}^{\dagger}c_{p}=\sum_{k}\mathrm{\hat{H}}_{\mathrm{knot}}(\vec{p})\hat
{n}_{p}%
\end{equation}
where $\hat{n}_{p}=c_{p}^{\dagger}c_{p}$. The definition of total Hamiltonian
$\mathcal{\hat{H}}$ means that for a given Kelvin wave there are $n_{p}$ knots.

Now, we consider the path integral formulation of multi-knot. The probability
amplitude
\begin{align}
&  K(\vec{x}_{M}^{\prime},...,\vec{x}_{2}^{\prime},\vec{x}_{1}^{\prime}%
,t_{f};\vec{x}_{M},...,\vec{x}_{2},\vec{x}_{1},t_{i})\\
&  =\langle t_{f},\vec{x}_{M}^{\prime},...,\vec{x}_{2}^{\prime},\vec{x}%
_{1}^{\prime}\left \vert t_{i},\vec{x}_{M},...,\vec{x}_{2},\vec{x}%
_{1}\right \rangle \nonumber
\end{align}
becomes a multi-variable function where $x_{j}^{\prime}$ and $x_{j}$ denote
the final position and initial position of $j$-th knot, respectively. For a
multi-knot system, quantum processes are described by
\begin{align}
&  K(\vec{x}_{M}^{\prime},...,\vec{x}_{2}^{\prime},\vec{x}_{1}^{\prime}%
,t_{f};\vec{x}_{M},...,\vec{x}_{2},\vec{x}_{1},t_{i})\nonumber \\
&  =\langle t_{f},\vec{x}_{M}^{\prime},...,\vec{x}_{2}^{\prime},\vec{x}%
_{1}^{\prime}\left \vert t_{i},\vec{x}_{M},...,\vec{x}_{2},\vec{x}%
_{1}\right \rangle \nonumber \\
&  =%
%TCIMACRO{\dprod \limits_{j}}%
%BeginExpansion
{\displaystyle \prod \limits_{j}}
%EndExpansion%
%TCIMACRO{\dsum \limits_{n}}%
%BeginExpansion
{\displaystyle \sum \limits_{n}}
%EndExpansion
e^{i\Delta \phi_{j,n}}=%
%TCIMACRO{\dprod \limits_{j}}%
%BeginExpansion
{\displaystyle \prod \limits_{j}}
%EndExpansion%
%TCIMACRO{\dsum \limits_{n}}%
%BeginExpansion
{\displaystyle \sum \limits_{n}}
%EndExpansion
e^{iS_{j,n}/\hbar_{\mathrm{knot}}}=%
%TCIMACRO{\dsum \limits_{n}}%
%BeginExpansion
{\displaystyle \sum \limits_{n}}
%EndExpansion
e^{i%
%TCIMACRO{\dsum \limits_{j}}%
%BeginExpansion
{\displaystyle \sum \limits_{j}}
%EndExpansion
S_{j,n}/\hbar_{\mathrm{knot}}}\nonumber \\
&  =%
%TCIMACRO{\dprod \limits_{p}}%
%BeginExpansion
{\displaystyle \prod \limits_{p}}
%EndExpansion
\psi_{p}^{\dagger}(\vec{x},t)\psi_{p}(\vec{x},t)e^{iS_{p}/\hbar_{\mathrm{knot}%
}}=\int \mathcal{D}\psi^{\dagger}(\vec{x},t)\mathcal{D}\psi(\vec{x}%
,t)e^{i\mathcal{S}/\hbar_{\mathrm{knot}}}%
\end{align}
where
\begin{equation}
\mathcal{S}=%
%TCIMACRO{\dsum \limits_{\omega,\vec{p}}}%
%BeginExpansion
{\displaystyle \sum \limits_{\omega,\vec{p}}}
%EndExpansion
S_{\omega,\vec{p}}=\int \mathcal{L}dtd^{3}x
\end{equation}
with
\begin{equation}
S_{\omega,\vec{p}}=\psi_{\vec{p}}^{\dagger}(i\hbar \omega(\vec{p}%
)-\mathrm{\hat{H}}_{\mathrm{knot}}(\vec{p}))\psi_{\vec{p}}%
\end{equation}
and
\begin{equation}
\mathcal{L}=i\psi^{\dagger}\partial_{t}\psi-\mathcal{\hat{H}}.
\end{equation}
The symbol $%
%TCIMACRO{\dsum \limits_{n}}%
%BeginExpansion
{\displaystyle \sum \limits_{n}}
%EndExpansion
$ denotes the summation of different pathes and the symbol $%
%TCIMACRO{\dprod \limits_{j}}%
%BeginExpansion
{\displaystyle \prod \limits_{j}}
%EndExpansion
$ denotes the product of different volume changing with different zeros.

\subsubsection{Discrete spatial translation symmetry for knots on
knot-crystal}

Symmetry and symmetry breaking play profound roles in particle physics for
quantum field theory and quantum many-body physics for solid physics.

In solid physics, a crystal is the system of periodically arranged atoms with
spontaneous breaking of continuous spatial translation symmetry to discrete
spatial translation symmetry. On a crystal, due to spontaneous breaking of
continuous spatial translation symmetry to discrete spatial translation
symmetry, the eigenstates are quantum states $\left \vert \vec{k}\right \rangle
$. The Bloch vector $\left \vert \vec{k}\right \rangle $ is defined as a linear
superposition of the localized states $|\vec{R},\vec{a}_{j}\rangle$ in the
unit cell $\vec{R}$ at position $\vec{a}_{j}$,
\begin{equation}
\left \vert \vec{k}\right \rangle =\sum_{\vec{R},j}c_{j}(\vec{k})e^{i\vec
{k}\cdot \vec{x}}|\vec{R},\vec{a}_{j}\rangle,
\end{equation}
that depends on Bloch momentum $\vec{k}$ and type of orbital $j$ in the unit
cell. One cannot do a translation operation $\mathcal{T}(\Delta \vec{x})$ with
$\Delta \vec{x}=\alpha \vec{R}$ where $\alpha$ is a non-integer real number. The
Bloch theorem is very useful in description of a quasi-particle in crystals
that states that the basis vector $\left \vert \vec{k}\right \rangle $
translated for the Bravais vector $\vec{R}$ changes as
\begin{equation}
\mathcal{T}(\vec{R})\left \vert \vec{k}\right \rangle =e^{i\vec{k}\cdot \vec{R}%
}\left \vert \vec{k}\right \rangle
\end{equation}
where $T(\vec{R})$ is translation operation. Due to the existence of Brillouin
zone (BZ), we have $\left \vert \vec{k}\right \rangle =\left \vert \vec{k}%
+\vec{Q}\right \rangle $ where $\vec{Q}$ is reciprocal lattice vector.

For knot-crystal, the eigenstates of an arbitrary quantum state in
pseudo-quantum mechanics have generalized spatial translation symmetry
$\left[  \vec{k}\right \rangle $ where $\vec{k}$ is wave-vector. By defining a
translation operation $T(\Delta \vec{x}),$ we have
\begin{equation}
\mathcal{T}(\Delta \vec{x})\left[  \vec{k}\right \rangle =e^{i\vec{k}\cdot
\Delta \vec{x}}\left[  \vec{k}\right \rangle
\end{equation}
where $\Delta \vec{x}$ denotes the distance from generalized spatial
translation operation.

However, for emergent quantum mechanics, the situation changes. If we consider
the vortex-piece with a zero as the elementary physics object, there exists a
generalization of the Bloch theorem that incorporates generalized
translational symmetries for quantum states of knots on a knot-crystal in
emergent quantum mechanics.

For a plane wave state $\left \vert \vec{k}_{\mathrm{L}/\mathrm{R}%
}\right \rangle $\ of zeros on $\mathrm{L}/\mathrm{R}$-sublattice along the
direction $\vec{k}$, we do a translation operation $\mathcal{T}(\Delta \vec
{x})$ (that corresponds to $\mathcal{T}(\Delta \vec{x})$\ on knot-crystal) and
get
\begin{equation}
\mathcal{T}(\Delta \vec{x})\left \vert \vec{k}_{\mathrm{L}/\mathrm{R}%
}\right \rangle =\left \vert \vec{k}_{\mathrm{L}/\mathrm{R}}\right \rangle
\end{equation}
where $(\Delta \vec{x})=2a.$

For the case of $(\Delta \vec{x})=a$, we have
\begin{equation}
\hat{T}(\Delta \vec{x})\left \vert \vec{k}_{\mathrm{L}/\mathrm{R}}\right \rangle
=\left \vert \vec{k}_{\mathrm{R}/\mathrm{L}}\right \rangle
\end{equation}
that is reduced to the results in solid state physics. And we have a
generalized BZ along a given direction $\vec{e}$ as%
\begin{equation}
\left \vert \vec{k}_{\mathrm{L}/\mathrm{R}}\right \rangle =\left \vert \vec
{k}_{\mathrm{R}/\mathrm{L}}+\vec{k}_{0}\right \rangle
\end{equation}
where $\vec{k}_{0}=\frac{\pi}{a}\vec{e}$ is reciprocal lattice vector along
the given direction $\vec{e}$. The generalized Bloch theorem constrains the
formulation of the Hamiltonian which becomes manifestly invariant under
additional spatial rotating symmetry.

For the case of $(\Delta \vec{x})\operatorname{mod}a\neq0$, we can recover the
continuous spatial translation symmetry by changing projection angle
\begin{align}
\theta &  \rightarrow \theta^{\prime}=\theta \mp k_{0}\cdot \lbrack a[(\Delta
\vec{x})\operatorname{mod}a]]\nonumber \\
&  =\theta \mp \pi \lbrack(\Delta \vec{x})\operatorname{mod}a].
\end{align}

\subsubsection{Quantum field theory for knot-crystal}

In emergent quantum mechanics, a knot-crystal becomes multi-knot system
described by a tensor-network state, of which the effective theory becomes a
Dirac model in quantum field theory.

In emergent quantum mechanics, the Hamiltonian for a 3D SOC knot-crystal has
two terms -- the kinetic term and the mass term from leapfrogging motion. In
Fig.11.(a), we consider a knot-crystal to be a "two-sublattice" model with
discrete spatial translation symmetry. We then label the knots by Wannier
state $\left \vert i,\mathrm{L}\right \rangle ,$ $\left \vert i+1,\mathrm{R}%
\right \rangle ,$ $\left \vert i+2,\mathrm{L}\right \rangle $, ...According to
the definition of knot states $\left \vert \mathrm{L}\right \rangle $ and
$\left \vert \text{\textrm{R}}\right \rangle $, we use the Wannier states
$c_{\mathrm{L},i}^{\dagger}\left \vert \text{\textrm{vacuum}}\right \rangle $
and $c_{\mathrm{R},j}^{\dagger}\left \vert \text{\textrm{vacuum}}\right \rangle
$ to describe them.

\begin{figure}[ptb]
\includegraphics[clip,width=0.65\textwidth]{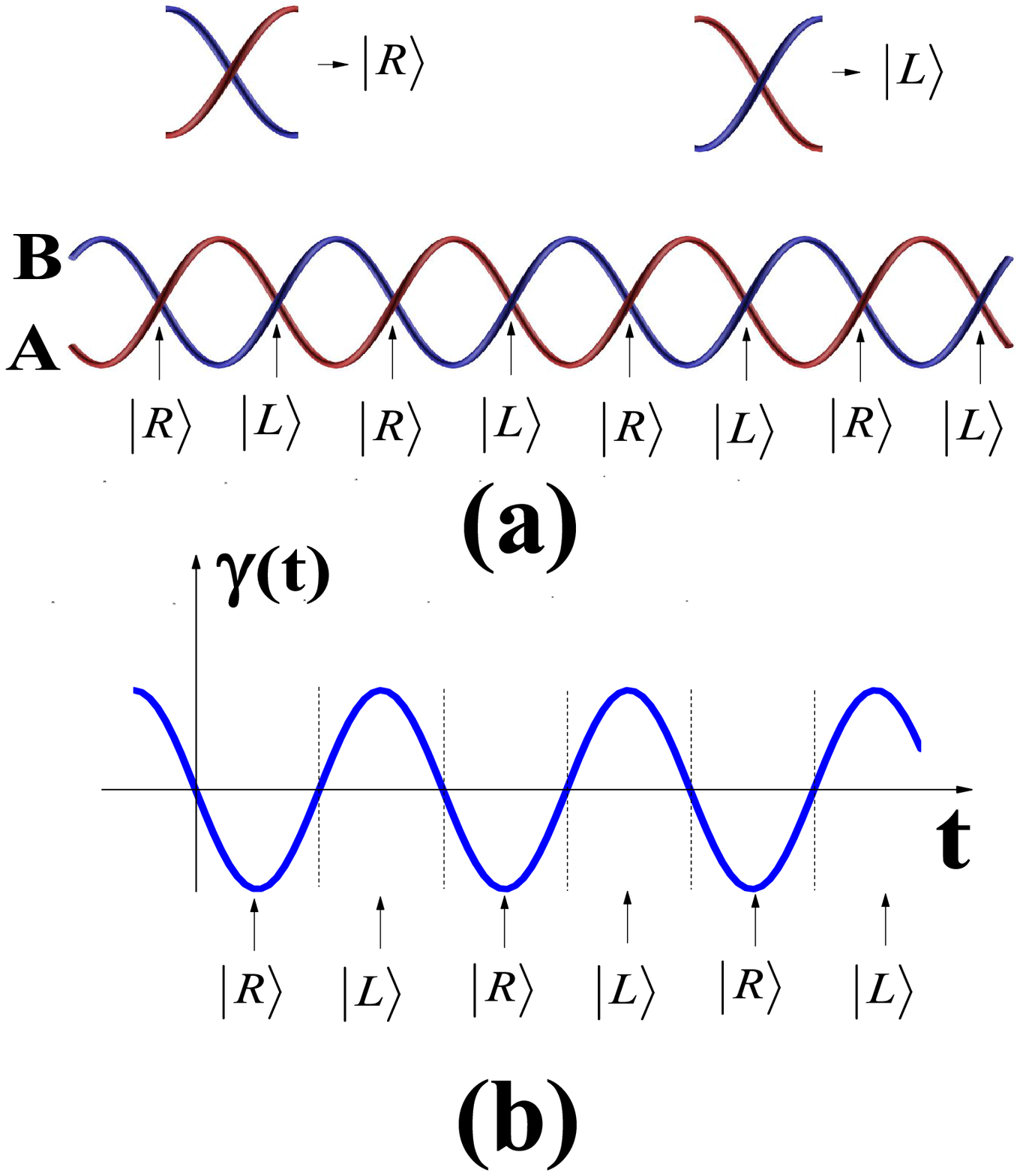}\caption{An illustration
of two "sublattices" of zero-lattice of projected knot-crystal: $\left \vert
\mathrm{L}\right \rangle $ and $\left \vert \mathrm{R}\right \rangle $ denotes
states on different sublattices; A and B denote the indices of
vortex-membranes; (b) An illustration of the periodic structure along
time-axis of knot states $\left \vert \mathrm{L}\right \rangle $ and $\left \vert
\mathrm{R}\right \rangle $ for leapfrogging knot-crystal. $\Upsilon
(x,t)=\eta_{\mathrm{A},\theta}(x,t)-\eta_{\mathrm{B},\theta}(x,t)$ is function
to characterize the oscillating effect from leapfrogging motion. We define the
states of \textrm{sgn}$[\Upsilon]=1$ by $\left \vert \mathrm{L}\right \rangle $
and the states \textrm{sgn}$[\Upsilon]=-1$ by $\left \vert \mathrm{R}%
\right \rangle $, respectively.}%
\end{figure}

We use the tight-binding formula to characterize the Hamiltonian that couples
two nearest neighbor states $\left \vert i,\mathrm{L}\right \rangle $ and
$\left \vert i+e_{I},\mathrm{R}\right \rangle $ ($I=X,Y,Z$),%
\begin{align}
&  -J(\left \vert i,\mathrm{L}\right \rangle \left \langle i+e_{I},\mathrm{R}%
\right \vert +\left \vert i+e_{I},\mathrm{R}\right \rangle \left \langle
i,\mathrm{L}\right \vert )\\
&  =-J(\left \vert i,\mathrm{L}\right \rangle \left \langle i+e_{I}%
,\mathrm{R}\right \vert +h.c.),\nonumber
\end{align}
where ${i}$ denotes a given knot and $J$ denotes the entanglement energy of
vortex-membranes that is equal to the global rotating energy of the two
entangled vortex-membranes. Then the total kinetic term is obtained as
\begin{equation}
-J\sum_{\left \langle i,j\right \rangle }(\left \vert i,\mathrm{L}\right \rangle
\left \langle j,\mathrm{R}\right \vert )+h.c.
\end{equation}
where $\left \langle {i,j}\right \rangle $ denotes the nearest-neighbor knots.
With help of $\hat{T}(a)$ and $\iota^{\pm}$, we have
\begin{equation}
\left \vert i+e_{I},\mathrm{L}\right \rangle =\mathcal{\hat{T}}(a)\cdot \iota
^{+}\left \vert i,\mathrm{R}\right \rangle
\end{equation}
where $\hat{T}(a)$ is the translation operator $\hat{T}(a)\left \vert
i\right \rangle =\left \vert i+e_{I}\right \rangle $ that is
\begin{equation}
\mathcal{\hat{T}}(a)=e^{ia(\hat{k}\cdot \vec{\sigma})}%
\end{equation}
and $\iota^{\pm}$ is an operator denoting switching of the chiral
(sub-lattice) degrees of freedom as $\iota^{-}=(\mathrm{L}\rightarrow
\mathrm{R})=$ $\left(
\begin{array}
[c]{cc}%
0 & 0\\
1 & 0
\end{array}
\right)  $ and $\iota^{+}=(\mathrm{R}\rightarrow \mathrm{L})=\left(
\begin{array}
[c]{cc}%
0 & 1\\
0 & 0
\end{array}
\right)  $. The kinetic term can also be described by a familiar formulation
\begin{equation}
\mathcal{\hat{H}}_{\mathrm{coupling}}=-J%
%TCIMACRO{\dsum \limits_{\left\langle i,j\right\rangle }}%
%BeginExpansion
{\displaystyle \sum \limits_{\left \langle i,j\right \rangle }}
%EndExpansion
c_{i}^{\dagger}c_{j}+h.c.
\end{equation}
where $\hat{c}_{i}^{\dagger}$ is the creation operator of knots at the site
$i$.

We then use path-integral formulation to characterize the effective
Hamiltonian for a knot-crystal as
\begin{equation}
\int \mathcal{D}\psi^{\dagger}(t,\vec{x})\mathcal{D}\psi(t,\vec{x}%
)e^{i\mathcal{S}/\hbar_{\mathrm{knot}}}%
\end{equation}
where $\mathcal{S}=\int \mathcal{L}dt$ and $\mathcal{L}=i\hbar_{\mathrm{knot}}%
%TCIMACRO{\dsum \limits_{i}}%
%BeginExpansion
{\displaystyle \sum \limits_{i}}
%EndExpansion
\psi_{i}^{\dagger}\partial_{t}\psi_{i}-\mathcal{H}_{\mathrm{coupling}}$. In
emergent quantum mechanics, $\psi^{\dagger}(t,\vec{x})$ is a four-component
fermion field as
\begin{equation}
\psi^{\dagger}(t,\vec{x})=\left(
\begin{array}
[c]{cccc}%
\psi_{\uparrow L}^{\dagger}(t,\vec{x}) & \psi_{\uparrow R}^{\dagger}(t,\vec
{x}) & \psi_{\downarrow L}^{\dagger}(t,\vec{x}) & \psi_{\downarrow R}%
^{\dagger}(t,\vec{x})
\end{array}
\right)
\end{equation}
where $\uparrow,\downarrow$ label two spin degrees of freedom that denote the
two possible winding directions, $L,R$ label two chiral-degrees of freedom
that denote the two possible sub-lattices. After projection, there is no
vortex-degree of freedom and we have%
\begin{align}
\mathcal{H}_{\mathrm{coupling}}  &  =-J%
%TCIMACRO{\dsum \limits_{\left\langle i,j\right\rangle }}%
%BeginExpansion
{\displaystyle \sum \limits_{\left \langle i,j\right \rangle }}
%EndExpansion
\psi_{i}^{\dagger}\psi_{j}+h.c.\nonumber \\
&  =-J%
%TCIMACRO{\dsum \limits_{i}}%
%BeginExpansion
{\displaystyle \sum \limits_{i}}
%EndExpansion
\psi_{i}^{\dagger}(\mathcal{\hat{T}}(a)\otimes \iota^{-})\psi_{i}%
+h.c.\nonumber \\
&  =-J%
%TCIMACRO{\dsum \limits_{k}}%
%BeginExpansion
{\displaystyle \sum \limits_{k}}
%EndExpansion
\psi_{k}^{\dagger}(e^{ia(\vec{k}\cdot \vec{\sigma})}\otimes \iota^{-})\psi
_{k}+h.c.\nonumber \\
&  =-J%
%TCIMACRO{\dsum \limits_{k}}%
%BeginExpansion
{\displaystyle \sum \limits_{k}}
%EndExpansion
\psi_{k}^{\dagger}[\cos(a(\vec{k}\cdot \vec{\sigma}))\otimes \iota_{x}]\psi
_{k}\nonumber \\
&  -J%
%TCIMACRO{\dsum \limits_{k}}%
%BeginExpansion
{\displaystyle \sum \limits_{k}}
%EndExpansion
\psi_{k}^{\dagger}[\sin(a(\vec{k}\cdot \vec{\sigma}))\otimes \iota_{y}]\psi_{k}%
\end{align}
where $\vec{k}$ is the wave vector along the direction $\vec{e}$ and
$\iota_{y}=i\iota^{-}-i\iota^{+}$. In continuum limit $\vec{k}\rightarrow
\Delta \vec{k}\rightarrow0$, the effective Hamiltonian is reduced into
\begin{align}
\mathcal{H}_{\mathrm{coupling}}  &  \simeq-J%
%TCIMACRO{\dsum \limits_{k}}%
%BeginExpansion
{\displaystyle \sum \limits_{k}}
%EndExpansion
\psi_{k}^{\dagger}(\vec{1}\otimes \iota_{x})\psi_{k}\nonumber \\
&  -aJ%
%TCIMACRO{\dsum \limits_{k}}%
%BeginExpansion
{\displaystyle \sum \limits_{k}}
%EndExpansion
\psi_{k}^{\dagger}[\Delta \vec{k}\cdot \vec{\sigma}\otimes \iota_{y}]\psi_{k}.
\end{align}
Thus, the reduced BZ $\left \vert \vec{k}_{\mathrm{L}/\mathrm{R}}\right \rangle
$ is given $[-\frac{\pi}{2},\frac{\pi}{2})\otimes \lbrack-\frac{\pi}{2}%
,\frac{\pi}{2})\otimes \lbrack-\frac{\pi}{2},\frac{\pi}{2})$. There exists a
node at $\left(  0,0,0\right)  $ and the excitation modes have chiral
properties. This dispersion is physical consequence of "two-sublattice".

Next, we consider the term from global rotating motion and leapfrogging
motion, of which the angular frequency $\omega^{\ast}$. For leapfrogging
motion obtained by\cite{1}, the function of the two entangled vortex-membranes
is simplified by
\begin{equation}
\left(
\begin{array}
[c]{c}%
\mathrm{z}_{\mathrm{A}}(\vec{x}=0,t)\\
\mathrm{z}_{\mathrm{B}}(\vec{x}=0,t)
\end{array}
\right)  =\frac{r_{0}}{2}\left(
\begin{array}
[c]{c}%
1+e^{i\omega^{\ast}t}\\
1-e^{i\omega^{\ast}t}%
\end{array}
\right)  e^{-i\omega_{0}t}.
\end{equation}
At $t=0$, we have knot state $\left(
\begin{array}
[c]{c}%
1\\
0
\end{array}
\right)  $; At $t=\frac{\pi}{\omega^{\ast}}$, we have knot state $\left(
\begin{array}
[c]{c}%
0\\
1
\end{array}
\right)  .$ Thus, the leapfrogging knot-crystal leads to periodic varied knot
states, i.e. at $t=0$ we have $\mathrm{sgn}[\Upsilon(t)]>0$ that is denoted by
$\left \vert \mathrm{L}\right \rangle ;$ at $t=\frac{\pi}{\omega^{\ast}}$ we
have $\mathrm{sgn}[\Upsilon(t)]<0$ that is denoted by $\left \vert
\mathrm{R}\right \rangle $. From the global rotating motion denoted
$e^{-i\omega_{0}t},$ the projected states also change periodically.\ As a
result, we use the following formulation to characterize the global rotating
process and the leapfrogging process,
\begin{equation}
\psi_{i}\rightarrow \psi_{i}^{\prime}(t)=e^{(\vec{1}\otimes \iota_{x}%
)(2i(\omega_{0}+\omega^{\ast})t)}\cdot \psi_{i}.
\end{equation}
After considering the energy from the global rotating process and the
leapfrogging process, a corresponding term is given by%
\begin{align}
i\hbar_{\mathrm{knot}}%
%TCIMACRO{\dsum \limits_{i}}%
%BeginExpansion
{\displaystyle \sum \limits_{i}}
%EndExpansion
\psi_{i}^{\prime \dagger}\partial_{t}\psi_{i}^{\prime}  &  =i\hbar
_{\mathrm{knot}}%
%TCIMACRO{\dsum \limits_{i}}%
%BeginExpansion
{\displaystyle \sum \limits_{i}}
%EndExpansion
(\psi_{i}^{\dagger}e^{(\vec{1}\otimes \iota_{x})(-2i(\omega_{0}+\omega^{\ast
})t)})\partial_{t}(e^{(\vec{1}\otimes \iota_{x})(2i(\omega_{0}+\omega^{\ast
})t)}\psi_{i})\nonumber \\
&  =i\hbar_{\mathrm{knot}}%
%TCIMACRO{\dsum \limits_{i}}%
%BeginExpansion
{\displaystyle \sum \limits_{i}}
%EndExpansion
(\psi_{i}^{\dagger}\partial_{t}\psi_{i}+2i\psi_{i}^{\dagger}(\omega_{0}%
+\omega^{\ast})(\vec{1}\otimes \iota_{x})\psi_{i}).\nonumber
\end{align}

We write down the total Lagrangian of a 3D leapfrogging knot-crystal in
tight-binding formula as
\begin{align}
\mathcal{L}  &  =i\hbar_{\mathrm{knot}}%
%TCIMACRO{\dsum \limits_{i}}%
%BeginExpansion
{\displaystyle \sum \limits_{i}}
%EndExpansion
\psi_{i}^{\prime \dagger}\partial_{t}\psi_{i}^{\prime}-\mathcal{H}%
_{\mathrm{coupling}}\\
&  =i\hbar_{\mathrm{knot}}\int \psi^{\dagger}\partial_{t}\psi d^{3}%
x-\mathcal{H}_{\mathrm{knot}}\nonumber
\end{align}
where%
\begin{align}
\mathcal{H}_{\mathrm{knot}}  &  =\mathcal{H}_{\mathrm{coupling}}%
+2\hbar_{\mathrm{knot}}(\omega_{0}+\omega^{\ast})\int \psi^{\dagger}(\vec
{1}\otimes \iota_{x})\psi d^{3}x\\
&  =-J%
%TCIMACRO{\dsum \limits_{k}}%
%BeginExpansion
{\displaystyle \sum \limits_{k}}
%EndExpansion
\psi_{k}^{\dagger}(\vec{1}\otimes \iota_{x})\psi_{k}-aJ%
%TCIMACRO{\dsum \limits_{k}}%
%BeginExpansion
{\displaystyle \sum \limits_{k}}
%EndExpansion
\psi_{k}^{\dagger}[\Delta \vec{k}\cdot \vec{\sigma}\otimes \iota_{y}]\psi
_{k}\nonumber \\
&  +2\hbar_{\mathrm{knot}}(\omega_{0}+\omega^{\ast})%
%TCIMACRO{\dsum \nolimits_{k}}%
%BeginExpansion
{\displaystyle \sum \nolimits_{k}}
%EndExpansion
\psi_{k}^{\dagger}(\vec{1}\otimes \iota_{x})\psi_{k}.\nonumber
\end{align}
From the fact that the energy of vortex-membranes is equal to the global
rotating energy, $J-2\hbar_{\mathrm{knot}}\omega_{0}=0$, we derive the value
of $J$, i.e., $J=2\cdot \hbar_{\mathrm{knot}}\omega_{0}$. Here, the factor
"$2$" comes from the global rotating energy of the two entangled vortex-membranes.

Finally, in long wave-length limit $\Delta k\rightarrow0$ we obtain the low
energy effective Hamiltonian as
\begin{align}
\mathcal{H}_{\mathrm{knot}}  &  \simeq-aJ%
%TCIMACRO{\dsum \limits_{k}}%
%BeginExpansion
{\displaystyle \sum \limits_{k}}
%EndExpansion
\psi_{k}^{\dagger}[((\Delta \vec{k}\cdot \vec{\sigma})\otimes \iota_{y})]\psi
_{k}\nonumber \\
&  +2\hbar_{\mathrm{knot}}\omega^{\ast}%
%TCIMACRO{\dsum \nolimits_{k}}%
%BeginExpansion
{\displaystyle \sum \nolimits_{k}}
%EndExpansion
\psi_{k}^{\dagger}(\vec{1}\otimes \iota_{x})\psi_{k}\nonumber \\
&  =-c_{\mathrm{eff}}\int \psi^{\dagger}[(\vec{\sigma}\otimes \iota_{y}%
)\cdot \vec{p}_{\mathrm{knot}}]\psi d^{3}x\nonumber \\
&  +m_{\mathrm{knot}}c_{\mathrm{eff}}^{2}\int \psi^{\dagger}(\vec{1}%
\otimes \iota_{x})\psi d^{3}x
\end{align}
where $c_{\mathrm{eff}}=\frac{a\cdot J}{\hbar_{\mathrm{knot}}}=2a\omega_{0}$
play the role of light speed and $m_{\mathrm{knot}}c_{\mathrm{eff}}^{2}%
=2\hbar_{\mathrm{knot}}\omega^{\ast}$ plays role of the mass of knots.
$\vec{p}_{\mathrm{knot}}=\hbar_{\mathrm{knot}}\vec{k}$ is the momentum
operator. We then re-write the effective Hamiltonian to be
\begin{equation}
\mathcal{H}_{\mathrm{knot}}=\int(\psi^{\dagger}\mathrm{\hat{H}}_{\mathrm{knot}%
}\psi)d^{3}x
\end{equation}
and
\begin{equation}
\mathrm{\hat{H}}_{\mathrm{knot}}=-c_{\mathrm{eff}}\vec{\Gamma}\cdot \vec
{p}_{\mathrm{knot}}+m_{\mathrm{knot}}c_{\mathrm{eff}}^{2}\Gamma^{5}%
\end{equation}
where $\Gamma^{5}=\vec{1}\otimes \iota_{x}\mathbf{,}$ $\vec{\Gamma}=(\Gamma
^{1},\Gamma^{2},\Gamma^{3})$ and $\Gamma^{1}=\sigma^{x}\otimes \iota_{y},$
$\Gamma^{2}=\sigma^{y}\otimes \iota_{y},$ $\Gamma^{3}=\sigma^{z}\otimes
\iota_{y}.$

It is obvious that the total Hamiltonian $\mathcal{H}$ has translation
symmetry and rotation symmetry%
\begin{equation}
\left[  \mathcal{\hat{T}}(\Delta \vec{x}),\mathrm{\hat{H}}_{\mathrm{knot}%
}\right]  =0
\end{equation}
and
\begin{equation}
\left[  \hat{R},\mathrm{\hat{H}}_{\mathrm{knot}}\right]  =0,
\end{equation}
respectively. Here, $\mathcal{\hat{T}}(\Delta x)=e^{ia(\hat{k}\cdot
\mathbf{\vec{\Gamma}})}\cdot \hat{S}$ is translation operator. That means we
have generalized translation symmetry. Another important property of the Dirac
model for knot-crystal is spin-orbital coupling\textbf{.} The global spatial
rotation operator is defined by
\begin{equation}
\hat{R}=\hat{R}_{\mathrm{spin}}\cdot \hat{R}_{\mathrm{space}}%
\end{equation}
where $\hat{R}_{\mathrm{spin}}$ is \textrm{SO(3)} spin rotation operator
$\hat{R}_{\mathrm{spin}}\mathbf{\vec{\Gamma}}\hat{R}_{\mathrm{spin}}%
^{-1}=\mathbf{\vec{\Gamma}}^{\prime},$ and $\hat{R}_{\mathrm{space}}$ is
\textrm{SO(3)} spatial rotation operator, $\hat{R}_{\mathrm{space}}\vec{p}%
\hat{R}_{\mathrm{space}}^{-1}=\vec{p}^{\prime},$ $\hat{R}_{\mathrm{space}}%
\vec{x}\hat{R}_{\mathrm{space}}^{-1}=\vec{x}^{\prime}.$ After doing a global
spatial rotation operation, the motion direction changes from $\vec{e}$ to
$\vec{e}^{\prime}$. That means we have generalized rotation symmetry.

The ground state of knot-crystal is a filled state of knots, of which the
total number of knots is $N_{\mathrm{knot}}$. The excitation is a knot, of
which the energy dispersion is
\begin{equation}
E_{p}=\pm \sqrt{\hbar_{\mathrm{knot}}^{2}k^{2}c_{\mathrm{eff}}^{2}%
+m_{\mathrm{knot}}^{2}c_{\mathrm{eff}}^{4}}%
\end{equation}
where $k^{2}=k_{x}^{2}+k_{y}^{2}+k_{z}^{2}$ and $m_{\mathrm{knot}}$ is the
mass of knot. There are four energy bands for Dirac states. For a
particle-like excitation, the energy is $\sqrt{\hbar_{\mathrm{knot}}^{2}%
k^{2}c_{\mathrm{eff}}^{2}+m^{2}c_{\mathrm{eff}}^{4}}$; for a hole-like
excitation, the energy is $-\sqrt{\hbar_{\mathrm{knot}}^{2}k^{2}%
c_{\mathrm{eff}}^{2}+m_{\mathrm{knot}}^{2}c_{\mathrm{eff}}^{4}}$. For a
particle-like excitation, an extra knot with a zero is put on the knot-crystal
and the total volume of the deformed knot-crystal becomes $\left(
N_{\mathrm{knot}}+1\right)  V_{\mathrm{knot}}$; for a hole-like excitation, a
knot with a zero is taken off from the knot-crystal and the total volume of
the deformed knot-crystal is $\left(  N_{\mathrm{knot}}-1\right)
V_{\mathrm{knot}}.$ The situation is very similar to the quasi-particles in
solid physics. From $\hat{H}_{\mathrm{3D}}$, for a particle-like knot, we can
easily obtain the Schr\"{o}dinger equation in low velocity limit,
\begin{equation}
i\hbar_{\mathrm{knot}}\frac{d\psi(\vec{x},t)}{dt}=\mathrm{\hat{H}%
}_{\mathrm{knot}}\psi(\vec{x},t)
\end{equation}
where
\begin{equation}
\mathrm{\hat{H}}_{\mathrm{knot}}=\frac{\hat{p}_{\mathrm{knot}}^{2}%
}{2m_{\mathrm{knot}}}.
\end{equation}

Finally, the low energy effective Lagrangian of 3D SOC knot-crystal is derived
as
\begin{align}
\mathcal{L}_{\mathrm{3D}}  &  =i\psi^{\dagger}\partial_{t}\psi-\mathcal{H}%
_{\mathrm{knot}}\\
&  =\bar{\psi}(i\gamma^{\mu}\hat{\partial}_{\mu}-m_{\mathrm{knot}}%
)\psi \nonumber
\end{align}
where $\bar{\psi}=\psi^{\dagger}\gamma^{0},$ $\gamma^{\mu}$ are the reduced
Gamma matrices, $\gamma^{1}=\gamma^{0}\Gamma^{1}$, $\gamma^{2}=\gamma
^{0}\Gamma^{2},$ $\gamma^{3}=\gamma^{0}\Gamma^{3},$ and $\gamma^{0}=\Gamma
^{5}=\vec{1}\otimes \iota_{x}$, $\gamma^{5}=i\gamma^{0}\gamma^{1}\gamma
^{2}\gamma^{3}.$ This is Lagrangian for massive Dirac fermions with emergent
\textrm{SO(3,1)} Lorentz-invariance. Since the velocity $c_{\mathrm{eff}}$
only depends on the microscopic parameter $k_{0}$ and $\kappa,$ we may regard
$c_{\mathrm{eff}}$ to be "light-velocity" and the invariance of light-velocity
becomes a fundamental principle for the knot physics. In above equation, we
have set $c_{\mathrm{eff}}=1$.

As a result, we answer the question -- \emph{how to characterize the evolution
of the object with a zero?} The pseudo-quantum mechanics describes the
entanglement-fluctuation of two entangled vortex-membrane with fixed volume in
the 5D fluid; The emergent quantum field theory describes the volume-changing
entanglement-fluctuation of two entangled vortex-membrane in the 5D fluid.
From point view of information, the elementary volume-changing with a zero is
a knot. We consider a given pattern of knot-pieces $\psi(\vec{x})$ as the
initial condition. We expand $\psi(\vec{x})$ by the eigenstates of plane waves
as $\psi(\vec{x})=%
%TCIMACRO{\dsum \nolimits_{k}}%
%BeginExpansion
{\displaystyle \sum \nolimits_{k}}
%EndExpansion
c_{k}e^{i\vec{p}_{\mathrm{knot}}\cdot \vec{x}/\hbar_{\mathrm{knot}}}$. Thus,
under time evolution, the function of vortex-membranes becomes $\psi(\vec
{x},t)=%
%TCIMACRO{\dsum \nolimits_{k}}%
%BeginExpansion
{\displaystyle \sum \nolimits_{k}}
%EndExpansion
c_{k}e^{i\vec{p}_{\mathrm{knot}}\cdot \vec{x}/\hbar_{\mathrm{knot}}%
}e^{-iE_{\mathrm{knot}}\cdot t/\hbar_{\mathrm{knot}}}.$ Finally, the density
of knot-pieces is obtained as
\begin{equation}
\rho_{\mathrm{knot}}(\vec{x},t)=\psi^{\dagger}(\vec{x},t)\psi(\vec{x},t).
\end{equation}
We point out that under time evolution, the energy $H(\vec{p}_{\mathrm{knot}%
})$, the momentum $\vec{p}_{\mathrm{knot}}$ (the Lamb impulse) and the (Lamb
impulse) angular momentum (the effective Planck constant $\hbar_{\mathrm{knot}%
}$) are all conserved. And owning to the the conservation conditions of the
volume of the knot in 5D space, the shape of knot is never changed and the
corresponding Kelvin waves of knots cannot evolve smoothly. Instead, the knot
can only split and knot-pieces evolves following the equation of motion of
Biot-Savart equation (Schr\"{o}dinger equation). This is the fundamental
principle of quantum mechanics.

So for 3D SOC knot, there exist two types of Hamiltonian -- one type on flat
vortex-membranes with continuum spatial translation symmetry, the other on
knot-crystal with lattice translation symmetry. For, a knot on flat
vortex-membranes, the equation of motion is determined by the Schr\"{o}dinger
equation with the Hamiltonian
\begin{equation}
\mathrm{\hat{H}}_{\mathrm{knot}}=%
%TCIMACRO{\dsum \limits_{I}}%
%BeginExpansion
{\displaystyle \sum \limits_{I}}
%EndExpansion
[\frac{(\hat{p}_{\mathrm{knot}}^{I})^{2}}{2m_{\mathrm{pseudo}}}]+\frac
{\hbar_{\mathrm{knot}}\omega^{\ast}}{2}(\vec{1}\otimes(\tau_{x}-\vec{1})).
\end{equation}
There doesn't exist Lorentz invariance; For the knots on 3D SOC knot-crystal,
the equation of motion is determined by the Schr\"{o}dinger equation with the
Hamiltonian
\begin{equation}
\mathrm{\hat{H}}_{\mathrm{knot}}=-c_{\mathrm{eff}}\vec{\Gamma}\cdot \vec
{p}_{\mathrm{knot}}+m_{\mathrm{knot}}c_{\mathrm{eff}}^{2}\Gamma^{5}{.}%
\end{equation}

\subsubsection{Possible physical realization}

We address the issue of the physical realization of a 1D knot-crystal based on
entangled vortex-lines in $^{4}$He superfluid.

We consider two rectilinear vortex-lines in SF that is stretched between
opposite points on the system. Then we rotate one vortex-line around the other
with a rotating velocity $\omega_{0}$. According to the theory of Kelvin
waves, the winding vortex-line becomes a helical one that is described by
$r_{0}e^{ik_{0}\cdot x-i\omega_{0}t+i\phi_{0}}$ where $\omega_{0}\simeq
(\frac{\kappa}{4\pi}\ln \frac{1}{k_{0}a_{0}})k_{0}^{2}$. As a result, in
principle, we realize a knot-crystal.

We estimate the two physical constants, $\hbar_{\mathrm{knot}}$ and
$c_{\mathrm{eff}}$. The emergent Planck constant $\hbar_{\mathrm{knot}}$ is
defined by $\hbar_{\mathrm{knot}}=\frac{\pi r_{0}^{2}\rho_{0}\kappa}{2k_{0}}$.
We set the length of the half pitch of the windings $a=\frac{\pi}{k_{0}}$ to
be $10^{-5}$\textrm{cm}, and the distance between two vortex-lines $r_{0}$ as
$10^{-6}$\textrm{cm.} For $^{4}$He superfluid,\textrm{ }$\kappa$ is about
$10^{-3}$\textrm{cm}$^{2}$\textrm{/s}, Then the effective Planck constant is
estimated to be $\hbar_{\mathrm{knot}}=N_{\mathrm{He4}}\cdot \hbar \sim
10^{5}\hbar$ where $N_{\mathrm{He4}}=\pi r_{0}^{2}a\cdot \rho^{\ast}$ is the
atom's number inside a knot. $\rho^{\ast}$ is the superfluid density. Another
important parameter is the effective light speed $c_{\mathrm{eff}}$, which is
defined by $c_{\mathrm{eff}}=\frac{\kappa k_{0}}{2}(\ln \frac{1}{k_{0}a_{0}%
}-\frac{1}{2})$. The effective light speed is estimated to be $c_{\mathrm{eff}%
}\sim12\mathrm{m/s}$.

\subsection{Measurement theory}

Measurement is an important issue of quantum mechanics. P.A.M. Dirac (1958) in
"\emph{The Principles of Quantum Mechanics said, "A measurement always causes
the system to jump into an eigenstate of the dynamical variable that is being
measured, the eigenvalue this eigenstate belongs to being equal to the result
of the measurement.}" According to the Copenhagen interpretation, the state of
a system is assumed to "collapse" into an eigenstate of the operator during
measurement. That is the so-called measurement process of "wave-function
collapse". The wave-function collapse is random and indeterministic and the
predicted value of the measurement is described by a probability distribution.
The wave-function collapse raises "the measurement problem", as well as
questions of determinism and locality, as demonstrated in the
Einstein--Podolsky--Rosen (EPR) paradox\cite{epr}. In this section, we show
measurement theory in emergent quantum mechanics.

\subsubsection{Information measurement and observers}

In physics, people try to know the information of patterns for perturbative
Kelvin wave of knots described by $\mathbf{Z}_{\mathrm{knot}}(\vec{x},t)$,
i.e.,%
\begin{equation}
\text{Information }=\text{ }\mathbf{Z}_{\mathrm{knot}}(\vec{x},t)\text{.}%
\end{equation}
In mathematics, the information is the distribution of zeros between projected
vortex-membranes that is determined by knot density for a given quantum state
described by $\left \vert \psi(\vec{x},t)\right \rangle $, i.e.,%
\begin{equation}
\text{Information }=\text{ }\left \vert \psi(\vec{x},t)\right \rangle \text{.}%
\end{equation}
That means to identify the entanglement pattern of vortex-membranes
$\mathbf{Z}_{\mathrm{knot}}(\vec{x},t)$, people detect the distribution
function of knot-pieces $\rho_{\mathrm{knot}}(\vec{x},t)$, i.e.,
\begin{align}
\text{Measurement }  &  =\text{ to detect }\mathbf{Z}_{\mathrm{knot}}(\vec
{x},t)\\
&  \rightarrow \text{to detect }\rho_{\mathrm{knot}}(\vec{x},t)\text{.}%
\nonumber
\end{align}

To do measurement, there exist observers. To detect the distribution of
knot-pieces, there are two types of observers: one type is the inner observers
that are special multi-knot systems on vortex-membranes, the other is outer
observers that are objects out of vortex-membranes (or even out of the
inviscid incompressible fluid). For the outer observers, the rulers and clocks
are independent on the physical properties of the knot-crystal. In principle,
the outer observers have the ability to detect the every detail of the
perturbative Kelvin waves on the knot-crystal. As a result, the theory for
outer observers is Biot-Savart mechanics. However, for the inner observers,
the rulers and clocks depend on the physical properties of the knot-crystal.
In general, the inner observers have very very slow clocks and very very large
rules and have no the ability to detect the every detail of the perturbative
Kelvin waves on the knot-crystal. By using a huge ruler, the inner observers
obtain a coarse-graining picture of a given system of multi-knots; By using a
slow clock, the inner observers obtain average results of a dynamic system of
multi-knots via time. So the inner observers will never obtain the complete
information. In particular, the information of absolute phase factor cannot be
observed for inner observers owing to a slow clock and a huge rule with very
large scale-distance. In this paper, we consider the measurement theory for
inner observers by considering slow clock. The situation for huge rule with
very large scale-distance is similar.

\begin{figure}[ptb]
\includegraphics[clip,width=0.63\textwidth]{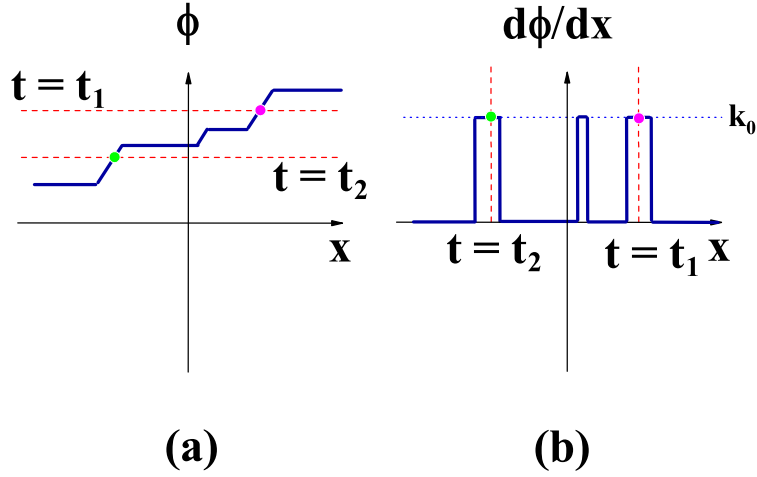}\caption{An illustration
of dynamic projection: (a) shows phase $\phi$ via $x$ and (b) shows phase
changing $\frac{d\phi}{dx}$ via $x$. There are two measurement processes at
time $t=t_{1}$ and $t=t_{2}$. The green spots in (a) and the purple spots in
(b) denote the position of knot-pieces under projection. In (a), people can
detect the phase changing and the information for a knot can be fully
observed; In (b), people can only detect the phase changing and the
information for a knot cannot be fully observed. }%
\end{figure}

\subsubsection{Dynamic projection}

In this part, we show the concept of\textbf{ }dynamic projection, that is a
process to detect $\rho_{\mathrm{knot}}(\vec{x},t)$.

Knot is topological object that leads to $\pi$-phase changing ($\Delta \phi
=\pi$) and knot-piece is tiny phase-changing ($\Delta \phi \rightarrow0$) with
fixed phase angle ($\phi=\phi_{0}$). To find a knot (or a knot-piece), people
need to detect the phase-changing $\Delta \phi$ that changes the density of
zeros between two vortex-membranes, $\rho_{\text{\textrm{zero}}}%
\rightarrow \rho_{\text{\textrm{zero}}}(x,t)$. So we may project the
vortex-membranes and solve the knot-function to look for extra solutions owing
to the knot-piece $\Delta \phi$. In experiments, people identify phase-changing
$\Delta \phi$ via dynamic projection%
\begin{equation}
\hat{P}_{\mathrm{dynamic}}=\text{ to capture a knot-piece }\Delta \phi \text{ at
}t=t_{0}.
\end{equation}
To find a fragmentized knot, we need to do projection%
\begin{equation}
\hat{P}_{\mathrm{dynamic}}(t_{0},\theta)\mathrm{z}_{\mathrm{B}}(\vec
{x},t)=\hat{P}_{\mathrm{dynamic}}(t_{0},\theta)\mathrm{z}_{\mathrm{A}}(\vec
{x}_{0},t_{0}).
\end{equation}
Under a projection, a fragmentized (or randomized) knot is fixed at $x_{0}$
and time $t_{0}$.

Because a knot is really a sharp phase-changing that could interact with the
measuring devices,
\begin{align}
\mathbf{\tilde{Z}}_{\mathrm{fragment}}^{\prime}(\vec{x},t)  &  =\mathrm{\hat
{U}}_{\mathrm{fragment}}\mathbf{Z}_{\mathrm{knot-crystal}}(\vec{x}%
,t)\nonumber \\
&  =%
%TCIMACRO{\dprod \limits_{i=1,I}^{N\rightarrow\infty}}%
%BeginExpansion
{\displaystyle \prod \limits_{i=1,I}^{N\rightarrow \infty}}
%EndExpansion
\mathrm{\hat{U}}_{\mathrm{fragment}}^{I}(\Delta \phi_{\frac{1}{N}%
\text{-}\mathrm{knot}}\left(  X_{0}\right)  _{i})\mathbf{Z}%
_{\mathrm{knot-crystal}}(\vec{x},t).
\end{align}
Thus, according to the different effects, we can decompose the dynamic
operator into three parts:
\begin{equation}
\hat{P}_{\mathrm{dynamic}}(t_{0},\theta)=\hat{P}_{\mathrm{fragment}}\hat
{P}_{\mathrm{Klevin-wave}}\hat{P}(t_{0})
\end{equation}
where $\hat{P}(t_{0})$ denotes the projection at time $t_{0}$, $\hat
{P}_{\mathrm{Klevin-wave}}$ denotes the projection on a given eigenstate of
perturbative Kelvin waves $e^{-i\Delta \omega \cdot t_{0}+i\Delta \vec{k}%
\cdot \vec{x}_{0}}\left(
\begin{array}
[c]{c}%
u_{\vec{k}_{0},\mathrm{A}}\\
u_{\vec{k}_{0},\mathrm{B}}\\
u_{-\vec{k}_{0},\mathrm{A}}\\
u_{-\vec{k}_{0},\mathrm{B}}%
\end{array}
\right)  $ with fixed spin and vortex degrees of freedom, $\hat{P}%
_{\mathrm{fragment}}$ denotes the projection on a given knot-piece for the
projected pertubative Kelvin waves. As a result, the dynamic projection
operator is defined by
\begin{align}
&  \hat{P}_{\mathrm{dynamic}}(t_{0},\theta)z(\vec{x},t_{0})\\
&  =\hat{P}_{\mathrm{dynamic}}(t_{0},\theta)%
%TCIMACRO{\dprod \limits_{i=1,I}^{N\rightarrow\infty}}%
%BeginExpansion
{\displaystyle \prod \limits_{i=1,I}^{N\rightarrow \infty}}
%EndExpansion
\mathrm{\hat{U}}_{\mathrm{fragment}}^{I}(\Delta \phi_{\frac{1}{N}%
\text{-}\mathrm{knot}}\left(  X_{0}\right)  _{i})\nonumber \\
&  \cdot \lbrack%
%TCIMACRO{\dsum \limits_{\Delta k}}%
%BeginExpansion
{\displaystyle \sum \limits_{\Delta k}}
%EndExpansion
a_{\Delta k}e^{\pm i\Delta \omega t_{0}\pm i\Delta \vec{k}\cdot \vec{x}_{0}%
}\mathbf{Z}_{\mathrm{knot-crystal}}(\vec{x},t)]\nonumber \\
&  =\mathrm{\hat{U}}_{\mathrm{fragment}}^{I}(\Delta \phi_{\frac{1}{N}%
\text{-}\mathrm{knot}}\left(  X_{0}\right)  _{i})\nonumber \\
&  \cdot \lbrack a_{\Delta k}e^{-i\Delta \omega t_{0}+i\Delta k\cdot x_{0}%
}\mathbf{Z}_{\mathrm{knot-crystal}}(\vec{x}_{0},t_{0})]\nonumber \\
&  \Rightarrow \lbrack z(\phi(x_{0}))_{\mathrm{fragment}}]\nonumber \\
&  \cdot \lbrack e^{-i\Delta \omega t_{0}+i\Delta k\cdot x_{0}}\mathbf{Z}%
_{\mathrm{knot-crustal}}(\vec{x}_{0},t_{0})].
\end{align}

After repeating the measurement processes $\hat{P}_{\mathrm{dynamic}}$, a lot
of knot-pieces is observed and people know their distribution $\rho
_{\mathrm{knot}}(\vec{x},t)$. Finally, the information of perturbative Kelvin
waves on vortex-membrane $\mathbf{Z}_{\mathrm{knot}}(\vec{x},t)$ is explored.
During measurement processes, the phase angle of knot-pieces is fixed to be
$\phi_{0}(t_{0})$ that depends on the time of measurement. This gives an
explanation of "wave-function collapse" in quantum mechanics.\emph{ }

\subsubsection{Fast-clock effect}

We point out that the probability of emergent quantum mechanics comes from
dynamically projecting entangled vortex-membranes via stochastic projected
time $t_{0}$ that leads to stochastic projected angle.

It is known that the knots are phase-changing of knot-crystal. For a knot
crystal, $\mathbf{Z}_{\mathrm{knot-crystal}}(\vec{x},t)\sim e^{-i(\omega
_{0}+\Delta \omega)t+\phi_{0}},$ the period $T_{\mathrm{clock}}$ of the phase
angle changing $2\pi$ is $T_{\mathrm{clock}}=\frac{2\pi}{\omega_{0}%
+\Delta \omega}$. So we have an intrinsic clock for knots, of which the phase
is just the phase of the wave-function. The clock for inner observers
$T_{\mathrm{inner}}$ must be very very slower than the intrinsic clock,
$T_{\mathrm{inner}}\gg T_{\mathrm{clock}}$. The inner observers can do
measurement at the scheduled time $\Delta t=nT_{\mathrm{inner}}$ ($n$ is an
arbitrary integer number), but cannot accurately do measurement at the
scheduled time $\Delta t=nT_{\mathrm{clock}}$. In the limit of $\frac
{T_{\mathrm{inner}}}{T_{\mathrm{clock}}}\rightarrow \infty$, $\left[
\frac{\Delta t}{T_{\mathrm{clock}}}-\operatorname{mod}(\frac{\Delta
t}{T_{\mathrm{clock}}})\right]  $ is a random value and the projection angles
$\left[  \frac{\Delta t}{T_{\mathrm{clock}}}-\operatorname{mod}(\frac{\Delta
t}{T_{\mathrm{clock}}})\right]  \omega_{0}$ become stochastic for the
different projection processes. As a result, owning to the stochastic
projecting time $\hat{P}(t_{0})$, the projecting angle $\theta$ is a random
value, i.e.,
\begin{align}
\theta &  \rightarrow \theta+\left[  \frac{\Delta t}{T_{\mathrm{clock}}%
}-\operatorname{mod}(\frac{\Delta t}{T_{\mathrm{clock}}})\right]  \omega_{0}\\
&  =a\text{ random value}.\nonumber
\end{align}
We call it "fast-clock" effect.

As shown in Fig.12.(b), the detecting process can be exactly demonstrated by
pulse sample figures -- people can detect the phase-changing from knots but
cannot detect the absolute phase angle.

For stochastic projection process due to "fast-clock" effect, the probability
for a knot-piece is defined by the average of the projection angle. If there
exists a knot solution at $\bar{x}(\bar{\theta})$ for a given projection angle
$\bar{\theta}$, we call $\bar{\theta}$ the permitted projection angle; If
there doesn't exist knot solution for a given projected angle $\tilde{\theta}%
$, we call $\tilde{\theta}$ the unpermitted projection angle. Thus, the
probability of a knot-piece is obtained as $\frac{1}{2\pi}\int d\bar{\theta
}=\frac{1}{N}$ where $\int d\bar{\theta}=\frac{2\pi}{N}$ denotes the summation
of all permitted projected angles $\bar{\theta}$. For a system with a
knot-piece, the probability for finding a knot (a zero between two projected
vortex-membranes) is just its knot number, i.e., $\frac{1}{2\pi}\int
d\bar{\theta}=\left \langle \hat{K}\right \rangle =\frac{1}{N}.$ As a result, if
we ignore the contribution from background (the knot crystal), the probability
density for finding a knot $n_{\mathrm{knot}}(\vec{x},t)=\frac{1}{\Delta
V_{p}}\cdot \frac{1}{2\pi}\int d\bar{\theta}$ is obtained as
\begin{equation}
n_{\mathrm{knot}}(\vec{x},t)=\left \langle \frac{\Delta \hat{K}}{\Delta V_{p}%
}\right \rangle \equiv \rho_{\mathrm{knot}}(\vec{x},t).
\end{equation}
This result indicates that the probability density for finding a knot is equal
to the knot density $\rho_{\mathrm{knot}}(\vec{x},t)$.

\subsubsection{Probability interpretation for wave-functions}

We point out that the function of a Kelvin wave with a fragmentized knot
describes the distribution of the knot-pieces and plays the role of the
wave-function in quantum mechanics as
\begin{equation}
\psi(\vec{x},t)\Longleftrightarrow \sqrt{\rho_{\mathrm{knot}}(\vec{x}%
,t)}e^{i\Delta \phi(\vec{x},t)},
\end{equation}
and
\begin{equation}
\rho_{\mathrm{knot}}(\vec{x},t)\Longleftrightarrow n_{\mathrm{knot}}(\vec
{x},t)
\end{equation}
where the function of Kelvin wave with a fragmentized knot becomes the
wave-function $\psi(\vec{x},t)$ in emergent quantum mechanics, the angle
$\Delta \phi_{\mathrm{B}}(x,t)$ becomes the quantum phase angle of
wave-function, the knot density $\rho_{\mathrm{knot}}=\left \langle
\frac{\Delta \hat{K}}{\Delta V_{P}}\right \rangle $ becomes the probability
density of knots $n_{\mathrm{knot}}(\vec{x})$. Thus, the measurement is to
find the zero between two projected vortex-membranes that occurs at certain
fragmentized knot with probability in a given region, $\psi^{\ast}(\vec
{x},t)\psi(\vec{x},t)\Delta V_{P}$.

In particular, we emphasize knot has holographic property. Although, a
knot-piece has a probability of $\rho_{\mathrm{knot}}=\frac{1}{N}$ to find a
knot, not only its tensor state are same to that of a knot, but also its mass
is $m_{\mathrm{knot}}$ rather than $m_{\mathrm{knot}}/N$.

\subsubsection{Complementarity principle}

In emergent quantum mechanics, complementarity principle comes from
complementarity property of knots. On the one hand, a knot piece is
phase-changing -- a sharp, time-independent, topological phase-changing,
$\Delta \phi \neq0$. To count the knot number, we had introduced knot-number
operator $\hat{K}=-i\frac{d}{d\phi};$ On the other hand, a knot-piece has a
fixed phase angle, $\phi_{0}$ that is determined by perturbative Kelvin waves
on the vortex-membranes. One cannot exactly determine the phase angle of a
knot-piece by observing its phase-changing. We call this property to be
complementarity principle in emergent quantum mechanics that leads to
uncertainty principle. Based on the complementarity principle, we discuss
wave-particle duality and uncertainty principle.

Wave--particle duality is the fact that elementary particles exhibit both
particle-like behavior and wave-like behavior. As Einstein wrote: "\emph{It
seems as though we must use sometimes the one theory and sometimes the other,
while at times we may use either. We are faced with a new kind of difficulty.
We have two contradictory pictures of reality; separately neither of them
fully explains the phenomena of light, but together they do}". Here, we point
out that wave--particle duality of quantum particles is really
"information-motion duality" that is property of complementarity principle. On
one hand, a knot is information unit that is a sharp, fragmentized,
topological phase-changing in physics and would become a "point" after
projection in mathematics. Thus, it shows particle-like behavior; On the other
hand, the dynamic, smooth, non-topological phase-changing from perturbative
Kelvin waves shows wave-like behavior that is characterized by wave-functions.
In other words, the Kelvin waves look like pilot-waves\cite{de brogile}, on
which knot-pieces stay.

For emergent quantum mechanics, the uncertainty principle is related to the
information nature of knots. From the point view of information, the momentum
denotes the spatial distribution of knot-pieces (or information), the energy
denotes the temporal distribution of knot-pieces (or information). On one
hand, a uniform distribution of knot-pieces generated by $\mathrm{\hat{U}%
}_{\mathrm{fragment}}$ on knot-crystal with an excited Kelvin wave
$\psi(x,t)=e^{-i\omega t+i\vec{k}\cdot \vec{x}}$ is described by a
wave-function of a plane wave with a fixed projected momentum $\vec
{p}_{\mathrm{knot}}=\hbar_{\mathrm{knot}}\vec{k}$. For this case, we know
momentum information of the knot but we do not know its position information;
On the other hand, a unified knot generated by $\mathrm{\hat{U}}%
_{\mathrm{Unified-knot}}$ is really a special distribution of knot-pieces and
can be regarded as a superposition state of $\psi(x,t)\sim%
%TCIMACRO{\dsum \limits_{k}}%
%BeginExpansion
{\displaystyle \sum \limits_{k}}
%EndExpansion
e^{-i\omega t+i\vec{k}\cdot \vec{x}}.$ For this case, we know the position
information of the knot but we do not know its momentum information.

\subsubsection{Applications: the Schr\"{o}dinger's cat paradox and the
double-slit experiment}

From above discussion, we show the answer to the Schr\"{o}dinger's cat
paradox. The stochastic and uncertain come from dynamic projection and
fragmentized knot. A cat is a classical object with slow inner clock. The time
scale of a cat is the same order of the detecting objects. There is no dynamic
projection process and all information of a cat can be obtained. So there is
no Schr\"{o}dinger's cat paradox at all.

We then give an explanation on the double-slit experiment. Before measurement
in double-slit experiment, the knot can be regarded as fragmentized knot on a
perturbative Kelvin wave, of which the probability distribution is described
by the wave-function. It has no classical path. There exists particular
interference pattern on the screen that agrees to the prediction from quantum
mechanics. However, after measurement, the phase angle of knot-pieces becomes
randomized. As a result, the phase coherence is destructed and the
interference disappears.

\subsection{Quantum entanglement theory}

In the end of the section, we address the issue of quantum entanglement.
Quantum entanglement is a physical phenomenon for two knots that the quantum
state of each knot cannot be described independently of the others, even when
the particles are separated by a large distance. An entangled state for
$n$-body quantum system comes from new type of knots -- \emph{composite knot}
with $n$ zeros. We call it (composite) $m\pi$-knots that change the volume of
vortex-membranes via $\pm mV_{\mathrm{knot}}$. Here, $m$ may be not equal to
$n$. For a fragmentized $m\pi$-knot, the distribution of its pieces is
described by its "wave-functions". In general, the quantum states for a
fragmentized $m\pi$-knot cannot be reduced into a product state of the
wave-function for $n$ knots.

Firstly, we study entangled states for two knots.

We consider two spin degrees of freedom for a knot $\left(
\begin{array}
[c]{c}%
\left \vert \uparrow \right \rangle _{\mathrm{knot}}\\
\left \vert \downarrow \right \rangle _{\mathrm{knot}}%
\end{array}
\right)  .$ On the basis of
\begin{equation}
\left(
\begin{array}
[c]{c}%
\left \vert \uparrow \right \rangle _{\mathrm{knot,}1}\otimes \left \vert
\uparrow \right \rangle _{\mathrm{knot,}2}\\
\left \vert \uparrow \right \rangle _{\mathrm{knot,}1}\otimes \left \vert
\downarrow \right \rangle _{\mathrm{knot,}2}\\
\left \vert \downarrow \right \rangle _{\mathrm{knot,}1}\otimes \left \vert
\uparrow \right \rangle _{\mathrm{knot,}2}\\
\left \vert \downarrow \right \rangle _{\mathrm{knot,}1}\otimes \left \vert
\downarrow \right \rangle _{\mathrm{knot,}2}%
\end{array}
\right)  ,
\end{equation}
an arbitrary entangled state for two knots is given by%
\begin{equation}
\left(
\begin{array}
[c]{c}%
b_{1}^{1,2}\\
b_{2}^{1,2}\\
b_{3}^{1,2}\\
b_{4}^{1,2}%
\end{array}
\right)
\end{equation}
where $b_{i}^{1,2}$ ($i=1,2,3,4$) denotes the weights for different elements.
For example, Bell states are defined as
\begin{equation}
\frac{1}{\sqrt{2}}(\left \vert \uparrow \right \rangle _{\mathrm{knot,}1}%
\otimes \left \vert \uparrow \right \rangle _{\mathrm{knot,}2}\pm \left \vert
\downarrow \right \rangle _{\mathrm{knot,}1}\otimes \left \vert \downarrow
\right \rangle _{\mathrm{knot,}2}),
\end{equation}
or
\begin{equation}
\frac{1}{\sqrt{2}}(\left \vert \uparrow \right \rangle _{\mathrm{knot,}1}%
\otimes \left \vert \downarrow \right \rangle _{\mathrm{knot,}2}\pm \left \vert
\downarrow \right \rangle _{\mathrm{knot,}1}\otimes \left \vert \uparrow
\right \rangle _{\mathrm{knot,}2}).
\end{equation}

\begin{figure}[ptb]
\includegraphics[clip,width=0.63\textwidth]{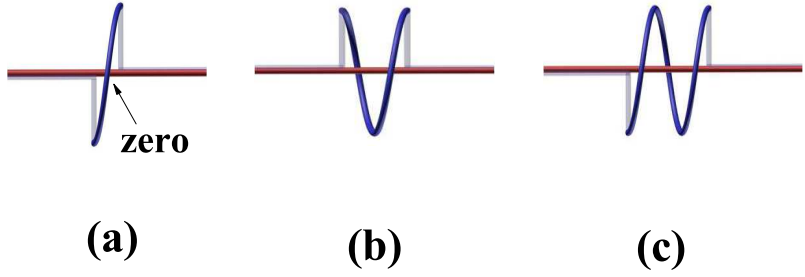}\caption{(a) A unified
knot with 1 zero; (b) A unified $2\pi$-knot with 2 correlated zeros; (c) A
unified $3\pi$-knot with correlated 3 zeros. Each crossing between two
vortex-lines corresponds to a zero.}%
\end{figure}

We then define the composite knots for each state of the basis $\left \vert
\uparrow \right \rangle _{\mathrm{knot,}1}\otimes \left \vert \uparrow
\right \rangle _{\mathrm{knot,}2},$ $\left \vert \uparrow \right \rangle
_{\mathrm{knot,}1}\otimes \left \vert \downarrow \right \rangle _{\mathrm{knot,}%
2},$ $\left \vert \downarrow \right \rangle _{\mathrm{knot,}1}\otimes \left \vert
\uparrow \right \rangle _{\mathrm{knot,}2},$ $\left \vert \downarrow \right \rangle
_{\mathrm{knot,}1}\otimes \left \vert \downarrow \right \rangle _{\mathrm{knot,}%
2}$:

1. The function for unified $\left \vert \uparrow \right \rangle _{\mathrm{knot,}%
1}\otimes \left \vert \uparrow \right \rangle _{\mathrm{knot,}2}$ $2\pi$-knot is
given by
\begin{equation}
\mathrm{z}_{\uparrow \uparrow,2\pi-\mathrm{knot}}=r(x)\exp[i\phi_{\uparrow
\uparrow,2\pi-\mathrm{knot}}(x,t)],
\end{equation}
with%
\begin{equation}
r(x)\rightarrow \left \{
\begin{array}
[c]{c}%
0,\text{ }x\in(-\infty,x_{0}]\\
r_{0},\text{ }x\in(x_{0},x_{0}+2a]\\
0,\text{ }x\in(x_{0}+2a,\infty)
\end{array}
\right \}
\end{equation}
and
\begin{equation}
\phi_{\uparrow \uparrow,2\pi-\mathrm{knot}}(x)=\left \{
\begin{array}
[c]{c}%
\phi_{0},\text{ }x\in(-\infty,x_{0}]\\
\phi_{0}+k_{0}(x-x_{0}),\text{ }x\in(x_{0},x_{0}+2a]\\
\phi_{0}+2\pi,\text{ }x\in(x_{0}+2a,\infty)
\end{array}
\right \}
\end{equation}
where $k_{0}=\frac{\pi}{a}$ and $\phi_{0}$ is constant. We denote this state
by $\left \vert \uparrow \uparrow \right \rangle _{2\pi-\mathrm{knot}}$. See the
illustration of a $2\pi$-knot in Fig.13.(b).

2. The function for unified $\left \vert \downarrow \right \rangle
_{\mathrm{knot,}1}\otimes \left \vert \downarrow \right \rangle _{\mathrm{knot,}%
2}$ $2\pi$-knot is given by
\begin{equation}
\mathrm{z}_{\downarrow \downarrow,2\pi-\mathrm{knot}}=r(x)\exp[i\phi
_{\downarrow \downarrow,2\pi-\mathrm{knot}}(x,t)],
\end{equation}
with%
\begin{equation}
r(x)\rightarrow \left \{
\begin{array}
[c]{c}%
0,\text{ }x\in(-\infty,x_{0}]\\
r_{0},\text{ }x\in(x_{0},x_{0}+2a]\\
0,\text{ }x\in(x_{0}+2a,\infty)
\end{array}
\right \}
\end{equation}
and
\begin{equation}
\phi_{\downarrow \downarrow,2\pi-\mathrm{knot}}(x)=\left \{
\begin{array}
[c]{c}%
\phi_{0},\text{ }x\in(-\infty,x_{0}]\\
\phi_{0}-k_{0}(x-x_{0}),\text{ }x\in(x_{0},x_{0}+2a]\\
\phi_{0}-2\pi,\text{ }x\in(x_{0}+a,\infty)
\end{array}
\right \}  .
\end{equation}
We denote this state by $\left \vert \downarrow \downarrow \right \rangle
_{2\pi-\mathrm{knot}}$.

3. The function for unified $\left \vert \uparrow \right \rangle _{\mathrm{knot,}%
1}\otimes \left \vert \downarrow \right \rangle _{\mathrm{knot,}2}$ $0$-knot is
given by
\begin{equation}
\mathrm{z}_{\uparrow \downarrow,0-\mathrm{knot}}=r(x)\exp[i\phi_{\uparrow
\downarrow,0-\mathrm{knot}}(x,t)],
\end{equation}
with%
\begin{equation}
r(x)\rightarrow \left \{
\begin{array}
[c]{c}%
0,\text{ }x\in(-\infty,x_{0}]\\
r_{0},\text{ }x\in(x_{0},x_{0}+2a]\\
0,\text{ }x\in(x_{0}+2a,\infty)
\end{array}
\right \}
\end{equation}
and
\begin{equation}
\phi_{\uparrow \downarrow,0-\mathrm{knot}}(x)=\left \{
\begin{array}
[c]{c}%
\phi_{0},\text{ }x\in(-\infty,x_{0}]\\
\phi_{0}+k_{0}(x-x_{0}),\text{ }x\in(x_{0},x_{0}+a]\\
\phi_{0}+\pi-k_{0}(x-x_{0}-a),\text{ }x\in(x_{0}+a,x_{0}+2a]\\
\phi_{0},\text{ }x\in(x_{0}+2a,\infty)
\end{array}
\right \}  .
\end{equation}
We denote this state by $\left \vert \uparrow \downarrow \right \rangle
_{0-\mathrm{knot}}$.

4. The function for unified $\left \vert \downarrow \right \rangle
_{\mathrm{knot,}1}\otimes \left \vert \uparrow \right \rangle _{\mathrm{knot,}2}$
$0$-knot is given by
\begin{equation}
\mathrm{z}_{\downarrow \uparrow,0-\mathrm{knot}}=r(x)\exp[i\phi_{\downarrow
\uparrow,0-\mathrm{knot}}(x,t)],
\end{equation}
with%
\begin{equation}
r(x)\rightarrow \left \{
\begin{array}
[c]{c}%
0,\text{ }x\in(-\infty,x_{0}]\\
r_{0},\text{ }x\in(x_{0},x_{0}+2a]\\
0,\text{ }x\in(x_{0}+2a,\infty)
\end{array}
\right \}
\end{equation}
and
\begin{equation}
\phi_{\downarrow \uparrow,0-\mathrm{knot}}(x)=\left \{
\begin{array}
[c]{c}%
\phi_{0},\text{ }x\in(-\infty,x_{0}]\\
\phi_{0}-k_{0}(x-x_{0}),\text{ }x\in(x_{0},x_{0}+a]\\
\phi_{0}-\pi+k_{0}(x-x_{0}-a),\text{ }x\in(x_{0}+a,x_{0}+2a]\\
\phi_{0},\text{ }x\in(x_{0}+2a,\infty)
\end{array}
\right \}  .
\end{equation}
We denote this state by $\left \vert \downarrow \uparrow \right \rangle
_{0-\mathrm{knot}}$.

We can see that the entangled knots can be regarded as a unified object with
two correlated zeros and fixed volume of vortex-membranes $\pm
2V_{\mathrm{knot}}$. For example, the Bell state $\left \vert \uparrow
\right \rangle _{1}\otimes \left \vert \uparrow \right \rangle _{2}+\left \vert
\downarrow \right \rangle _{1}\otimes \left \vert \downarrow \right \rangle _{2}$ is
just $\frac{1}{\sqrt{2}}(\left \vert \uparrow \right \rangle _{2\pi
-\mathrm{knot}}+\left \vert \downarrow \right \rangle _{2\pi-\mathrm{knot}})$. So
after each dynamic projection process, there exist two zeros. When the
properties of a knot (a zero) is detected, the other is exactly known. The
non-locality of quantum entangled states comes from the non-locality of a
perturbative Kelvin wave and the fixed phase difference between the
knot-pieces of two knots. We point out that there exist different choices for
an entangled state for two knots by considering different types of composite knots.

The concept of composite knot can be generalized to the entangled states for
$n$ knots. For entangled states of $n$ knots, on the basis of
\begin{equation}
\left(
\begin{array}
[c]{c}%
\left \vert \uparrow \right \rangle _{\mathrm{knot,}1}\otimes \left \vert
\uparrow \right \rangle _{\mathrm{knot,}2}\otimes \left \vert \uparrow
\right \rangle _{\mathrm{knot,}3}...\otimes \left \vert \uparrow \right \rangle
_{\mathrm{knot,}n}\\
\left \vert \downarrow \right \rangle _{\mathrm{knot,}1}\otimes \left \vert
\uparrow \right \rangle _{\mathrm{knot,}2}\otimes \left \vert \uparrow
\right \rangle _{\mathrm{knot,}3}...\otimes \left \vert \uparrow \right \rangle
_{\mathrm{knot,}n}\\
\left \vert \uparrow \right \rangle _{\mathrm{knot,}1}\otimes \left \vert
\downarrow \right \rangle _{\mathrm{knot,}2}\otimes \left \vert \uparrow
\right \rangle _{\mathrm{knot,}3}...\otimes \left \vert \uparrow \right \rangle
_{\mathrm{knot,}n}\\
...\\
\left \vert \downarrow \right \rangle _{\mathrm{knot,}1}\otimes \left \vert
\downarrow \right \rangle _{\mathrm{knot,}2}\otimes \left \vert \downarrow
\right \rangle _{\mathrm{knot,}3}...\otimes \left \vert \downarrow \right \rangle
_{\mathrm{knot,}n}%
\end{array}
\right)  ,
\end{equation}
an arbitrary entangled state for $n$ knots is given by%
\begin{equation}
\left(
\begin{array}
[c]{c}%
b_{1}^{1,n}\\
b_{2}^{1,n}\\
b_{3}^{1,n}\\
...\\
b_{2^{n}}^{1,n}%
\end{array}
\right)  ,
\end{equation}
where $b_{i}^{1,n}$ ($i=1,2,3,...2^{n}$) denotes the weights for different
elements. We could define $2^{n}$ different types of composite knots to
characterize an arbitrary entangled state of $n$ knot.

We take the states $\left \vert \uparrow \right \rangle _{\mathrm{knot,}1}%
\otimes \left \vert \uparrow \right \rangle _{\mathrm{knot,}2}\otimes \left \vert
\uparrow \right \rangle _{\mathrm{knot,}3}...\otimes \left \vert \uparrow
\right \rangle _{\mathrm{knot,}n}$ and $\left \vert \downarrow \right \rangle
_{\mathrm{knot,}1}\otimes \left \vert \downarrow \right \rangle _{\mathrm{knot,}%
2}\otimes \left \vert \downarrow \right \rangle _{\mathrm{knot,}3}...\otimes
\left \vert \downarrow \right \rangle _{\mathrm{knot,}n}$ as examples: The
function for the unified $\left \vert \uparrow \right \rangle _{\mathrm{knot,}%
1}\otimes \left \vert \uparrow \right \rangle _{\mathrm{knot,}2}\otimes \left \vert
\uparrow \right \rangle _{\mathrm{knot,}3}...\otimes \left \vert \uparrow
\right \rangle _{\mathrm{knot,}n}$ $n\pi$-knot is given by
\begin{equation}
\mathrm{z}_{\uparrow \uparrow...\uparrow,n\pi-\mathrm{knot}}=r_{0}\exp
[i\phi_{\uparrow \uparrow...\uparrow,n\pi-\mathrm{knot}}(x,t)],
\end{equation}
with%
\begin{equation}
r(x)\rightarrow \left \{
\begin{array}
[c]{c}%
0,\text{ }x\in(-\infty,x_{0}]\\
r_{0},\text{ }x\in(x_{0},x_{0}+na]\\
0,\text{ }x\in(x_{0}+na,\infty)
\end{array}
\right \}
\end{equation}
and
\begin{equation}
\phi_{\uparrow \uparrow...\uparrow,n\pi-\mathrm{knot}}(x)=\left \{
\begin{array}
[c]{c}%
\phi_{0},\text{ }x\in(-\infty,x_{0}]\\
\phi_{0}+k_{0}(x-x_{0}),\text{ }x\in(x_{0},x_{0}+na]\\
\phi_{0}+n\pi,\text{ }x\in(x_{0}+na,\infty)
\end{array}
\right \}  .
\end{equation}
We denote this state by $\left \vert \uparrow \uparrow...\uparrow \right \rangle
_{n\pi-\mathrm{knot}}$; The function for the unified $\left \vert
\downarrow \right \rangle _{\mathrm{knot,}1}\otimes \left \vert \downarrow
\right \rangle _{\mathrm{knot,}2}\otimes \left \vert \downarrow \right \rangle
_{\mathrm{knot,}3}...\otimes \left \vert \downarrow \right \rangle
_{\mathrm{knot,}n}$ $n\pi$-knot is given by
\begin{equation}
\mathrm{z}_{\downarrow \downarrow...\downarrow,n\pi-\mathrm{knot}}=r_{0}%
\exp[i\phi_{\downarrow \downarrow...\downarrow,n\pi-\mathrm{knot}}(x,t)],
\end{equation}
with%
\begin{equation}
r(x)\rightarrow \left \{
\begin{array}
[c]{c}%
0,\text{ }x\in(-\infty,x_{0}]\\
r_{0},\text{ }x\in(x_{0},x_{0}+na]\\
0,\text{ }x\in(x_{0}+na,\infty)
\end{array}
\right \}
\end{equation}
and
\begin{equation}
\phi_{\downarrow \downarrow...\downarrow,n\pi-\mathrm{knot}}(x)=\left \{
\begin{array}
[c]{c}%
\phi_{0},\text{ }x\in(-\infty,x_{0}]\\
\phi_{0}-k_{0}(x-x_{0}),\text{ }x\in(x_{0},x_{0}+na]\\
\phi_{0}-n\pi,\text{ }x\in(x_{0}+na,\infty)
\end{array}
\right \}  .
\end{equation}
We denote this state by $\left \vert \downarrow \downarrow...\downarrow
\right \rangle _{n\pi-\mathrm{knot}}$.

The GHZ state for $n$ knots
\begin{align}
&  \frac{1}{\sqrt{2}}(\left \vert \uparrow \right \rangle _{\mathrm{knot,}%
1}\otimes \left \vert \uparrow \right \rangle _{\mathrm{knot,}2}\otimes \left \vert
\uparrow \right \rangle _{\mathrm{knot,}3}...\otimes \left \vert \uparrow
\right \rangle _{\mathrm{knot,}n}\\
&  \pm \left \vert \downarrow \right \rangle _{\mathrm{knot,}1}\otimes \left \vert
\downarrow \right \rangle _{\mathrm{knot,}2}\otimes \left \vert \downarrow
\right \rangle _{\mathrm{knot,}3}...\otimes \left \vert \downarrow \right \rangle
_{\mathrm{knot,}n}).\nonumber
\end{align}
is just $\frac{1}{\sqrt{2}}(\left \vert \uparrow \uparrow...\uparrow
\right \rangle _{n\pi-\mathrm{knot}}\pm \left \vert \downarrow \downarrow
...\downarrow \right \rangle _{n\pi-\mathrm{knot}})$. From the point view of
entanglement, a 1D knot-crystal can be regarded as a big unified
$(N_{\mathrm{knot}}\pi)$-knots with $N_{\mathrm{knot}}$ correlated zeros and
fixed volume of vortex-membranes $\pm N_{\mathrm{knot}}V_{\mathrm{knot}}$, of
which the entangled state can be $\left \vert \uparrow \uparrow...\uparrow
\right \rangle _{N_{\mathrm{knot}}\pi-\mathrm{knot}}.$ This interpretation
gives another explanation on information scaling from knot-crystal to knot.

\begin{figure}[ptb]
\includegraphics[clip,width=0.75\textwidth]{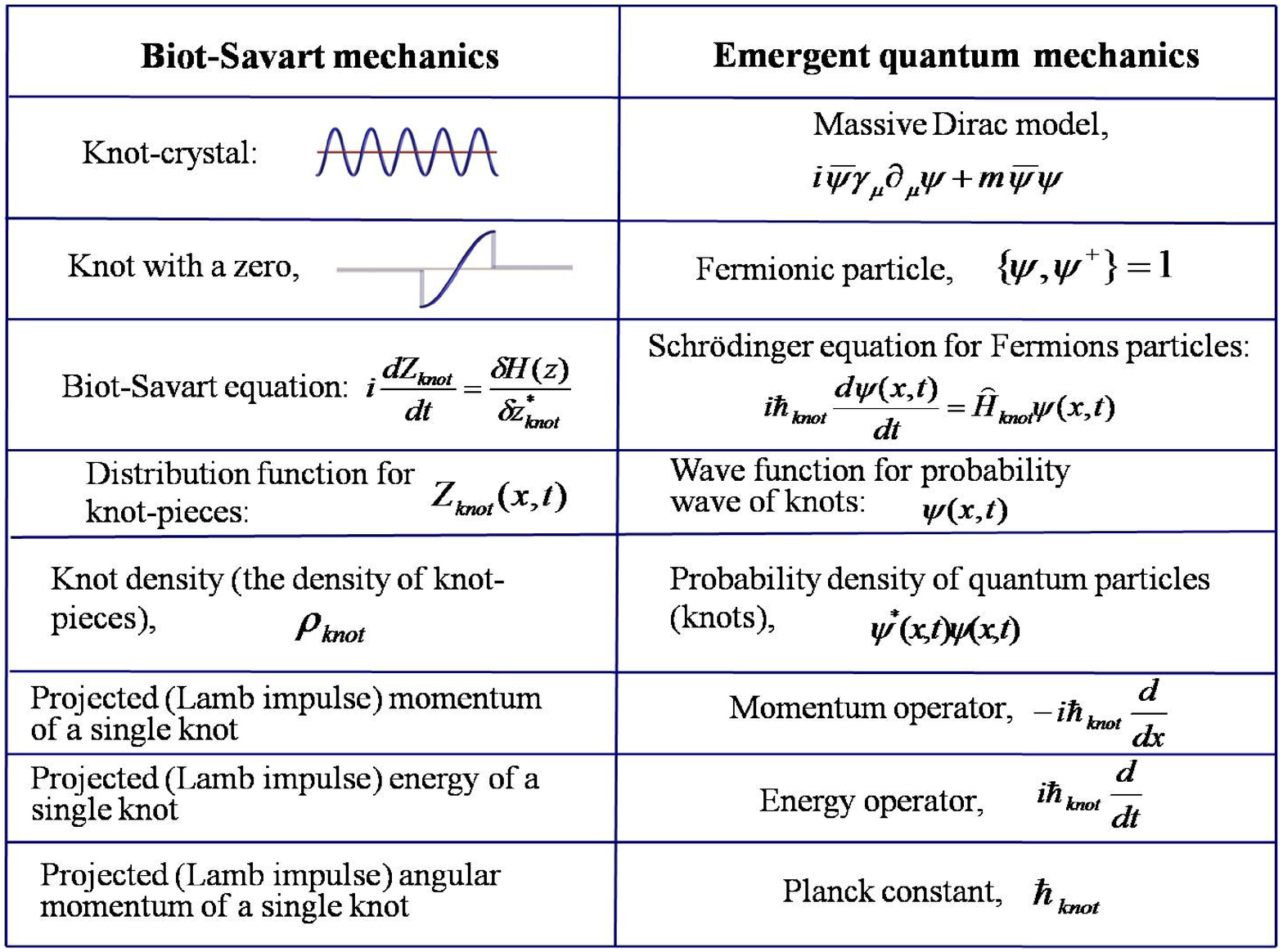}\caption{Correspondence
between Biot-Savart mechanics and emergent-quantum mechanics}%
\end{figure}

\section{Conclusion and summary}

In the end, we draw the conclusion. Fig.14 shows the correspondence between
the Biot-Savart mechanics for knots and emergent quantum mechanics for
particles. Owning to the the conservation conditions of the volume of the knot
in 5D space, the shape and the volume of knot is never changed and the knot
can only split and knot-pieces evolves following the equation of motion of
Schr\"{o}dinger equation. Three dimensional quantum Dirac model is obtained to
effectively describe the deformed vortex-membranes: The elementary excitations
are knots with information unit; The physics quality to describe local
deformation is knot density that is defined by the density of zeros between
two projected vortex-membranes; The Biot-Savart equation for Kelvin waves
becomes Schr\"{o}dinger equation for probability waves of knots; the classical
functions for perturbative Kelvin waves become wave-functions for knots; The
probability of quantum mechanics comes from dynamic projection for inner
observers owing to "fast-clock" effect; The angular frequency for leapfrogging
motion plays the role of the mass of knots; etc.\emph{ }This work would help
people to understand the mystery of quantum mechanics.

Finally, there are several unsolved problems. In this paper, we only consider
the perturbative effect on the Kelvin waves. According to Hasimoto, an
important result after considering the nonlinear effect is the existence of
soliton solutions by mapping the Biot-Savart equation to 1D non-linear
Schr\"{o}dinger equation via Hasimoto representation\cite{h}. The
generalization of Hasimoto representation to 3D knot-crystal may be
interesting. In addition, after considering the dissipation effect, the Kelvin
waves are subject to Kolmogorov-like turbulence (even in quantum
fluid\cite{S95,V00}). The stability of 3D knot-crystals and the possible
cascade of Kelvin waves to turbulence are also open issues. From the point
view of emergent quantum field theory, the emergence of gauge fields and gauge
interaction on a knot-crystal is a critical problem. In the future, we will
study these problems and develop a complete theory for emergent quantum physics.

\acknowledgments This work is supported by NSFC Grant No. 11674026. Because it
was Kelvin (Sir W. Thomson) who firstly proposed the two key concepts in this
manuscript (dynamic knots for elementary particles and Kelvin waves), I wish
to express my gratitude to this wonderful physicist who worked during the 19th century.

\section{Appendix 1: Theory of Kelvin waves for 1D vortex-lines}

vortex-lines are one-dimensional topological objects in three-dimensional
superfluid (SF). For SF with a vortex-line, a rotationless superfluid
component flow $\mathbf{\nabla \times v}=0$ is violated on one-dimensional
singularities $\mathbf{s}(¦Î ,t)$, which depends on the variables -- arc
length $¦Î $ and the time $t$. Away from the singularities, the velocity
increases to infinity so that the circulation $¦Ê $ of the superfluid
velocity remains constant,
\[%
%TCIMACRO{\doint }%
%BeginExpansion
{\displaystyle \oint}
%EndExpansion
\mathbf{v}\cdot d\mathbf{l}=\kappa
\]
where $\mathbf{v=\dot{s}}$. As a result, the superfluid vortex filament could
be described by the Biot-Savart equation
\begin{equation}
\dot{\mathbf{r}}={\frac{\kappa}{4\pi}}\int{\frac{d\mathbf{s}\times
(\mathbf{r}-\mathbf{s})}{|\mathbf{r}-\mathbf{s}|^{3}}},
\end{equation}
where $\kappa=h/m_{4}$ (for $^{4}$He) or $h/m_{3}$ (for $^{3}$He) is the
discreteness of circulation. $\mathbf{r}$ is the vector that denotes the
position of vortex filament and $\mathbf{s}$ denotes the integral variable.
This means that the singularity of $\mathbf{v}(\mathbf{r})$ comes from
integrating around the vicinity part of $\mathbf{s}_{0}$.

\subsection{Local induction approximation}

In Cartesian coordinate system, the position vector of vortex filament is
$\mathbf{s}=(x,$ $y,$ $z)$. By using the natural parameterizations, it is
convenient to express $\mathbf{s}$ in form of $\mathbf{s}(\xi)$, in which
$\xi$ is the (algebraic) arc length measured from the point $\mathbf{s}_{0}$.
In the vicinity of $\mathbf{s}_{0}$, we can expand the function $\mathbf{s}%
(\xi)$ as
\begin{equation}
\mathbf{s}(\xi)=\mathbf{s}_{0}+\mathbf{s}^{\prime}\xi+\mathbf{s}^{\prime
\prime}\xi^{2}/2+\cdots
\end{equation}
and
\begin{equation}
d\mathbf{s=s}^{\prime}d\xi+\mathbf{s}^{\prime \prime}\xi d\xi+\cdots
\end{equation}
in which $\mathbf{s}^{\prime}=\frac{\partial S}{\partial \xi}$ and
$\mathbf{s}^{\prime \prime}=\frac{\partial^{2}S}{\partial^{2}\xi}$,
respectively. From the geometrical point of view, $\mathbf{s}^{\prime}$ is the
unit vector tangent to vortex filament, $\mathbf{s}^{\prime \prime}$ is the
vector normal to vortex filament.

When $\mathbf{r}\rightarrow \mathbf{s}_{0}$, we can separate the integral into
two parts: one is in the vicinity of $\mathbf{s}_{0}$, the other is regular
part without singularity, i.e.,
\begin{align}
\mathbf{v}(\mathbf{r}  &  \rightarrow \mathbf{s}_{0})=\frac{\kappa}{4\pi}%
\int_{-\xi_{0}}^{\xi_{0}}d\xi \frac{(\mathbf{s}^{\prime \prime}\xi
+\mathbf{s}^{\prime})\times(\mathbf{s}_{0}-\mathbf{s})}{|\mathbf{s}%
_{0}-\mathbf{s}|^{3}}+\text{regular}\\
&  \approx \frac{\kappa}{4\pi}\int_{-\xi_{0}}^{\xi_{0}}d\xi \frac{(\mathbf{s}%
^{\prime \prime}\xi+\mathbf{s}^{\prime})\times(\mathbf{s}_{0}-\mathbf{s}%
_{0}-\mathbf{s}^{\prime}\xi-\mathbf{s}^{\prime \prime}\xi^{2}/2)}%
{|\mathbf{s}_{0}-\mathbf{s}_{0}-\mathbf{s}^{\prime}\xi|^{3}}+\text{regular}%
\nonumber \\
&  =\frac{\kappa}{4\pi}\int_{-\xi_{0}}^{\xi_{0}}d\xi \frac{(\mathbf{s}%
^{\prime \prime}\xi+\mathbf{s}^{\prime})\times(-\mathbf{s}^{\prime}%
\xi-\mathbf{s}^{\prime \prime}\xi^{2}/2)}{|-\mathbf{s}^{\prime}\xi|^{3}%
}+\text{regular}\nonumber \\
&  =\frac{\kappa}{4\pi}\int_{-\xi_{0}}^{\xi_{0}}d\xi \frac{-\mathbf{s}%
^{\prime \prime}\times \mathbf{s}^{\prime}\xi^{2}-\mathbf{s}^{\prime}%
\times \mathbf{s}^{\prime \prime}\xi^{2}/2}{|\mathbf{s}^{\prime}\xi|^{3}%
}+\text{regular}\nonumber \\
&  =\frac{\kappa}{8\pi|\mathbf{s}^{\prime}|^{3}}\mathbf{s}^{\prime}%
\times \mathbf{s}^{\prime \prime}\int_{-\xi_{0}}^{\xi_{0}}\frac{d\xi}{|\xi
|}+\text{regular.}\nonumber
\end{align}
where $\xi_{0}$ is the length of the order of the curvature radius (or
inter-vortex distance when the considered vortex filament is a part of a
vortex tangle).

By using $\int_{-\xi_{0}}^{\xi_{0}}\frac{d\xi}{|\xi|}\mathbf{\rightarrow2}%
\ln \mathbf{(}\xi_{0}/a_{0}\mathbf{),}$ we reduce the Biot-Savart equation by
local induction approximation,
\begin{align}
\mathbf{v}(\mathbf{r}\mathbf{\rightarrow s}  &  _{0})\approx \frac{\kappa}%
{4\pi|\mathbf{s}^{\prime}|^{3}}\mathbf{s}^{\prime}\times \mathbf{s}%
^{\prime \prime}\ln(\xi_{0}/a_{0})+\text{regular}\\
&  \approx \frac{\kappa}{4\pi}\mathbf{s}^{\prime}\times \mathbf{s}^{\prime
\prime}\ln(\xi_{0}/a_{0})\nonumber \\
&  =A\mathbf{s}^{\prime}\times \mathbf{s}^{\prime \prime}\nonumber
\end{align}
or
\begin{equation}
\frac{\partial \mathbf{s}}{\partial t}=A\mathbf{s}^{\prime}\times
\mathbf{s}^{\prime \prime}%
\end{equation}
where
\begin{equation}
A=\frac{\kappa}{4\pi}\ln(\xi_{0}/a_{0})
\end{equation}
$\ $and $a_{0}$ denotes the vortex filament radius that is much smaller than
any other characteristic size in the system.

\subsection{Physical quantities of Kelvin waves}

Next, we discuss the physical quantities of the Kelvin waves along winding
direction (z-direction), i.e., the projected momentum $p_{z}$ and the
projected angular momentum $J_{z}$. For a plane Kelvin wave of a lightly
deformed straight vortex-line, the function is described by
\begin{equation}
x=a\cos(kz-\omega t),\text{ }y=a\sin(kz-\omega t).
\end{equation}

In SF with a vortex-line, the conventional momentum of the superfluid motion
cannot be well defined. Instead, the hydrodynamic impulse (the Lamb impulse)
plays the role of the effective momentum that denotes the total mechanical
impulse of the non-conservative body force applied to a limited fluid volume
to generate instantaneously from rest the given motion of the whole of the
fluid at time $t$. In general, the (effective) momentum for a vortex-line from
Lamb impulse density is defined by
\[
\mathbf{P}_{\mathrm{Lamb}}=\frac{1}{2}\rho_{0}\int(\mathbf{s}\times
\mathbf{\omega)}dV.
\]
where $\mathbf{\omega=\nabla \times v}$.

By using the definition of a vortex-line
\[
\mathbf{\omega}=\kappa%
%TCIMACRO{\dint }%
%BeginExpansion
{\displaystyle \int}
%EndExpansion
\mathbf{s}^{\prime}(\xi)\delta(\mathbf{r}-\mathbf{s(}\xi,t\mathbf{)})d\xi,
\]
we have
\begin{align}
\mathbf{P}_{\mathrm{Lamb}}  &  =\frac{\rho_{0}\kappa}{2}\int \mathbf{s}\times
d\mathbf{s}=\frac{\rho_{0}\kappa}{2}\int \mathbf{s\times s}^{\prime}%
d\xi \nonumber \\
&  =\frac{\rho_{0}\kappa}{2}\int[x(z,t)\mathbf{e}_{x}+y(z,t)\mathbf{e}%
_{y}+z(t)\mathbf{e}_{z}]\nonumber \\
&  \times(x_{z}\mathbf{e}_{x}+y_{z}\mathbf{e}_{y}+\mathbf{e}_{z})z_{\xi}%
d\xi \nonumber \\
&  =\frac{\rho_{0}\kappa}{2}\int[(y-zy_{z})\mathbf{e}_{x}+(zx_{z}%
-x)\mathbf{e}_{y}\nonumber \\
&  +(xy_{z}-yx_{z})\mathbf{e}_{z}]z_{\xi}d\xi.\nonumber
\end{align}
The projected (Lamb impulse) momentum along z-direction $p_{z}$ of a
vortex-line with a Kelvin wave is obtained as
\begin{align}
p_{z}  &  =\mathbf{P}_{\mathrm{Lamb}}\cdot \mathbf{e}_{z}\\
&  =\frac{\rho_{0}\kappa}{2}\int(xy_{z}-yx_{z})z_{\xi}d\xi \nonumber \\
&  =\frac{1}{2}\rho_{0}\kappa la^{2}k\nonumber
\end{align}
where $l$ is length of the system along z-direction.

In general, the (Lamb impulse) angular momentum for a vortex-line is defined
by
\begin{align}
\mathbf{J}_{\mathrm{Lamb}}  &  =\frac{\rho_{0}\kappa}{2}\int \mathbf{s\times
(s\times}d\mathbf{\mathbf{s})}\nonumber \\
&  =\frac{\rho_{0}\kappa}{2}\int \mathbf{s\times(s\times s}^{\prime}%
\mathbf{)}d\xi \nonumber \\
&  =\frac{\rho_{0}\kappa}{2}\int(xyy_{z}-y^{2}x_{z}-z^{2}x_{z}+zx)\mathbf{e}%
_{x}\nonumber \\
&  +(yz-z^{2}y_{z}-x^{2}y_{z}+xyx_{z})\mathbf{e}_{y}\nonumber \\
&  +(xzx_{z}-x^{2}-y^{2}+yzy_{z})\mathbf{e}_{z}dz\nonumber
\end{align}
where $\rho_{0}$ is the superfluid mass density. Along z-direction, the
projected (Lamb impulse) angular momentum of a vortex-line with a Kelvin wave
is given by
\begin{align}
J_{z}  &  =\mathbf{J}_{\mathrm{Lamb}}\cdot \mathbf{e}_{z}\\
&  =\frac{\rho_{0}\kappa}{2}\int(xzx_{z}-x^{2}-y^{2}+yzy_{z})dz\nonumber \\
&  =\frac{\rho_{0}\kappa}{2}\int(-a^{2}kz\cos(kz-\omega t)\sin(kz-\omega
t)\nonumber \\
&  -a^{2}+a^{2}kz\cos(kz-\omega t)\sin(kz-\omega t))dz\nonumber \\
&  =-\frac{1}{2}\rho_{0}\kappa la^{2}.\nonumber
\end{align}

\end{document}